\documentclass[a4paper]{article}

\usepackage{graphicx}
\usepackage{amsmath}
\usepackage{amssymb}

\newcommand{\nat}{I\!\!N}
\newcommand{\real}{I\!\!R}
\newcommand{\eop}{\sqcap\!\!\!\!\sqcup}
\newcommand{\bis}{\protect\underline{\leftrightarrow}}

\newcommand{\doublera}{\rightarrow\!\!\!\!\!\rightarrow}
\newcommand{\cho}{[]}
\newcommand{\rs}{\ {\sf rs}\ }
\newcommand{\sy}{\ {\sf sy}\ }

\newtheorem{definition}{Definition}[section]
\newtheorem{example}{Example}[section]
\newtheorem{proposition}{Proposition}[section]
\newtheorem{theorem}{Theorem}[section]

\date{}

\title{Discrete time stochastic and deterministic Petri box calculus\thanks{The work was partially supported by
Deutsche Forschungsgemeinschaft (DFG) under grant BE 1267/14-1}}

\author{{\sc Igor V. Tarasyuk}\\
A.P. Ershov Institute of Informatics Systems,\\
Siberian Branch of the Russian Academy of Sciences,\\
Acad. Lavrentiev pr. 6, 630090 Novosibirsk, Russian Federation\\
{\tt itar@iis.nsk.su}}

\begin{document}

\maketitle

\begin{abstract}
We propose an extension with deterministically timed multiactions of discrete time stochastic and immediate Petri box
calculus (dtsiPBC), previously presented by I.V. Tarasyuk, H. Maci\`a and V. Valero. In dtsdPBC, non-negative integers
specify multiactions with fixed (including zero) time delays. The step operational semantics is constructed via labeled
probabilistic transition systems. The denotational semantics is defined on the basis of a subclass of labeled discrete
time stochastic Petri nets with deterministic transitions. The consistency of both semantics is demonstrated. In order
to evaluate performance, the corresponding semi-Markov chains and (reduced) discrete time Markov chains are
analyzed.\bigskip
\end{abstract}

\section{Introduction}
\label{introduction.sec}

Algebraic process calculi like CSP \cite{Hoa85}, ACP \cite{BK85} and CCS \cite{Mil89} are well-known formal models for
specification of computing systems and analysis of their behaviour. In such process algebras (PAs), systems and
processes are specified by formulas, and verification of their properties is accomplished at a syntactic level via
equivalences, axioms and inference rules. In recent decades, stochastic extensions of PAs were proposed, such as MTIPP
\cite{HR94}, PEPA \cite{Hil96} and EMPA \cite{BGo98}. Unlike standard PAs, stochastic process algebras (SPAs) do not
just specify actions which can occur (qualitative features), but they associate with the actions the distribution
parameters of their random time delays (quantitative characteristics).

\subsection{Petri box calculus}

PAs specify concurrent systems in a compositional way via an expressive formal syntax. On the other hand, Petri nets
(PNs) provide a graphical representation of such systems and capture explicit asynchrony in their behaviour. To combine
the advantages of both models, a semantics of algebraic formulas
via PNs was defined.

Petri box calculus (PBC) \cite{BDH92,BKo95,BDK01} is a flexible and expressive process algebra developed as a tool for
specification of the PNs structure and their interrelations. Its goal was also to propose a compositional semantics for
high level constructs of concurrent programming languages in terms of elementary PNs. Formulas of PBC are combined not
from single (visible or invisible) actions and variables, like in CCS, but from multisets of elementary actions and
their conjugates, called multiactions ({\em basic formulas}). The empty multiset of actions is interpreted as the
silent multiaction specifying some invisible activity. In contrast to CCS, synchronization is separated from
parallelism ({\em concurrent constructs}). Synchronization is a unary multi-way stepwise operation, based on
communication of actions and their conjugates. This extends the CCS approach with conjugate matching labels.
Synchronization in PBC is asynchronous, unlike that in Synchronous CCS (SCCS) \cite{Mil89}. Other operations are
sequence and choice ({\em sequential constructs}). The calculus includes also restriction and relabeling ({\em
abstraction constructs}). To specify infinite processes, refinement, recursion and iteration operations were added
({\em hierarchical constructs}). Thus, unlike CCS, PBC has an additional iteration operation to specify infinite
behaviour when the semantic interpretation in finite PNs is possible. PBC has a step operational semantics in terms of
labeled transition systems, based on the rules of structural operational semantics (SOS).
The operational semantics of PBC is of step type, since its SOS rules have transitions with (multi)sets of activities,
corresponding to simultaneous executions of activities (steps). A denotational semantics of PBC was proposed via a
subclass of PNs equipped with an interface and considered up to isomorphism, called Petri boxes. For more detailed
comparison of PBC with other process algebras and the reasoning about importance of non-interleaving semantics see
\cite{BDH92,BDK01}.

The extensions of PBC with a deterministic, a nondeterministic or a stochastic model of time were presented.

\subsection{Time extensions of Petri box calculus}

To specify systems with time constraints,
deterministic (fixed) or nondeterministic (interval)
delays are used.

A time extension of PBC with a nondeterministic time model, called time Petri box calculus (tPBC), was proposed in
\cite{Kou00}. In tPBC, timing information is added by associating time intervals (the earliest and the latest firing
time) with instantaneous {\em actions}. Its denotational semantics was defined in terms of a subclass of labeled time
Petri nets (LtPNs), based on tPNs \cite{MFa76} and called time Petri boxes (ct-boxes). tPBC has a step time operational
semantics in terms of labeled transition systems.

Another time enrichment of PBC, called Timed Petri box calculus (TPBC), was defined in \cite{MF00,MF01}, it
accommodates a deterministic model of time. In contrast to tPBC, multiactions of TPBC are not instantaneous, but have
time durations. Additionally, in TPBC there exist no ``illegal'' multiaction occurrences, unlike tPBC. The complexity
of ``illegal'' occurrences mechanism was one of the main intentions to construct TPBC though this calculus appeared to
be more complicated than tPBC. The denotational semantics of TPBC was defined in terms of a subclass of labeled Timed
Petri nets (LTPNs), based on TPNs \cite{Ram73} and called Timed Petri boxes (T-boxes). TPBC has a step timed
operational semantics in terms of labeled transition systems. tPBC and TPBC differ in ways they capture time
information, and they are not in competition but complement each~other.

The third time extension of PBC, called arc time Petri box calculus (atPBC), was constructed in \cite{Nia05,NK05}, and
it implements a nondeterministic time. In atPBC, multiactions are associated with time delay intervals. Its
denotational semantics was defined on a subclass of labeled arc time Petri nets (atPNs), based of those from
\cite{BLT90,Hani93}, where time restrictions are associated with the arcs, called arc time Petri boxes (at-boxes).
atPBC possesses a step time operational semantics in terms of labeled transition systems.

tPBC, TPBC and atPBC, all
adopt the discrete time approach, but
TPBC has no immediate (multi)actions.

\subsection{Stochastic extensions of Petri box calculus}

The set of states for the systems with deterministic or nondeterministic delays often differs drastically from that for
the timeless systems, hence, the analysis results for untimed systems may be not valid for the time ones. To solve this
problem, stochastic delays are considered, which are the random variables with a (discrete or continuous) probability
distribution. If the random variables governing delays have an infinite support then the corresponding SPA can exhibit
all the same behaviour as its underlying untimed PA.

A stochastic extension of PBC, called stochastic Petri box calculus (sPBC), was proposed in \cite{MVF01,MVCC04}. In
sPBC, multiactions have stochastic delays that follow (negative) exponential distribution. Each multiaction is equipped
with a rate that is a parameter of the corresponding exponential distribution. The instantaneous execution of a
stochastic multiaction is possible only after the corresponding stochastic time delay. The calculus has an interleaving
operational semantics defined via transition systems labeled with multiactions and their rates. Its denotational
semantics was defined in terms of a subclass of labeled continuous time stochastic PNs, based on CTSPNs
\cite{Mar90,Bal01} and called stochastic Petri boxes (s-boxes). In sPBC, performance of the processes is evaluated by
analyzing their underlying continuous time Markov chains (CTMCs). In \cite{MVCF08}, a number of new equivalence
relations were proposed for regular terms of sPBC to choose later a suitable candidate for a congruence. sPBC was
enriched with immediate multiactions having zero delay in \cite{MVCR08}. We call such an sPBC extension generalized
sPBC or gsPBC. An interleaving operational semantics of gsPBC was constructed via transition systems labeled with
stochastic or immediate multiactions together with their rates or probabilities. A denotational semantics of gsPBC was
defined via a subclass of labeled generalized stochastic PNs, based on GSPNs \cite{Mar90,Bal01,Bal07} and called
generalized stochastic Petri boxes (gs-boxes). The performance analysis in gsPBC is based on the semi-Markov chains
(SMCs).

PBC has a step operational semantics, whereas sPBC has an interleaving one. In step semantics, parallel executions of
activities (steps) are permitted while in interleaving semantics, we can execute only single activities. Hence, a
stochastic extension of PBC with a step semantics was needed to keep the concurrency degree of behavioural analysis at
the same level as in PBC. As mentioned in \cite{Mol81,Mol85}, in contrast to continuous time approach (used in sPBC),
discrete time approach allows for constructing models of common clock systems and clocked devices. In such models,
multiple transition firings (or executions of multiple activities) at time moments (ticks of the central clock) are
possible, resulting in a step semantics. Moreover, employment of discrete stochastic time fills the gap between the
models with deterministic (fixed) time delays and those with continuous stochastic time delays. As argued in
\cite{AHR00}, arbitrary delay distributions are much easier to handle in a discrete time domain. In
\cite{MVi08,MVi09,MABV12}, discrete stochastic time was preferred to enable simultaneous expiration of multiple delays.

In \cite{Tar05,Tar06,Tar07,Tar14}, a discrete time stochastic extension dtsPBC of finite PBC was presented. In dtsPBC,
the residence time in the process states is geometrically distributed. A step operational semantics of dtsPBC was
constructed via labeled probabilistic transition systems. Its denotational semantics was defined in terms of a subclass
of labeled discrete time stochastic PNs (LDTSPNs), based on DTSPNs \cite{Mol81,Mol85} and called discrete time
stochastic Petri boxes (dts-boxes). The performance evaluation in dtsPBC is accomplished via the underlying discrete
time Markov chains (DTMCs) of the algebraic processes. Since dtsPBC has a discrete time semantics and geometrically
distributed sojourn time in the process states, unlike sPBC with continuous time semantics and exponentially
distributed delays, the calculi apply two different approaches to the stochastic extension of PBC, in spite of some
similarity of their syntax and semantics inherited from PBC. The main advantage of dtsPBC is that concurrency is
treated like in PBC having step semantics, whereas in sPBC parallelism is simulated by interleaving, obliging one to
collect the information on causal independence of activities before constructing the semantics.

In \cite{TMV10,TMV13,TMV14,TMV15,TMV18}, we presented an enhanced calculus dtsiPBC, an extension with immediate
multiactions of dtsPBC. Immediate multiactions increase the specification capability: they can model logical
conditions, probabilistic branching, instantaneous probabilistic choices and activities whose durations are negligible
in comparison with those of others. They are also used to specify urgent activities and the ones that are not relevant
for performance evaluation. Thus, immediate multiactions can be considered as a kind of instantaneous dynamic state
adjustment and, in many cases, they result in a simpler and more clear system representation.

\subsection{Our contributions}

In this paper, we present dtsiPBC, extended with deterministic multiactions, called {\em discrete time stochastic and
deterministic Petri box calculus} (dtsdPBC). In dtsdPBC, besides the probabilities from the real-valued interval
$(0;1)$ that are used to calculate discrete-time delays of stochastic multiactions, also non-negative integers are used
to specify fixed time delays of deterministic multiactions (including zero delay, which is the case of immediate
multiactions). To resolve conflicts among deterministic multiactions, they are additionally equipped with positive
real-valued weights. As argued in \cite{Zij95,Zij97}, a combination of deterministic and stochastic delays fits well to
model
technical systems with constant (fixed) durations of the regular non-random activities and probabilistically
distributed (stochastic) durations of the randomly occurring activities. The step operational semantics of dtsdPBC is
constructed with the use of labeled probabilistic transition systems. The denotational semantics of dtsdPBC is defined
in terms of a subclass of labeled discrete time stochastic and deterministic Petri nets (LDTSPNs with deterministic
transitions, LDTSDPNs), based on the extension of DTSPNs with transition labeling and deterministic transitions, called
dtsd-boxes.
The consistency of both semantics is demonstrated. The corresponding stochastic process, which is a semi-Markov chain
(SMC), is constructed and investigated, with the purpose of performance evaluation, which is the same for both
semantics. In addition, the alternative solution methods are developed, based on the underlying discrete time Markov
chain (DTMC) and its reduction (RDTMC) by eliminating vanishing states with zero sojourn (residence) times. The theory
developed is illustrated with a series of the interesting and non-trivial examples that include the travel system case
study, used to demonstrate application of the performance analysis methods within dtsdPBC.

Thus, the main contributions of the paper are the following.
\begin{itemize}

\item New powerful and expressive discrete time SPA with deterministic activities called dtsdPBC.

\item Step operational semantics of dtsdPBC in terms of labeled probabilistic transition systems.

\item Petri net denotational semantics of dtsdPBC via discrete time stochastic and deterministic Petri nets.

\item Performance analysis via underlying semi-Markov chains and (reduced) discrete time Markov chains.

\end{itemize}

\subsection{Structure of the paper}

The paper is organized as follows. In Section \ref{syntax.sec}, the syntax of the extended calculus dtsdPBC is
presented. In Section \ref{opersem.sec}, we construct the operational semantics of the algebra in terms of labeled
probabilistic transition systems. In Section \ref{denosem.sec}, we propose the denotational semantics based on a
subclass of LDTSDPNs. In Section \ref{perfeval.sec}, the corresponding stochastic process is defined and analyzed.
Finally, Section \ref{conclusion.sec} summarizes the results obtained and outlines research perspectives in this area.

\section{Syntax}
\label{syntax.sec}

In this section, we propose the syntax of dtsdPBC. First, we recall a definition of multiset that is an extension of
the set notion by allowing several identical elements.

\begin{definition}
Let $X$ be a set. A finite {\em multiset (bag)} $M$ over $X$ is a mapping $M:X\rightarrow\nat$ such that $|\{x\in X\mid
M(x)>0\}|<\infty$, i.e. it can contain a finite number of elements only.
\end{definition}

We denote the {\em set of all finite multisets} over a set $X$ by $\nat_{fin}^X$. Let $M,M'\in\nat_{fin}^X$. The {\em
cardinality} of $M$ is defined as $|M|=\sum_{x\in X}M(x)$. We write $x\in M$ if $M(x)>0$ and $M\subseteq M'$ if
$\forall x\in X\ M(x)\leq M'(x)$. We define $(M+M')(x)=M(x)+M'(x)$ and $(M-M')(x)=\max\{0,M(x)-M'(x)\}$. When $\forall
x\in X,\ M(x)\leq 1,\ M$ can be interpreted as a proper set and denoted by $M\subseteq X$. The {\em set of all subsets
(powerset)} of $X$ is denoted by $2^X$.

Let $Act=\{a,b,\ldots\}$ be the set of {\em elementary actions}. Then $\widehat{Act}=\{\hat{a},\hat{b},\ldots\}$ is the
set of {\em conjugated actions (conjugates)} such that $\hat{a}\neq a$ and $\hat{\hat{a}}=a$. Let ${\cal
A}=Act\cup\widehat{Act}$ be the set of {\em all actions}, and ${\cal L}=\nat_{fin}^{\cal A}$ be the set of {\em all
multiactions}. Note that $\emptyset\in{\cal L}$, this corresponds to an internal move, i.e. the execution of a
multiaction that contains no visible action names. The {\em alphabet} of $\alpha\in{\cal L}$ is defined as ${\cal
A}(\alpha )=\{x\in{\cal A}\mid\alpha (x)>0\}$.

A {\em stochastic multiaction} is a pair $(\alpha ,\rho )$, where $\alpha\in{\cal L}$ and $\rho\in (0;1)$ is the {\em
probability} of the multiaction $\alpha$. This probability is interpreted as that of independent execution of the
stochastic multiaction at the next discrete time moment. Such probabilities are used to calculate those to execute
(possibly empty) sets of stochastic multiactions after one time unit delay. The probabilities of stochastic
multiactions are required not to be equal to $1$ to avoid extra model complexity, since in this case one should assign
with them weights, needed to make a choice when several stochastic multiactions with probability $1$ can be executed
from a state. The difficulty is that when the stochastic multiactions with probability $1$ occur in a step (parallel
execution), all other with the less probabilities do not. In this case, the conflicts resolving requires a special
attention, as discussed in \cite{Mol81,Mol85} within SPNs. This decision also allows us to avoid technical difficulties
related to conditioning events with probability $0$.
The probability $1$ is left for (implicitly assigned to) waiting multiactions (positively delayed deterministic
multiactions, to be defined later), which are delayed for at least one time unit before their execution and have
weights to resolve conflicts with other waiting multiactions. On the other hand, there is no sense to allow probability
$0$ of stochastic multiactions, since they would never be performed in this case. Let ${\cal SL}$ be the set of {\em
all stochastic multiactions}.

A {\em deterministic multiaction} is a pair $(\alpha ,\natural_l^\theta )$, where $\alpha\in{\cal L},\ \theta\in\nat$
is the non-negative integer-valued {\em (fixed) delay} and $l\in\real_{>0}=(0;+\infty )$ is the positive real-valued
{\em weight} of the multiaction $\alpha$. This weight is interpreted as a measure of importance (urgency, interest) or
a bonus reward associated with execution of the deterministic multiaction at the current discrete time moment. Such
weights are used to calculate the probabilities to execute sets of deterministic multiactions at the same moment of
time. An {\em immediate multiaction} is a deterministic multiaction with the delay $0$ while a {\em waiting
multiaction} is a deterministic multiaction with a positive delay. In case of no conflicts among waiting multiactions,
whose remaining times to execute (RTEs, to be explained later in more detail) are equal to one time unit, they are
executed with probability $1$ at the next time moment. Deterministic multiactions have a priority over stochastic ones,
and there is also difference in priorities between immediate and waiting multiactions. One can assume that all
immediate multiactions have (the highest) priority $2$ and all waiting multiactions have (the intermediate) priority
$1$, whereas all stochastic multiactions have (the lowest) priority $0$. This means that in a state where all kinds of
multiactions can occur, immediate multiactions always occur before waiting ones that, in turn, are always executed
before stochastic ones. Different types of multiactions cannot participate together in some step (parallel execution),
i.e. just the steps consisting only of immediate multiactions or waiting ones, or those including only stochastic
multiactions, are allowed. Let ${\cal DL}$ be the set of {\em all deterministic multiactions}, ${\cal IL}$ be the set
of {\em all immediate multiactions} and ${\cal WL}$ be the set of {\em all waiting multiactions}. Obviously, we have
${\cal DL}={\cal IL}\cup{\cal WL}$.

Let us note that the same multiaction $\alpha\in{\cal L}$ may have different probabilities, (fixed) delays and weights
in the same specification. An {\em activity} is a stochastic or a deterministic multiaction. Let ${\cal SDL}={\cal
SL}\cup{\cal DL}={\cal SL}\cup{\cal IL}\cup{\cal WL}$ be the set of {\em all activities}. The {\em alphabet} of an
activity $(\alpha ,\kappa )\in{\cal SDL}$ is defined as ${\cal A}(\alpha ,\kappa )={\cal A}(\alpha )$. The {\em
alphabet} of a multiset of activities $\Upsilon\in\nat_{fin}^{\cal SDL}$ is defined as ${\cal A}(\Upsilon
)=\cup_{(\alpha ,\kappa )\in\Upsilon}{\cal A}(\alpha )$. For an activity $(\alpha ,\kappa )\in{\cal SDL}$, we define
its {\em multiaction part} as ${\cal L}(\alpha ,\kappa )=\alpha$ and its {\em probability} or {\em weight part} as
$\Omega (\alpha ,\kappa )=\kappa$ if $\kappa\in (0;1)$; or $\Omega (\alpha ,\kappa )=l$ if $\kappa =\natural_l^\theta,\
\theta\in\nat ,\ l\in\real_{>0}$. The {\em multiaction part} of a multiset of activities $\Upsilon\in\nat_{fin}^{\cal
SIL}$ is defined as ${\cal L}(\Upsilon )=\sum_{(\alpha ,\kappa )\in\Upsilon}\alpha$.

Activities are combined into formulas (process expressions) by the following operations: {\em sequential execution}
$;$, {\em choice} $\cho$, {\em parallelism} $\|$, {\em relabeling} $[f]$ of actions, {\em restriction} $\!\!\rs\!\!$
over a single action, {\em synchronization} $\!\!\sy\!\!$ on an action and its conjugate, and {\em iteration}
$[\,*\,*\,]$ with three arguments: initialization, body and termination.

Sequential execution and choice have a standard interpretation, like in other process algebras, but parallelism does
not include synchronization, unlike the corresponding operation in CCS \cite{Mil89}.

Relabeling functions $f:{\cal A}\rightarrow{\cal A}$ are bijections preserving conjugates, i.e. $\forall x\in{\cal A}\
f(\hat{x})=\widehat{f(x)}$. Relabeling is extended to multiactions in the usual way: for $\alpha\in{\cal L}$ we define
$f(\alpha )=\sum_{x\in\alpha}f(x)$. Relabeling is extended to activities: for $(\alpha ,\kappa )\in{\cal SDL}$, we
define $f(\alpha ,\kappa )=(f(\alpha ),\kappa )$. Relabeling is extended to the multisets of activities as follows: for
$\Upsilon\in\nat_{fin}^{\cal SDL}$ we define $f(\Upsilon )=\sum_{(\alpha ,\kappa )\in\Upsilon}(f(\alpha ),\kappa )$.

Restriction over an elementary action $a\in Act$ means that, for a given expression, any process behaviour containing
$a$ or its conjugate $\hat{a}$ is not allowed.

Let $\alpha ,\beta\in{\cal L}$ be two multiactions such that for some elementary action $a\in Act$ we have $a\in\alpha$
and $\hat{a}\in\beta$, or $\hat{a}\in\alpha$ and $a\in\beta$. Then, synchronization of $\alpha$ and $\beta$ by $a$ is
defined as $\alpha\oplus_a\beta =\gamma$, where

$$\gamma (x)=\left\{
\begin{array}{ll}
\alpha (x)+\beta (x)-1, & x=a\mbox{ or }x=\hat{a};\\
\alpha (x)+\beta (x), & \mbox{otherwise}.
\end{array}
\right.$$
In other words, we require that $\alpha\oplus_a\beta =\alpha +\beta -\{a,\hat{a}\}$, i.e. we remove one exemplar of $a$
and one exemplar of $\hat{a}$ from the multiset sum $\alpha +\beta$, since the synchronization of $a$ and $\hat{a}$
produces $\emptyset$. Activities are synchronized with the use of their multiaction parts, i.e. the synchronization by
$a$ of two activities, whose multiaction parts $\alpha$ and $\beta$ possess the properties mentioned above, results in
the activity with the multiaction part $\alpha\oplus_a\beta$. We may synchronize activities of the same type only:
either both stochastic multiactions or both immediate ones, since immediate multiactions have a priority over
stochastic ones, hence, stochastic and immediate multiactions cannot be executed together (note also that the execution
of immediate multiactions takes no time, unlike that of stochastic ones). Synchronization by $a$ means that, for a
given expression with a process behaviour containing two concurrent activities that can be synchronized by $a$, there
exists also the process behaviour that differs from the former only in that the two activities are replaced by the
result of their synchronization.

In the iteration, the initialization subprocess is executed first, then the body is performed zero or more times, and
finally, the termination subprocess is executed.

Static expressions specify the structure of processes. As we shall see, the expressions correspond to unmarked LDTSDPNs
(LDTSDPNs are marked by definition). Remember that a marking is the allocation of tokens in the places of a PN and
markings are used to describe dynamic behaviour of PNs in terms of transition firings.

We assume that every waiting multiaction has a countdown timer associated, whose value is the discrete time amount left
till the moment when the waiting multiaction can be executed. Therefore, besides standard (unstamped) waiting
multiactions in the form of $(\alpha ,\natural_l^\theta )\in{\cal WL}$, a special case of the {\em stamped} waiting
multiactions should be considered in the definition of static expressions. Each (time) stamped waiting multiaction in
the form of $(\alpha ,\natural_l^\theta )^\delta$ has an extra superscript $\delta\in\{1,\ldots ,\theta\}$ assigned
that specifies a time stamp indicating the {\em latest} value of the countdown timer associated with that multiaction.
The standard waiting multiactions have no time stamps, to demonstrate irrelevance of the timer values for them (for
example, their timers have not yet started or have already finished their operation). The notions of the alphabet,
multiaction part,
weight part for (the multisets of) stamped waiting multiactions are defined, respectively, like those for (the
multisets of) unstamped waiting multiactions.

By reasons of simplicity, we do not assign the timer value superscripts $\delta$ to immediate multiactions, which are a
special case of deterministic multiactions $(\alpha ,\natural_l^\theta )$ with the delay $\theta =0$ in the form of
$(\alpha ,\natural_l^0)$, since their timer values can only be equal to $0$. Analogously, the superscript $\delta$
might be omitted for the waiting multiactions $(\alpha ,\natural_l^\theta )$ with the delay $\theta =1$ in the form of
$(\alpha ,\natural_l^1)$, since the corresponding timer can only have a single value $1$. Nevertheless, to maintain
syntactic uniformity among waiting multiactions, we leave the timer value superscripts for those that are $1$-delayed.

%
%
\begin{definition}
Let $(\alpha ,\kappa )\in{\cal SDL},\ (\alpha ,\natural_l^\theta )\in{\cal WL},\ \delta\in\{1,\ldots ,\theta\}$ and
$a\in Act$. A {\em static expression} of dtsdPBC is defined as

$$E::=\ (\alpha ,\kappa )\mid (\alpha ,\natural_l^\theta )^\delta\mid E;E\mid E\cho E\mid E\| E\mid E[f]\mid E\rs a\mid
E\sy a\mid [E*E*E].$$

\end{definition}

Let $StatExpr$ denote the set of {\em all static expressions} of dtsdPBC.

To make the grammar above unambiguous, one can add parentheses in the productions with binary operations: $(E;E),\
(E\cho E),\ (E\| E)$. However, here and further we prefer the PBC approach and add them to resolve ambiguities only.

To avoid technical difficulties with the iteration operator, we should not allow any concurrency at the highest level
of the second argument of iteration. This is not a severe restriction though, since we can always prefix parallel
expressions by an activity with the empty multiaction part. Later on, in Example \ref{nrboxrg.exm}, we shall
demonstrate that relaxing the restriction can result in nets which are not safe. Alternatively, we can use a different,
safe, version of the iteration operator, but its net translation has six arguments. See also \cite{BDK01} for
discussion on this subject. Remember that a PN is {\em $n$-bounded} ($n\in\nat$) if for all its reachable (from the
initial marking by the sequences of transition firings) markings there are at most $n$ tokens in every place, and a PN
is {\em safe} if it is $1$-bounded.

%
\begin{definition}
Let $(\alpha ,\kappa )\in{\cal SDL},\ (\alpha ,\natural_l^\theta )\in{\cal WL},\ \delta\in\{1,\ldots ,\theta\}$ and
$a\in Act$. A {\em regular static expression} of dtsdPBC is defined as

$$\begin{array}{c}
E::=\ (\alpha ,\kappa )\mid (\alpha ,\natural_l^\theta )^\delta\mid E;E\mid E\cho E\mid E\| E\mid E[f]\mid E\rs a\mid
E\sy a\mid [E*D*E],\\
\mbox{where }D::=\ (\alpha ,\kappa )\mid (\alpha ,\natural_l^\theta )^\delta\mid D;E\mid D\cho D\mid D[f]\mid
D\rs a\mid D\sy a\mid [D*D*E].
\end{array}$$

\end{definition}

Let $RegStatExpr$ denote the set of {\em all regular static expressions} of dtsdPBC.

Let $E$ be a regular static expression. The {\em underlying timer-free regular static expression}
$\downharpoonleft\!\!E$ of $E$ is obtained by removing from it all timer value superscripts.

Further, the set of {\em all stochastic multiactions (from the syntax) of} $E$ is ${\cal SL}(E)=\{(\alpha ,\rho )\mid
(\alpha ,\rho )\mbox{ is a}\\
\mbox{subexpression of }E\}$. The set of {\em all immediate multiactions (from the syntax) of} $E$ is ${\cal
IL}(E)=\{(\alpha ,\natural_l^0)\mid (\alpha ,\natural_l^0)\mbox{ is a subexpression of }E\}$. The set of {\em all
waiting multiactions (from the syntax) of} $E$ is ${\cal WL}(E)=\{(\alpha ,\natural_l^\theta )\mid (\alpha
,\natural_l^\theta )\mbox{ or }(\alpha ,\natural_l^\theta )^\delta\mbox{ is a}\mbox{subexpression of }E\mbox{ for
}\delta\in\{1,\ldots ,\theta\}\}$. Thus, the set of {\em all deterministic multiactions (from the syntax) of} $E$ is
${\cal DL}(E)={\cal IL}(E)\cup{\cal WL}(E)$ and the set of {\em all activities (from the syntax) of} $E$ is ${\cal
SDL}(E)={\cal SL}(E)\cup{\cal DL}(E)={\cal SL}(E)\cup{\cal IL}(E)\cup{\cal WL}(E)$.

Dynamic expressions specify the states of processes. As we shall see, the expressions correspond to LDTSDPNs (which are
marked by default). Dynamic expressions are obtained from static ones, by annotating them with upper or lower bars
which specify the active components of the system at the current moment of time. The dynamic expression with upper bar
(the overlined one) $\overline{E}$ denotes the {\em initial}, and that with lower bar (the underlined one)
$\underline{E}$ denotes the {\em final} state of the process specified by a static expression $E$.

For every overlined stamped waiting multiaction in the form of $\overline{(\alpha ,\natural_l^\theta )^\delta}$, the
superscript $\delta\in\{1,\ldots ,\theta\}$ specifies the {\em current} value of the {\em running} countdown timer
associated with the waiting multiaction. That decreasing discrete timer is started with the {\em initial} value
$\theta$ (equal to the delay of the waiting multiaction) at the moment when the waiting multiaction becomes overlined.
Then such a newly overlined stamped waiting multiaction $\overline{(\alpha ,\natural_l^\theta )^\theta}$ may be seen
similar to the freshly overlined unstamped waiting multiaction $\overline{(\alpha ,\natural_l^\theta )}$. Such
similarity will be captured by the structural equivalence, to be defined later.

While the stamped waiting multiaction stays overlined with the specified process execution, the timer decrements by one
discrete time unit with each global time tick until the timer value becomes $1$. This fact indicates that one unit of
time remains till execution of that multiaction (the remaining time to execute, RTE, equals one) that should follow in
the next moment with probability $1$, in case the stamped waiting multiaction is still overlined and there are no
conflicting with it waiting multiactions, whose RTEs equal to one.

%
%
\begin{definition}
Let
$E\in StatExpr$ and $a\in Act$. A {\em dynamic expression} of dtsdPBC is defined as

$$G::=\
\overline{E}\mid\underline{E}\mid G;E\mid E;G\mid G\cho E\mid E\cho G\mid G\| G\mid G[f]\mid G\rs a\mid G\sy a\mid
[G*E*E]\mid [E*G*E]\mid [E*E*G].$$

\end{definition}

Let $DynExpr$ denote the set of {\em all dynamic expressions} of dtsdPBC.

Let $G$ be a dynamic expression. The {\em underlying static (line-free) expression} $\lfloor G\rfloor$ of $G$ is
obtained by removing from it all upper and lower bars. Note that if the underlying static expression of a dynamic one
is not regular, the corresponding LDTSDPN can be non-safe (though, it is $2$-bounded in the worst case \cite{BDK01}).

\begin{definition}
A dynamic expression $G$ is {\em regular} if its underlying static expression $\lfloor G\rfloor$ is regular.
\end{definition}

Let $RegDynExpr$ denote the set of {\em all regular dynamic expressions} of dtsdPBC.

Let $G$ be a regular dynamic expression. The {\em underlying timer-free regular dynamic expression}
$\downharpoonleft\!\!G$ of $G$ is obtained by removing from it all timer value superscripts.

Further, the set of {\em all stochastic multiactions (from the syntax) of} $G$ is ${\cal SL}(G)={\cal SL}(\lfloor
G\rfloor )$. The set of {\em all immediate multiactions (from the syntax) of} $G$ is ${\cal IL}(G)={\cal IL}(\lfloor
G\rfloor )$. The set of {\em all waiting multiactions (from the syntax) of} $G$ is ${\cal WL}(G)={\cal WL}(\lfloor
G\rfloor )$. Thus, the set of {\em all deterministic multiactions (from the syntax) of} $G$ is ${\cal DL}(G)={\cal
IL}(G)\cup{\cal WL}(G)$ and the set of {\em all activities (from the syntax) of} $G$ is ${\cal SDL}(G)={\cal
SL}(G)\cup{\cal DL}(G)={\cal SL}(G)\cup{\cal IL}(G)\cup{\cal WL}(G)$.

\section{Operational semantics}
\label{opersem.sec}

In this section, we define the step operational semantics in terms of labeled transition systems.

\subsection{Inaction rules}

The inaction rules for dynamic expressions describe their structural transformations in the form of
$G\Rightarrow\widetilde{G}$ which do not change the states of the specified processes. The goal of those syntactic
transformations is to obtain the well-structured resulting expressions called operative ones to which no inaction rules
can be further applied. As we shall see, the application of an inaction rule to a dynamic expression does not lead to
any discrete time tick or any transition firing in the corresponding LDTSDPN, hence, its current marking
stays unchanged.

Thus, an application of every inaction rule does not require any discrete time delay, i.e. the dynamic expression
transformation described by the rule is accomplished instantly.

In Table \ref{inactrulesdm1.tab}, we define inaction rules for regular dynamic expressions
being overlined and underlined static ones. In this table, $(\alpha ,\natural_l^\theta )\in{\cal WL},\
\delta\in\{1,\ldots ,\theta\},\ E,F,K\in RegStatExpr$ and $a\in Act$. The first inaction rule suggests that the timer
value of each newly overlined waiting multiaction is set to the delay of that waiting multiaction.

\begin{table}[h]
\caption{Inaction rules for overlined and underlined regular static expressions}
\label{inactrulesdm1.tab}
\begin{center}
$\small\begin{array}{|llll|}
\hline
\rule{0mm}{4mm}
\overline{(\alpha ,\natural_l^\theta )}\Rightarrow\overline{(\alpha ,\natural_l^\theta )^\theta} &
\overline{E;F}\Rightarrow\overline{E};F &
\underline{E};F\Rightarrow E;\overline{F} &
E;\underline{F}\Rightarrow\underline{E;F}\\[1mm]

\overline{E\cho F}\Rightarrow\overline{E}\cho F &
\overline{E\cho F}\Rightarrow E\cho\overline{F} &
\underline{E}\cho F\Rightarrow\underline{E\cho F} &
E\cho\underline{F}\Rightarrow\underline{E\cho F}\\[1mm]

\overline{E\| F}\Rightarrow\overline{E}\|\overline{F} &
\underline{E}\|\underline{F}\Rightarrow\underline{E\| F} &
\overline{E[f]}\Rightarrow\overline{E}[f] &
\underline{E}[f]\Rightarrow\underline{E[f]}\\[1mm]

\overline{E\rs a}\Rightarrow\overline{E}\rs a &
\underline{E}\rs a\Rightarrow\underline{E\rs a} &
\overline{E\sy a}\Rightarrow\overline{E}\sy a &
\underline{E}\sy a\Rightarrow\underline{E\sy a}\\[1mm]

\overline{[E*F*K]}\Rightarrow [\overline{E}*F*K] &
[\underline{E}*F*K]\Rightarrow [E*\overline{F}*K] &
[E*\underline{F}*K]\Rightarrow [E*\overline{F}*K] &
[E*\underline{F}*K]\Rightarrow [E*F*\overline{K}]\\[1mm]

[E*F*\underline{K}]\Rightarrow\underline{[E*F*K]} & & & \\[1mm]
\hline
\end{array}$
\end{center}
\end{table}

In Table \ref{inactrulesdm2.tab}, we introduce inaction rules for regular dynamic expressions in the arbitrary form. In
this table, $E,F\in RegStatExpr,\ G,H,\widetilde{G},\widetilde{H}\in RegDynExpr$ and $a\in Act$. By reason of brevity,
two distinct inaction rules with the same premises are collated in some cases, resulting in the inaction rules with
double conclusion.
\begin{table}[h]
\caption{Inaction rules for arbitrary regular dynamic expressions}
\label{inactrulesdm2.tab}
\begin{center}
$\small\begin{array}{|llll|}
\hline
\rule{0mm}{6.5mm}
\dfrac{G\Rightarrow\widetilde{G},\ \circ\in\{;,\cho\}}{G\circ E\Rightarrow\widetilde{G}\circ E,\ E\circ G\Rightarrow
E\circ\widetilde{G}} &
\dfrac{G\Rightarrow\widetilde{G}}{G\| H\Rightarrow\widetilde{G}\| H,\ H\| G\Rightarrow H\|\widetilde{G}} &
\dfrac{G\Rightarrow\widetilde{G}}{G[f]\Rightarrow\widetilde{G}[f]} &
\dfrac{G\Rightarrow\widetilde{G},\ \circ\in\{\!\!\rs\!\!,\!\!\sy\!\!\}}{G\circ a\Rightarrow\widetilde{G}\circ a}\\[4mm]

\dfrac{G\Rightarrow\widetilde{G}}{[G*E*F]\Rightarrow [\widetilde{G}*E*F]} &
\multicolumn{3}{l|}{\dfrac{G\Rightarrow\widetilde{G}}{[E*G*F]\Rightarrow [E*\widetilde{G}*F],\ [E*F*G]\Rightarrow
[E*F*\widetilde{G}]}}\\[4mm]
\hline
\end{array}$
\end{center}
\end{table}

\begin{example}
Let $E=(\{a\},\natural_1^3)\cho (\{b\},\frac{1}{3})$. The following inferences by the inaction rules are possible from
$\overline{E}$:

$$\begin{array}{ll}
\overline{(\{a\},\natural_1^3)\cho (\{b\},\frac{1}{3})}\Rightarrow
\overline{(\{a\},\natural_1^3)}\cho (\{b\},\frac{1}{3})\Rightarrow
\overline{(\{a\},\natural_1^3)^3}\cho (\{b\},\frac{1}{3}), &

\overline{(\{a\},\natural_1^3)\cho (\{b\},\frac{1}{3})}\Rightarrow
(\{a\},\natural_1^3)\cho\overline{(\{b\},\frac{1}{3})}.
\end{array}$$

\label{overlines.exm}
\end{example}

%
%
%
%
\begin{definition}
A regular dynamic expression $G$ is {\em operative} if no inaction rule can be applied to it.
\end{definition}

Let $OpRegDynExpr$ denote the set of {\em all operative regular dynamic expressions} of dtsdPBC.

Note that any dynamic expression can be always transformed into a (not necessarily unique) operative one by using the
inaction rules.

In the following, we consider regular expressions only and omit the word ``regular''.

\begin{definition}
The relation $\approx\ =(\Rightarrow\cup\Leftarrow )^*$ is a {\em structural equivalence} of dynamic expressions in
dtsdPBC. Thus, two dynamic expressions $G$ and $G'$ are {\em structurally equivalent}, denoted by $G\approx G'$, if
they can be reached from each other by applying the inaction rules in a forward or a backward direction.
\end{definition}

Let $X$ be some set. We denote the Cartesian product $X\times X$ by $X^2$. Let ${\cal E}\subseteq X^2$ be an
equivalence relation on $X$. Then the {\em equivalence class} (with respect to ${\cal E}$) of an element $x\in X$ is
defined by $[x]_{\cal E}=\{y\in X\mid (x,y)\in{\cal E}\}$. The equivalence ${\cal E}$ partitions $X$ into the {\em set
of equivalence classes} $X/_{\cal E}=\{[x]_{\cal E}\mid x\in X\}$.

Let $G$ be a dynamic expression. Then $[G]_\approx =\{H\mid G\approx H\}$ is the equivalence class of $G$ with respect
to the structural equivalence, called the (corresponding) {\em state}. Next, $G$ is an {\em initial} dynamic
expression, denoted by $init(G)$, if $\exists E\in RegStatExpr\ G\in [\overline{E}]_\approx$. Further, $G$ is a {\em
final} dynamic expression, denoted by $final(G)$, if $\exists E\in RegStatExpr\ G\in [\underline{E}]_\approx$.

Let $G$ be a dynamic expression and $s=[G]_\approx$. The set of {\em all enabled stochastic multiactions of $s$} is
$EnaSto(s)=\{(\alpha ,\rho )\in{\cal SL}\mid\exists H\in s\cap OpRegDynExpr\ \overline{(\alpha ,\rho )}\mbox{ is a
subexpression of }H\}$, i.e. it consists of all stochastic multiactions that, being overlined, are the subexpressions
of some operative dynamic expression from the state $s$. Analogously, the set of {\em all enabled immediate
multiactions of $s$} is $EnaImm(s)=\{(\alpha ,\natural_l^0)\in{\cal IL}\mid\exists H\in s\cap OpRegDynExpr\
\overline{(\alpha ,\natural_l^0)}\mbox{ is a subexpression of }H\}$.
The set of {\em all enabled waiting multiactions of $s$} is $EnaWait(s)=\{(\alpha ,\natural_l^\theta )\in{\cal
WL}\mid\exists H\in s\cap OpRegDynExpr\ \overline{(\alpha ,\natural_l^\theta )^\delta},\ \delta\in\{1,\ldots
,\theta\},\mbox{ is a subexpression of }H\}$, i.e. it consists of all waiting multiactions that, being superscribed
with the values of their timers and overlined, are the subexpressions of some operative dynamic expression from the
state $s$.
The set of {\em all newly enabled waiting multiactions of $s$} is $EnaWaitNew(s)=\{(\alpha ,\natural_l^\theta )\in{\cal
WL}\mid\exists H\in s\cap OpRegDynExpr\ \overline{(\alpha ,\natural_l^\theta
)^\theta}\mbox{ is a }\\
\mbox{subexpression of }H\}$, i.e. it consists of all waiting multiactions that, being superscribed with the initial
values of their timers (delays of those waiting multiactions) and overlined, are the subexpressions of some operative
dynamic expression from the state $s$.

Thus, the set of {\em all enabled deterministic multiactions of $s$} is $EnaDet(s)=EnaImm(s)\cup EnaWait(s)$ and the
set of {\em all enabled activities of $s$} is defined as $Ena(s)=EnaSto(s)\cup EnaDet(s)=EnaSto(s)\cup EnaImm(s)\cup
EnaWait(s)$. As we shall see, $Ena(s)=Ena([G]_\approx )$ is an algebraic analogue of the set of all transitions enabled
at the initial marking of the LDTSDPN corresponding to $G$.
Note that the activities, resulted from synchronization, are not present explicitly in the syntax of the dynamic
expressions. Nevertheless, their enabledness status can be recovered by observing that of the pair of synchronized
activities from the syntax (they both should be enabled for enabling their synchronized product), even if they are
affected by restriction after the synchronization. An activity is said to be affected by restriction, if it is within
the scope of a restriction operation with the argument action, such that it or its conjugate is contained in the
multiaction part of that activity.

\begin{definition}
An operative dynamic expression $G$ is {\em saturated (with the values of timers)}, if each enabled wai\-ting
multiaction of $[G]_\approx$, being (certainly) superscribed with the value of its timer and possibly overlined, is the
subexpression of $G$.
\end{definition}

Let $SaOpRegDynExpr$ denote the set of {\em all saturated operative dynamic expressions} of dtsdPBC.

\begin{proposition}
Any operative dynamic expression can be always transformed into the saturated one by using the inaction rules.
\label{saturation.pro}
\end{proposition}
{\em Proof.} Let $G$ be a dynamic expression, $(\alpha ,\natural_l^\theta )\in EnaWait([G]_\approx )$
and there exists $H\in [G]_\approx\cap OpRegDynExpr$ that contains a subexpression $\overline{(\alpha
,\natural_l^\theta )^\delta},\ \delta\in\{1,\ldots ,\theta -1\}$. Then all
operative dynamic expressions from $[G]_\approx\cap OpRegDynExpr$ contain a subexpression $\overline{(\alpha
,\natural_l^\theta )^\delta}$ or $(\alpha ,\natural_l^\theta )^\delta$, i.e. the (possibly overlined) enabled waiting
multiaction $(\alpha ,\natural_l^\theta )$ with the (non-initial) timer value superscript $\delta\leq\theta -1$. Note
that the timer value superscript $\delta$ is the same for all such structurally equivalent operative dynamic
expressions. Indeed, all inaction rules, besides the first one, do not change the values of timers, but those rules
just modify the overlines and underlines of dynamic expressions.
The first inaction rule just sets up the timer of each overlined waiting multiaction $\overline{(\alpha
,\natural_l^\theta )}$ with the initial value $\delta =\theta$, equal to the delay of that waiting multiaction, as
follows: $\overline{(\alpha ,\natural_l^\theta )^\theta}$. Then the remaining inaction rules can shift out the overline
of that enabled waiting multiaction before setting up its timer, which results in a non-overlined enabled waiting
multiaction without timer value superscript $(\alpha ,\natural_l^\theta )$. Thus, for
$(\alpha ,\natural_l^\theta )\in EnaWait([G]_\approx )$, it can happen that $\overline{(\alpha ,\natural_l^\theta
)^\theta}$ a subexpression of some $H\in [G]_\approx\cap OpRegDynExpr$ and $(\alpha ,\natural_l^\theta )$ is a
subexpression of a different $H'\in [G]_\approx\cap OpRegDynExpr$.

Let now $G$ be an operative dynamic expression that is not saturated. By the arguments above, the saturation can be
violated only if $G$ contains as a subexpression at least one newly enabled waiting multiaction $(\alpha
,\natural_l^\theta )$ of $[G]_\approx$
that is not superscribed with the timer value.
By the definition of the new-enabling, there exists $H\in [G]_\approx\cap OpRegDynExpr$ such that $\overline{(\alpha
,\natural_l^\theta )^\theta}$ is a subexpression of $H$.
%
Since $G\approx H$, there is a sequence of the inaction rules applications (in a forward or a backward direction) that
transforms $G$ into $H$. Then the reverse sequence transforms $H$ into $G$. Let us remove from that reverse sequence
the following backward application of the first inaction rule: $\overline{(\alpha ,\natural_l^\theta
)}\Leftarrow\overline{(\alpha ,\natural_l^\theta )^\theta}$. Then such a reduced reverse sequence will turn $H$ into
$G_1\in [G]_\approx\cap OpRegDynExpr$, obtained from $G$ by replacing $(\alpha ,\natural_l^\theta )$ with $(\alpha
,\natural_l^\theta )^\theta$.

Let us start from $G_1$ and apply the above procedure to the remaining not superscribed with the timer values newly
enabled waiting multiactions of $[G]_\approx$ (which are also those of such kind of $[G_1]_\approx$). After repeated
application of the mentioned procedure for all $n\geq 1$ non-superscribed newly enabled waiting multiactions of $G$, we
shall get from it the saturated operative dynamic expression $G_n=\widetilde{G}\in [G]_\approx\cap OpRegDynExpr$. Note
that the presented transformation of $G$ into $\widetilde{G}$ does not change the enabling, since it does not change
any overlines or underlines in the syntax of the traversed operative dynamic expressions, but only iteratively assigns
the timer value superscripts to all newly enabled waiting multiactions of $G$. Hence, $EnaWait([G]_\approx
)=EnaWait([G_1]_\approx )=\cdots =EnaWait([G_n]_\approx )=EnaWait([\widetilde{G}]_\approx )$. \hfill $\eop$

Thus, any dynamic expression can be always transformed into a (not necessarily unique) saturated operative one by using
the inaction rules.

\begin{example}
Let $E$ be from Example \ref{overlines.exm}. Consider the sequence of inaction rules, applied (in a forward or a
backward direction) in the following trans\-formation of a non-saturated $G\in [\overline{E}]_\approx\cap OpRegDynExpr$
with the non-superscribed with the timer value (unstamped) enabled waiting multiaction $(\{a\},\natural_1^3)$ into (a
saturated) $H\in [\overline{E}]_\approx\cap OpRegDynExpr$, in which $(\{a\},\natural_1^3)$ is stamped:

$$\begin{array}{c}
G=(\{a\},\natural_1^3)\cho\overline{(\{b\},\frac{1}{3})}\approx
\overline{(\{a\},\natural_1^3)\cho (\{b\},\frac{1}{3})}\approx
\overline{(\{a\},\natural_1^3)}\cho (\{b\},\frac{1}{3})\approx
\overline{(\{a\},\natural_1^3)^3}\cho (\{b\},\frac{1}{3})=H.
\end{array}$$

The reduced reverse sequence of inaction rules induces the following transformations of $H$ that result in a saturated
$G_1=\widetilde{G}\in [\overline{E}]_\approx\cap OpRegDynExpr$, in which $(\{a\},\natural_1^3)$ is stamped:

$$\begin{array}{c}
H=\overline{(\{a\},\natural_1^3)^3}\cho (\{b\},\frac{1}{3})\approx
\overline{(\{a\},\natural_1^3)^3\cho (\{b\},\frac{1}{3})}\approx
(\{a\},\natural_1^3)^3\cho\overline{(\{b\},\frac{1}{3})}=G_1=\widetilde{G}.
\end{array}$$

\label{saturation.exm}
\end{example}

Let $G$ be a saturated operative
dynamic expression. Then $\circlearrowleft\!G$ is written for the {\em timer decrement} operator $\circlearrowleft$,
applied to $G$. It denotes a saturated operative
dynamic expression, obtained from $G$ via decrementing by one time unit all greater than $1$ values of the timers
associated with all (if any)
stamped waiting multiactions from the syntax of $G$.
Thus, each such
stamped waiting multiaction changes its timer value from $\delta$ in $G$ to $\max\{1,\delta -1\}$ in
$\circlearrowleft\!G$, where $\delta\in\nat_{\geq 1}$. More formally,
the timer decrement operator affects the (possibly overlined)
stamped waiting multiactions being the subexpressions of $G$ as follows. The overlined
stamped waiting multiaction
$\overline{(\alpha ,\natural_l^\theta )^\delta}$
is replaced with $\overline{(\alpha ,\natural_l^\theta )^{\max\{1,\delta -1\}}}$ while
the
stamped waiting multiaction without overline or underline $(\alpha ,\natural_l^\theta )^\delta$
is replaced with $(\alpha ,\natural_l^\theta )^{\max\{1,\delta -1\}}$.

Note that when $\delta =1$, we have $\max\{1,\delta -1\}=\max\{1,0\}=1$, hence, the timer value $\delta =1$ may remain
unchanged for a
stamped
waiting multiaction that is not executed by some reason at the next time moment, but stays
stamped.
For example, that
stamped
waiting multiaction may be affected by restriction.
If the timer values cannot be decremented with a time tick for all
stamped waiting multiactions (if any) from $G$ then $\circlearrowleft\!G=G$ and we obtain so-called {\em empty loop}
transition that will be formally defined later.

Observe that the timer decrement operator keeps
stamping of the waiting multiactions, since it does not change any overlines or underlines, but it may only decrease
their timer values, so that the stamped enabled waiting multiactions stay stamped (with their timer values, possibly
decremented by one).

\begin{example}
Let $E$ be from Example \ref{overlines.exm}. We have $Ena([\overline{E}]_\approx
)=\{(\{a\},\natural_1^3),(\{b\},\frac{1}{3})\}$ and $Ena([\overline{E}]_\approx )\cap{\cal
WL}=\{(\{a\},\natural_1^3)\}$. The following one time unit timer decrements are possible from the saturated operative
dynamic expressions belonging to $[\overline{E}]_\approx$:

$$\begin{array}{ll}
\circlearrowleft\!(\overline{(\{a\},\natural_1^3)^3}\cho (\{b\},\frac{1}{3}))=
\overline{(\{a\},\natural_1^3)^2}\cho (\{b\},\frac{1}{3}), &

\circlearrowleft\!((\{a\},\natural_1^3)^3\cho\overline{(\{b\},\frac{1}{3})})=
(\{a\},\natural_1^3)^2\cho\overline{(\{b\},\frac{1}{3})}.
\end{array}$$

\label{decrements.exm}
\end{example}

Note that, similar in \cite{Kou00}, we are mainly interested in the dynamic expressions, derived from the overlined
static expressions, such that no stamped waiting multiaction is a subexpression of them, to ensure that time proceeds
uniformly. Therefore, we consider a dynamic expression $G=\overline{(\{a\},\natural_1^2)^1}\cho (\{b\},\natural_2^3)^1$
as ``illegal'' and that $H=\overline{(\{a\},\natural_1^2)^1}\cho (\{b\},\natural_2^3)^2$ as ``legal'', since the latter
is obtained from the overlined static expression without timer value superscripts
$\overline{E}=\overline{(\{a\},\natural_1^2)\cho (\{b\},\natural_2^3)}$ after one time tick. On the other hand, $G$ is
``illegal'' only when it is intended to specify a complete process, but it may become ``legal'' as a part of such
complete specification, like $G\rs a$, since after two time ticks from $\overline{E\rs a}$, the timer values cannot be
decreased further when approaching the value $1$. Thus, we should allow the dynamic expressions like $G$, by assuming
that they are incomplete specifications, to be further composed.

Let $G$ be a dynamic expression. Then $I_G:{\cal WL}(G)\rightarrow\nat_{\geq 1}$ is the {\em timer valuation function}
of the waiting multiactions of $G$, defined as follows. For $(\alpha ,\natural_l^\theta )\in{\cal WL}(G)$, let
$I_G((\alpha ,\natural_l^\theta ))=\delta\in\{1,\ldots\theta\}$, if $(\alpha ,\natural_l^\theta )\in
EnaWait([G]_\approx )$ and $\exists H\in [G]_\approx\cap SatOpRegDynExpr\ \overline{(\alpha ,\natural_l^\theta
)^\delta}$ or $(\alpha ,\natural_l^\theta )^\delta$ is a subexpression of $H$. We let $I_G(t)=*$, if $(\alpha
,\natural_l^\theta )\not\in EnaWait([G]_\approx )$,
where `$*$' denotes the undefined value. The defi\-nition is correct by the argumentation from the proof of Proposition
\ref{saturation.pro}. Indeed, for each
waiting multiaction of $G$, its timer value superscript (if any) is the same for every $H\in [G]_\approx\cap
SatOpRegDynExpr$, in which that waiting multiaction, possibly being superscribed with the value of its timer and
overlined or underlined, is a subexpression.

Let $G\in SatOpRegDynExpr$. Then for all $(\alpha ,\natural_l^\theta )\in{\cal WL}(G)$, we have
$V_{\circlearrowleft\!G}((\alpha ,\natural_l^\theta ))=\max\{1,I_G((\alpha ,\natural_l^\theta ))-1\}$.

\subsection{Action and empty move rules}

The action rules are applied when some activities are executed. With these rules we capture the prioritization among
different types of multiactions. We also have the empty move rule which is used to capture a delay of one discrete time
unit
when no immediate or waiting multiactions are executable. In this case, the empty multiset of activities is executed.
The action and empty move rules will be used later to determine all multisets of activities which can be executed from
the structural equivalence class of every dynamic expression (i.e. from the state of the corresponding process). This
information together with that about probabilities or delays and weights of the activities to be executed from the
current process state will be used to calculate the probabilities of such executions.

The action rules with stochastic (immediate or waiting, respectively) multiactions describe dynamic expression
transformations in the form of $G\stackrel{\Gamma}{\rightarrow}\widetilde{G}$
($G\stackrel{I}{\rightarrow}\widetilde{G}$ or $G\stackrel{W}{\rightarrow}\widetilde{G}$, respectively) due to execution
of non-empty multisets $\Gamma$ of stochastic ($I$ of immediate or $W$ of waiting, respectively) multiactions. The
rules represent possible state changes of the specified processes when some non-empty multisets of stochastic
(immediate or waiting, respectively) multiactions are executed. As we shall see, the application of an action rule with
stochastic (immediate or waiting, respectively) multiactions to a dynamic expression leads in the corresponding LDTSDPN
to a discrete time tick at which some stochastic or waiting transitions fire (or to the instantaneous firing of some
immediate transitions) and possible change of the current marking. The current marking
stays unchanged only if there is a self-loop produced by the iterative execution of a non-empty multiset, which must be
one-element, i.e. a single stochastic (immediate or waiting, respectively) multiaction. The reason is the regularity
requirement that allows no concurrency at the highest level of the second argument of iteration.

The empty move rule (applicable only when no immediate or waiting multiactions can be executed from the current state)
describes dynamic expression transformations in the form of $G\stackrel{\emptyset}{\rightarrow}\circlearrowleft\!G$, called
the {\em empty moves}, due to execution of the empty multiset of activities at a discrete time tick.
When no timer values are decremented within $G$ with the empty multiset execution at the next moment (for example, if
$G$ contains no
enabled waiting multiactions), we have $\circlearrowleft\!G=G$. In such a case, the empty move from $G$ is in the form of
$G\stackrel{\emptyset}{\rightarrow}G$, called the {\em empty loop}. As we shall see, the application of the empty move
rule to a dynamic expression leads to a discrete time tick in the corresponding LDTSDPN at which no transitions fire
and the current marking is not changed, but the timer values of the waiting transitions enabled at the marking (if any)
are decremented by one. This is a new rule that has no prototype among inaction rules of PBC, since it represents a
time delay, but no notion of time exists in PBC. The PBC rule $G\stackrel{\emptyset}{\rightarrow}G$ from
\cite{BKo95,BDK01} in our setting would correspond to the rule $G\Rightarrow G$ that describes staying in the current
state when no time elapses. Since we do not need the latter rule to transform dynamic expressions into operative ones
and it can even destroy the definition of operative expressions, we do not introduce it in dtsdPBC.

Thus, an application of every action rule with stochastic or waiting multiactions or the empty move rule requires one
discrete time unit delay, i.e. the execution of a (possibly empty) multiset of stochastic or (non-empty) multiset of
waiting multiactions leading to the dynamic expression transformation described by the rule is accomplished instantly
after one time unit. An application of every action rule with immediate multiactions does not take any time, i.e. the
execution of a (non-empty) multiset of immediate multiactions is accomplished instantly at the current moment of time.

Note that expressions of dtsdPBC can contain identical activities. To avoid technical difficulties, such as the proper
calculation of the state change probabilities for multiple transitions, we can always enumerate coinciding activities
from left to right in the syntax of expressions. The new activities, resulted from synchronization will be annotated
with concatenation of numberings of the activities they come from, hence, the numbering should have a tree structure to
reflect the effect of multiple synchronizations. We now define the numbering which encodes a binary tree with the
leaves labeled by natural numbers.

\begin{definition}
The {\em numbering} of expressions is defined as $\iota ::=\ n\mid (\iota )(\iota )$, where $n\in\nat$.
\end{definition}

Let $Num$ denote the set of {\em all numberings} of expressions.

\begin{figure}
\begin{center}
\includegraphics{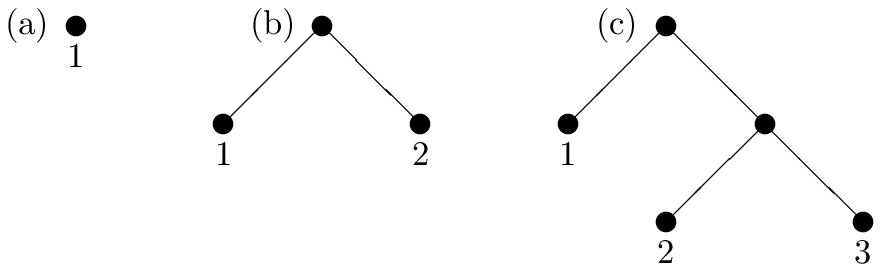}
\end{center}
\caption{The binary trees encoded with the numberings $1,\ (1)(2)$ and $(1)((2)(3))$}
\label{bintrnum.fig}
\end{figure}

\begin{example}
The numbering $1$ encodes the binary tree depicted in Figure \ref{bintrnum.fig}(a) with the root labeled by $1$. The
numbering $(1)(2)$ corresponds to the binary tree depicted in Figure \ref{bintrnum.fig}(b) without internal nodes and
with two leaves labeled by $1$ and $2$. The numbering $(1)((2)(3))$ represents the binary tree depicted in Figure
\ref{bintrnum.fig}(c) with one internal node, which is the root for the subtree $(2)(3)$, and three leaves labeled by
$1,2$ and $3$.
\end{example}

The new activities resulting from synchronizations in different orders should be considered up to permutation of their
numbering. In this way, we shall recognize different instances of the same activity. If we compare the contents of
different numberings, i.e. the sets of natural numbers in them, we shall be able to identify the mentioned instances.

The {\em content} of a numbering $\iota\in Num$ is

$$Cont(\iota )=\left\{
\begin{array}{ll}
\{\iota\}, & \iota\in\nat ;\\
Cont(\iota_1)\cup Cont(\iota_2), & \iota =(\iota_1)(\iota_2).
\end{array}
\right.$$

After the enumeration, the multisets of activities from the expressions will become the proper sets. In the following,
we suppose that the identical activities are enumerated when needed to avoid ambiguity.
This enumeration is considered to be implicit.

%

\begin{definition}
Let $G\in OpRegDynExpr$. We now define the {\em set of all non-empty multisets of activities which can be potentially
executed from $G$}, denoted by $Can(G)$. Let $(\alpha ,\kappa )\in{\cal SDL},\ E,F\in RegStatExpr,\ H\in OpRegDynExpr$
and $a\in Act$.
\begin{enumerate}

\item If $final(G)$ then $Can(G)=\emptyset$.

\item If $G=\overline{(\alpha ,\kappa )^\delta}$ and $\kappa =\natural_l^\theta ,\ \theta\in\nat_{\geq 2},\
l\in\real_{>0},\ \delta\in\{2,\ldots ,\theta\}$, then $Can(G)=\emptyset$.

\item If $G=\overline{(\alpha ,\kappa )}$ and $\kappa\in (0;1)$ or $\kappa =\natural_l^0 ,\ l\in\real_{>0}$, then
$Can(G)=\{\{(\alpha ,\kappa )\}\}$.

\item If $G=\overline{(\alpha ,\kappa )^1}$ and $\kappa =\natural_l^\theta ,\ \theta\in\nat_{\geq 1},\ l\in\real_{>0}$,
then $Can(G)=\{\{(\alpha ,\kappa )\}\}$.

\item If $\Upsilon\in Can(G)$ then $\Upsilon\in Can(G\circ E),\ \Upsilon\in Can(E\circ G)\ (\circ\in\{;,\cho\}),\
\Upsilon\in Can(G\| H),\ \Upsilon\in Can(H\| G),\\
f(\Upsilon )\in Can(G[f]),\ \Upsilon\in Can(G\rs a)\ (\mbox{when }a,\hat{a}\not\in{\cal A}(\Upsilon )),\
\Upsilon\in Can(G\sy a),\ \Upsilon\in Can([G*E*F]),\\
\Upsilon\in Can([E*G*F]),\ \Upsilon\in Can([E*F*G])$.

\item If $\Upsilon\in Can(G)$ and $\Xi\in Can(H)$ then $\Upsilon +\Xi\in Can(G\| H)$.

\item If $\Upsilon\in Can(G\sy a)$ and $(\alpha ,\kappa ),(\beta ,\lambda )\in\Upsilon$ are different activities such
that $a\in\alpha ,\ \hat{a}\in\beta$, then
\begin{enumerate}

\item $(\Upsilon +\{(\alpha\oplus_a\beta ,\kappa\cdot\lambda )\})\setminus\{(\alpha ,\kappa ),(\beta ,\lambda )\}\in
Can(G\sy a)$ if $\kappa ,\lambda\in (0;1)$;

\item $(\Upsilon +\{(\alpha\oplus_a\beta ,\natural_{l+m}^\theta )\})\setminus\{(\alpha ,\kappa ),(\beta ,\lambda )\}\in
Can(G\sy a)$ if $\kappa =\natural_l^\theta ,\ \lambda =\natural_m^\theta ,\ \theta\in\nat ,\ l,m\in\real_{>0}$.

When we synchronize the same multiset of activities in different orders, we obtain several activities with the
same multiaction and probability or delay and weight parts, but with different numberings having the same
content. Then we only consider a single one of the resulting activities to avoid introducing redundant ones.

For example, the synchronization of stochastic multiactions $(\alpha ,\rho )_1$ and $(\beta ,\chi )_2$ in
different orders generates the activities $(\alpha\oplus_a\beta ,\rho\cdot\chi )_{(1)(2)}$ and
$(\beta\oplus_a\alpha ,\chi\cdot\rho )_{(2)(1)}$. Similarly, the synchronization of deterministic multiactions
$(\alpha ,\natural_l^\theta )_1$ and $(\beta ,\natural_m^\theta )_2$ in different orders generates the
activities $(\alpha\oplus_a\beta ,\natural_{l+m}^\theta )_{(1)(2)}$ and $(\beta\oplus_a\alpha
,\natural_{m+l}^\theta )_{(2)(1)}$. Since $Cont((1)(2))=\{1,2\}=Cont((2)(1))$, in both cases, only the first
activity (or, symmetrically, the second one) resulting from synchronization will appear in a multiset from
$Can(G\sy a)$.

%
\end{enumerate}
\end{enumerate}
\end{definition}

Note that if $\Upsilon\in Can(G)$ then by definition of $Can(G),\ \forall\Xi\subseteq\Upsilon ,\ \Xi\neq\emptyset$, we
have $\Xi\in Can(G)$.

Let $G\in OpRegDynExpr$ and $Can(G)\neq\emptyset$. Obviously, if there are only stochastic (immediate or waiting,
respectively) multiactions in the multisets from $Can(G)$ then these stochastic (immediate or waiting, respectively)
multiactions can be executed from $G$. Otherwise, besides stochastic ones, there are also deterministic (immediate
and/or waiting) multiactions in the multisets from $Can(G)$. By the note above, there are non-empty multisets of
deterministic multiactions in $Can(G)$ as well, i.e. $\exists\Upsilon\in Can(G)\ \Upsilon\in\nat_{fin}^{\cal
DL}\setminus\{\emptyset\}$. In this case, no stochastic can be executed from $G$, even if $Can(G)$ contains non-empty
multisets of stochastic multiactions, since deterministic multiactions have a priority over stochastic ones, and should
be executed first. Further, if there are no stochastic, but both waiting and immediate multiactions in the multisets
from $Can(G)$, then, analogously, no waiting multiactions can be executed from $G$, since immediate multiactions have a
priority over waiting ones (besides that over stochastic ones).

When there are only waiting and, possibly, stochastic multiactions in the multisets from $Can(G)$ then, from above,
only waiting ones can be executed from $G$. Then just {\em maximal} non-empty multisets of waiting multiactions can be
executed from $G$, since all non-conflicting waiting multiactions cannot wait anymore and they should occur at the next
time moment with probability $1$. The next definition formalizes these requirements.

\begin{definition}
Let $G\in OpRegDynExpr$. The {\em set of all non-empty multisets of activities which can be exe\-cuted from $G$} is

$$Now(G)=
\left\{
\begin{array}{ll}
Can(G)\cap\nat_{fin}^{\cal IL}, & Can(G)\cap\nat_{fin}^{\cal IL}\neq\emptyset ;\\
\{W\in Can(G)\cap\nat_{fin}^{\cal WL}\mid\forall V\in Can(G)\cap\nat_{fin}^{\cal WL}\ W\subseteq V\ \Rightarrow\
V\!=\!W\}, &
(Can(G)\cap\nat_{fin}^{\cal IL}=\emptyset )\wedge\\
 & (Can(G)\cap\nat_{fin}^{\cal WL}\neq\emptyset );\\
Can(G), & \mbox{otherwise}.
\end{array}
\right.$$

\end{definition}

Consider an operative dynamic expression $G\in OpRegDynExpr$. The expression $G$ is {\em s-tangible (stochastically
tangible)}, denoted by $stang(G)$, if $Now(G)\subseteq\nat_{fin}^{\cal SL}\setminus\{\emptyset\}$. In particular, we
have $stang(G)$, if $Now(G)=\emptyset$. The expression $G$ is {\em w-tangible (waitingly tangible)}, denoted by
$wtang(G)$, if $\emptyset\neq Now(G)\subseteq\nat_{fin}^{\cal WL}\setminus\{\emptyset\}$. The expression $G$ is {\em
tangible}, denoted by $tang(G)$, if $stang(G)$ or $wtang(G)$, i.e. $Now(G)\subseteq (\nat_{fin}^{\cal
SL}\cup\nat_{fin}^{\cal WL})\setminus\{\emptyset\}$. Again, we particularly have $tang(G)$, if $Now(G)=\emptyset$.
Otherwise, the expression $G$ is {\em vanishing}, denoted by $vanish(G)$, and in this case $\emptyset\neq
Now(G)\subseteq\nat_{fin}^{\cal IL}\setminus\{\emptyset\}$.

For a (possibly non-operative) dynamic expression $G\in RegDynExpr$, we write $stang(G)$ ($wtang(G)$ or $vanish(G)$,
respectively), if $\forall H\in [G]_\approx\cap OpRegDynExpr$, we have $stang(H)$ ($wtang(H)$ or $vanish(H)$,
respectively). We write $tang(G)$, if $stang(G)$ or $wtang(G)$. Note that the operative dynamic expressions from
$[G]_\approx$ may have different types in general. The following example demonstrates two operative dynamic expressions
$H$ and $H'$ with $H\approx H'$, such that $vanish(H)$, but $stang(H')$.

\begin{example}
Let $G=(\overline{(\{a\},\natural_1^0)}\cho (\{b\},\natural_2^0))\|\overline{(\{c\},\frac{1}{2})}$ and
$G'=((\{a\},\natural_1^0)\cho\overline{(\{b\},\natural_2^0)})\|\overline{(\{c\},\frac{1}{2})}$. Then $G\approx G'$,\\
since $G\Leftarrow G''\Rightarrow G'$ for $G''=\overline{((\{a\},\natural_1^0)\cho (\{b\},\natural_2^0))}\|
\overline{(\{c\},\frac{1}{2})}$, but $Can(G)=\{\{(\{a\},\natural_1^0)\},\{(\{c\},\frac{1}{2})\},\{(\{a\},\natural_1^0),\\
(\{c\},\frac{1}{2})\}\},\ Can(G')=\{\{(\{b\},\natural_2^0)\},\{(\{c\},\frac{1}{2})\},\{(\{b\},\natural_2^0),
(\{c\},\frac{1}{2})\}\}$ and $Now(G)=\{\{(\{a\},\natural_1^0)\}\},\ Now(G')=\{\{(\{b\},\natural_2^0)\}\}$. Clearly, we
have $vanish(G)$ and $vanish(G')$. The executions like that of $\{(\{c\},\frac{1}{2})\}$ (and all multisets including
it) from $G$ and $G'$ must be disabled using preconditions in the action rules, since immediate multiactions have a
priority over stochastic ones, hence, the former are always executed first.

Let $H=\overline{(\{a\},\natural_1^0)}\cho (\{b\},\frac{1}{2})$ and
$H'=(\{a\},\natural_1^0)\cho\overline{(\{b\},\frac{1}{2})}$. Then $H\approx H'$, since $H\Leftarrow H''\Rightarrow H'$
for $H''=\overline{(\{a\},\natural_1^0)\cho (\{b\},\frac{1}{2})}$, but $Can(H)=Now(H)=\{\{(\{a\},\natural_1^0)\}\}$ and
$Can(H')=Now(H')=\{\{(\{b\},\frac{1}{2})\}\}$. We have $vanish(H)$, but $stang(H')$. To make the action rules correct
under structural equivalence, the executions like that of $\{(\{b\},\frac{1}{2})\}$ from $H'$ must be disabled using
preconditions in the action rules, since immediate multiactions have a priority over stochastic ones, hence, the
choices between them are always resolved in favour of the former.
\label{cannow.exm}
\end{example}

In Table \ref{actrulesdm.tab}, we define the action and empty move rules. In the table, $(\alpha ,\rho ),(\beta ,\chi
)\!\in{\cal SL},\ (\alpha ,\natural_l^0),(\beta ,\natural_m^0)\in{\cal IL}$ and $(\alpha ,\natural_l^\theta ),(\beta
,\natural_m^\theta )\in{\cal WL}$
Further, $E,F,K\in RegStatExpr,\ G,H\in SatOpRegDynExpr,\ \widetilde{G},\widetilde{H}\!\in\!RegDynExpr$ and $a\in Act$.
Moreover, $\Gamma ,\Delta\in\nat_{fin}^{\cal SL}\setminus\{\emptyset\},\ \Gamma '\in\nat_{fin}^{\cal SL},\
I,J\in\nat_{fin}^{\cal IL}\setminus\{\emptyset\},\ I'\in\nat_{fin}^{\cal IL},\ V,W\in\nat_{fin}^{\cal
WL}\setminus\{\emptyset\},\ V'\in\nat_{fin}^{\cal WL}$ and $\Upsilon\in\nat_{fin}^{\cal SDL}\setminus\{\emptyset\}$.

We use the following abbreviations in the names of the rules from the table: ``{\bf E}'' for ``{\bf E}mpty move'',
``{\bf B}'' for ``{\bf B}asis case'', ``{\bf S}'' for ``{\bf S}equence'', ``{\bf C}'' for ``{\bf C}hoice'', ``{\bf P}''
for ``{\bf P}arallel'', ``{\bf L}'' for ``re{\bf L}abeling'', ``{\bf R}'' for ``{\bf R}estriction'', ``{\bf I}'' for
``{\bf I}teraton'' and ``{\bf Sy}'' for ``{\bf Sy}nchronization''. The first rule in the table is the empty move rule
{\bf E}. The other rules are the action rules, describing transformations of dynamic expressions, which are built using
particular algebraic operations. If we cannot merge the rules with stochastic, immediate ans waiting multiactions in
one rule for some operation then we get the coupled action rules. In such cases, the names of the action rules with
stochastic multiactions have a suffix `{\bf s}', those with immediate multiactions have a suffix `{\bf i}', and those
with waiting multiactions have a suffix `{\bf w}'.
To make presentation more compact, the action rules with double conclusion are combined from two distinct action rules
with the same premises.

\begin{table}[h]
\caption{Action and empty move rules}
\label{actrulesdm.tab}
\begin{center}
$\small\begin{array}{|ll|}
\hline
\multicolumn{2}{|l|}{\rule{0mm}{5.5mm}
{\bf E}\ \dfrac{stang(G)}{G\stackrel{\emptyset}{\rightarrow}\circlearrowleft\!G}\hspace{15mm}
{\bf Bs}\ \overline{(\alpha ,\rho )}\stackrel{\{(\alpha ,\rho )\}}{\longrightarrow}
\underline{(\alpha ,\rho )}\hspace{15mm}
{\bf Bi}\ \overline{(\alpha ,\natural_l^0)}\stackrel{\{(\alpha ,\natural_l^0)\}}{\longrightarrow}
\underline{(\alpha ,\natural_l^0)}\hspace{15mm}
{\bf Bw}\ \overline{(\alpha ,\natural_l^\theta )^1}\stackrel{\{(\alpha ,\natural_l^\theta )\}}{\longrightarrow}
\underline{(\alpha ,\natural_l^\theta )}}\\[3mm]

{\bf S}\ \dfrac{G\stackrel{\Upsilon}{\rightarrow}\widetilde{G}}{G;E\stackrel{\Upsilon}{\rightarrow}\widetilde{G};E,\
E;G\stackrel{\Upsilon}{\rightarrow}E;\widetilde{G}} &
{\bf Cs}\ \dfrac{G\stackrel{\Gamma}{\rightarrow}\widetilde{G},\ \neg init(G)\vee
(init(G)\wedge stang(\overline{E}))}
{G\cho E\stackrel{\Gamma}{\rightarrow}\widetilde{G}\cho\!\downharpoonleft\!\!E,\
E\cho G\stackrel{\Gamma}{\rightarrow}\downharpoonleft\!\!E\cho\widetilde{G}}\\[3mm]

{\bf Ci}\ \dfrac{G\stackrel{I}{\rightarrow}\widetilde{G}}{G\cho E\stackrel{I}{\rightarrow}
\widetilde{G}\cho\!\downharpoonleft\!\!E,\ E\cho G\stackrel{I}{\rightarrow}\downharpoonleft\!\!E\cho\widetilde{G}} &
{\bf Cw}\ \dfrac{G\stackrel{V}{\rightarrow}\widetilde{G},\ \neg init(G)\vee
(init(G)\wedge tang(\overline{E}))}
{G\cho E\stackrel{V}{\rightarrow}\widetilde{G}\cho\!\downharpoonleft\!\!E,\
E\cho G\stackrel{V}{\rightarrow}\downharpoonleft\!\!E\cho\widetilde{G}}\\[3mm]

{\bf P1s}\ \dfrac{G\stackrel{\Gamma}{\rightarrow}\widetilde{G},\ stang(H)}{G\| H\stackrel{\Gamma}{\rightarrow}
\widetilde{G}\|\circlearrowleft\!H,\ H\| G\stackrel{\Gamma}{\rightarrow}\circlearrowleft\!H\|\widetilde{G}} &
{\bf P1i}\ \dfrac{G\stackrel{I}{\rightarrow}\widetilde{G}}{G\| H\stackrel{I}{\rightarrow}\widetilde{G}\| H,\
H\| G\stackrel{I}{\rightarrow}H\|\widetilde{G}}\\[3mm]

{\bf P1w}\ \dfrac{G\stackrel{V}{\rightarrow}\widetilde{G},\ stang(H)}{G\| H\stackrel{V}{\rightarrow}
\widetilde{G}\|\circlearrowleft\!H,\ H\| G\stackrel{V}{\rightarrow}\circlearrowleft\!H\|\widetilde{G}} &
{\bf P2s}\ \dfrac{G\stackrel{\Gamma}{\rightarrow}\widetilde{G},\ H\stackrel{\Delta}{\rightarrow}\widetilde{H}}
{G\| H\stackrel{\Gamma +\Delta}{\longrightarrow}\widetilde{G}\|\widetilde{H}}\\[3mm]

{\bf P2i}\ \dfrac{G\stackrel{I}{\rightarrow}\widetilde{G},\ H\stackrel{J}{\rightarrow}\widetilde{H}}
{G\| H\stackrel{I+J}{\longrightarrow}\widetilde{G}\|\widetilde{H}} &
{\bf P2w}\ \dfrac{G\stackrel{V}{\rightarrow}\widetilde{G},\ H\stackrel{W}{\rightarrow}\widetilde{H}}
{G\| H\stackrel{V+W}{\longrightarrow}\widetilde{G}\|\widetilde{H}}\\[3mm]

{\bf L}\ \dfrac{G\stackrel{\Upsilon}{\rightarrow}\widetilde{G}}{G[f]\stackrel{f(\Upsilon )}{\longrightarrow}
\widetilde{G}[f]} &
{\bf R}\ \dfrac{G\stackrel{\Upsilon}{\rightarrow}\widetilde{G},\ a,\hat{a}\not\in{\cal A}(\Upsilon )}
{G\rs a\stackrel{\Upsilon}{\rightarrow}\widetilde{G}\rs a}\\[3mm]

{\bf I1}\ \dfrac{G\stackrel{\Upsilon}{\rightarrow}\widetilde{G}}
{[G*E*F]\stackrel{\Upsilon}{\rightarrow}[\widetilde{G}*E*F]} &
{\bf I2s}\ \dfrac{G\stackrel{\Gamma}{\rightarrow}\widetilde{G},\ \neg init(G)\vee
(init(G)\wedge stang(\overline{E}))}
{[E*G*F]\stackrel{\Gamma}{\rightarrow}[E*\widetilde{G}*\!\downharpoonleft\!\!F],\
[E*F*G]\stackrel{\Gamma}{\rightarrow}[E*\!\downharpoonleft\!\!F*\widetilde{G}]}\\[3mm]

{\bf I2i}\ \!\!\dfrac{G\stackrel{I}{\rightarrow}\widetilde{G}}
{[E*G*F]\stackrel{I}{\rightarrow}[E*\widetilde{G}*\!\downharpoonleft\!\!F],\
[E*F*G]\stackrel{I}{\rightarrow}[E*\!\downharpoonleft\!\!F*\widetilde{G}]} &
{\bf I2w}\ \!\!\dfrac{G\stackrel{V}{\rightarrow}\widetilde{G},\ \neg init(G)\vee
(init(G)\wedge tang(\overline{E}))}
{[E*G*F]\stackrel{V}{\rightarrow}[E*\widetilde{G}*\!\downharpoonleft\!\!F],\
[E*F*G]\stackrel{V}{\rightarrow}[E*\!\downharpoonleft\!\!F*\widetilde{G}]}\\[3mm]

{\bf Sy1}\ \dfrac{G\stackrel{\Upsilon}{\rightarrow}\widetilde{G}}
{G\sy a\stackrel{\Upsilon}{\rightarrow}\widetilde{G}\sy a} &
{\bf Sy2s}\ \dfrac{G\sy a\xrightarrow{\Gamma '+\{(\alpha ,\rho )\}+\{(\beta ,\chi )\}}\widetilde{G}\sy a,\
a\in\alpha ,\ \hat{a}\in\beta}{G\sy a\xrightarrow{\Gamma '+\{(\alpha\oplus_a\beta,\rho\cdot\chi )\}}
\widetilde{G}\sy a}\\[3mm]

{\bf Sy2i}\ \dfrac{G\sy a\xrightarrow{I'+\{(\alpha ,\natural_l^0)\}+\{(\beta ,\natural_m^0)\}}\widetilde{G}\sy a,\
a\in\alpha ,\ \hat{a}\in\beta}{G\sy a\xrightarrow{I'+\{(\alpha\oplus_a\beta ,\natural_{l+m}^0)\}}
\widetilde{G}\sy a} &
{\bf Sy2w}\ \dfrac{G\sy a\xrightarrow{V'+\{(\alpha ,\natural_l^\theta )\}+\{(\beta ,\natural_m^\theta )\}}
\widetilde{G}\sy a,\ a\in\alpha ,\ \hat{a}\in\beta}
{G\sy a\xrightarrow{V'+\{(\alpha\oplus_a\beta ,\natural_{l+m}^\theta )\}}\widetilde{G}\sy a}\\[5mm]
\hline
\end{array}$
\end{center}
\end{table}

Almost all the rules in Table \ref{actrulesdm.tab} (excepting {\bf E}, {\bf Bw}, {\bf P2s}, {\bf P2i}, {\bf P2w}, {\bf
Sy2s}, {\bf Sy2i} and {\bf Sy2w}) resemble those of gsPBC, but the former correspond to execution of multisets of
activities, not of single activities, as in the latter, and our rules have simpler preconditions (if any), since all
immediate multiactions in dtsdPBC have the same priority level, unlike those of gsPBC.

The preconditions in rules {\bf E}, {\bf Cs}, {\bf P1s}, and {\bf I2s}
are needed to ensure that (possibly empty) multisets of stochastic multiactions are executed only from {\em s-tangible}
saturated operative dynamic expressions, such that all dynamic expressions structurally equivalent to them are
s-tangible as well. For example, if $init(G)$ in rule {\bf Cs} then $G\approx\overline{F}$ for some static expression
$F$ and $G\cho E\approx\overline{F}\cho E\approx\overline{F\cho E}\approx F\cho\overline{E}$. Hence, it should be
guaranteed that $stang(F\cho\overline{E})$, which holds iff $stang(\overline{E})$.
The case $E\cho G$ is treated similarly. In rule {\bf P1s}, assuming that $stang(G)$, it should be guaranteed that
$stang(G\| H)$ and $stang(H\| G)$, which holds iff $stang(H)$. The
precondition in rule {\bf I2s} is analogous to that in rule {\bf Cs}.

Analogously, the preconditions in rules {\bf Cw}, and {\bf I2w}
are needed to ensure that non-empty multisets of waiting multiactions are executed only from {\em tangible} saturated
operative dynamic expressions, such that all dynamic expressions structurally equivalent to them are tangible as well.
This requirement (about tangible expressions) means that only (possibly empty) multisets of stochastic multiactions or
non-empty multisets of waiting multiactions, and no immediate multiactions, can be executed from the subprocess $H$
that is composed {\em alternatively (in choice)} with the subprocess $G$. Hence, the multiset $W$ of waiting
multiactions, executed from $G$, can also be executed from the composition of $G$ and $H$, since immediate multiactions
cannot occur from $H$. Otherwise, it would prevent the execution of $W$ from $G$ in the composite process, by
disregarding the alternative choice of the branch specified by $H$, due to the zero delays and priority (captured by
all action rules) of immediate multiactions over all other multiaction types.

The precondition in rule {\bf P1w} is an exception from the above, and is needed to ensure that non-empty multisets of
waiting multiactions are executed only from {\em s-tangible} saturated operative dynamic expressions, such that all
dynamic expressions structurally equivalent to them are s-tangible as well. This stricter requirement (about
s-tangible, instead of just tangible, expressions) means that only (possibly empty) multisets of stochastic
multiactions, and no immediate or waiting multiactions, can be executed from the subprocess $H$ that is composed {\em
concurrently (in parallel)} with the subprocess $G$. Hence, the multiset $W$ of waiting multiactions, executed from
$G$, is also a maximal (by the inclusion relation) multiset that can be executed from the parallel composition of $G$
and $H$. The reason is that only the timers decrement by one time unit (by applying rule {\bf E}) is actually possible
in $H$ while executing $W$ from $G$, due to priority (captured by all action rules) of waiting multiactions over
stochastic ones. Thus, taking the rule precondition $stang(H)$ instead of $tang(H)$ preserves maximality of the steps
consisting of waiting multiactions while applying parallel composition.

In rules {\bf P1s} and {\bf P1w}, the timer value decrementing by one $\circlearrowleft\!H$, applied to the s-tangible
saturated operative dynamic expression $H$ that is composed in parallel with $G$, from which stochastic multiactions
are executed at the next time tick, is used to maintain the time progress uniformity in the composite expression.
Although rules {\bf P1s} and {\bf P1w} can be merged, we have not done it,
aiming to emphasize the exceptional precondition in rule {\bf P1w}.

In rules {\bf Cs}, {\bf Ci} and {\bf Cw}, the timer values discarding $\downharpoonleft\!\!E$, applied to the static
expression $E$ that is composed in choice with $G$, from which activities are executed, signifies that the timer values
of the non-chosen subexpression (branch) become irrelevant in the composite expression and thus may be removed.
Analogously, in rules {\bf I2s}, {\bf I2i} and {\bf I2w}, the timer values discarding $\downharpoonleft\!\!F$ is
applied to the static expression $F$ that is an alternative to $G$, from which activities are executed, since the
choice is always made between the body and termination subexpressions of the composite iteration expression (between
the second and third arguments of iteration).

Rule {\bf E} corresponds to one discrete time unit delay (passage of one unit of time) while executing no activities
and therefore it has no analogues among the rules of gsPBC that adopts the continuous time model. Rule {\bf E} is a
{\em global} one, i.e. it is applied only to the whole (topmost level of) expressions, rather than to their parts. The
reason is that all other action rules describe dynamic expressions transformations due to execution of {\em non-empty}
multisets of activities. Hence, the actionless time move described by rule {\bf E} cannot ``penetrate'' with action
rules through the expressions structure. This guarantees that time progresses uniformly in all their subexpressions.

Rule {\bf Bw} differs from the more standard ones {\bf Bs} and {\bf Bi} that both resemble rule {\bf B} in gsPBC. The
reason is that in {\bf Bw}, the overlined waiting multiaction has an extra superscript `$1$', indicating that one time
unit is remained until the multiaction's execution (RTE equals one) that should follow in the next moment.

Rules {\bf P2s}, {\bf P2i} and {\bf P2w} have no similar rules in gsPBC, since interleaving semantics of the algebra
allows no simultaneous execution of activities. On the other hand, {\bf P2s}, {\bf P2i} and {\bf P2w} have in PBC the
analogous rule {\bf PAR} that is used to construct step semantics of the calculus, but the former two rules correspond
to execution of multisets of activities, unlike that of multisets of multiactions in the latter rule. Rules {\bf P2s},
{\bf P2i} and {\bf P2w} cannot be merged, since otherwise simultaneous execution of different types of multiactions
would be allowed.

Rules {\bf Sy2s}, {\bf Sy2i} and {\bf Sy2w} differ from the corresponding synchronization rules in gsPBC, since the
probability or the weight of synchronization in the former rules and the rate or the weight of synchronization in the
latter rules are calculated in two distinct ways. Rules {\bf Sy2i} and {\bf Sy2w} cannot be merged, since otherwise
synchronous execution of immediate and waiting multiactions would be allowed.

Rule {\bf Sy2s} establishes that the synchronization of two stochastic multiactions is made by taking the product of
their probabilities, since we are considering that both must occur for the synchronization to happen, so this
corresponds, in some sense, to the probability of the independent event intersection, but the real situation is more
complex, since these stochastic multiactions can also be executed in parallel. Nevertheless, when scoping (the combined
operation consisting of synchronization followed by restriction over the same action \cite{BDK01}) is applied over a
parallel execution, we get as final result just the simple product of the probabilities, since no normalization is
needed there. Multiplication is an associative and commutative binary operation that is distributive over addition,
i.e. it fulfills all practical conditions imposed on the synchronization operator in \cite{Hil94}. Further, if both
arguments of multiplication are from $(0;1)$ then the result belongs to the same interval, hence, multiplication
naturally maintains probabilistic compositionality in our model. Our approach is similar to the multiplication of rates
of the synchronized actions in MTIPP \cite{HR94} in the case when the rates are less than $1$. Moreover, for the
probabilities $\rho$ and $\chi$ of two stochastic multiactions to be synchronized we have $\rho\cdot\chi <\min\{\rho
,\chi\}$, i.e. multiplication meets the performance requirement stating that the probability of the resulting
synchronized stochastic multiaction should be less than the probabilities of the two ones to be synchronized. While
performance evaluation, it is usually supposed that the execution of two components together require more system
resources and time than the execution of each single one. This resembles the {\em bounded capacity} assumption from
\cite{Hil94}. Thus, multiplication is easy to handle with and it satisfies the algebraic, probabilistic, time and
performance requirements. Therefore, we have chosen the product of the probabilities for the synchronization. See also
\cite{BKLL95,BrHe01} for a discussion about binary operations producing the rates of synchronization in the continuous
time setting.

In rules {\bf Sy2i} and {\bf Sy2w}, we sum the weights of two synchronized immediate (waiting, respectively)
multiactions, since the weights can be interpreted as the rewards \cite{Ros96}, thus, we collect the rewards. Moreover,
we express that the synchronized execution of immediate (waiting) multiactions has more importance than that of every
single one. The weights of immediate and waiting (i.e. deterministic) multiactions can also be seen as bonus rewards
associated with transitions \cite{BBr01}. The rewards are summed during synchronized execution of immediate (waiting)
multiactions, since in that case all the synchronized activities can be seen as participated in the execution. We
prefer to collect more rewards, thus, the transitions providing greater rewards will have a preference and they will be
executed with a greater probability. In particular, since execution of immediate multiactions takes no time, we prefer
to collect in a step (parallel execution) as many synchronized immediate multiactions as possible to get more
significant progress in behaviour. Under behavioural progress we understand an advance in executing activities, which
does not always imply a progress in time, as in the case when the activities are immediate multiactions. This aspect
will be used later, while evaluating performance via analysis of the embedded discrete time Markov chains (EDTMCs) of
expressions. Since every state change in EDTMC takes one unit of (its local) time, greater advance in operation of the
EDTMC allows one to calculate quicker many performance indices. As for waiting multiactions, only the maximal multisets
of them, executable from a state, occur with a time tick. The reason is that each waiting multiaction has a probability
$1$ to occur in the next moment, when the remaining time of its timer (RTE) equals one and there exist no conflicting
waiting multiactions. Hence, all waiting multiactions with the
RTE being one that are executable together from a state must participate in a step from that state. Since there may
exist different such maximal multisets of waiting multiactions, a probabilistic choice among all possible steps is
made, imposed by the weights of those multiactions. Thus, the steps of waiting multiactions always produce maximal
overall weights, but they are mainly used to calculate the probabilities of alternative maximal steps rather than the
cumulative bonus rewards.

We do not have self-synchronization, i.e. synchronization of an activity with itself, since all the (enumerated)
activities executed together are considered to be different. This allows us to avoid rather cumbersome and unexpected
behaviour, as well as many technical difficulties \cite{BDK01}.

Notice that the timers of all waiting multiactions that lose their overlines when a state change occurs become inactive
(turned off) and their values become irrelevant while the timers of all those preserving their overlines continue
running with their stored values. Hence, we adopt the {\em enabling memory} memory policy
\cite{MBCDF95,AHR00,Bal01,Bal07} when the process states are changed and the overlining of deterministic multiactions
is possibly modified (remember that immediate multiactions may be seen as those with the timers displaying a single
value $0$, so we do not need to store their values). Then the timer values of waiting multiactions are taken as the
enabling memory~variables.

In Table \ref{rulescompdm.tab}, inaction rules, action rules (with stochastic or immediate, or waiting multiactions)
and empty move rule are compared according to the three questions about their application: whether it changes the
current state, whether it leads to a time progress, and whether it results in execution of some activities. Positive
answers to the questions are denoted by the plus sign while negative ones are specified by the minus sign. If both
positive and negative answers can be given to some of the questions in different cases then the plus-minus sign is
written. Notice that the process states are considered up to structural equivalence of the corresponding expressions,
and time progress is not regarded as a state change.

\begin{table}
\caption{Comparison of inaction, action and empty move rules}
\label{rulescompdm.tab}
\begin{center}
\begin{tabular}{|c||c|c|c|}
\hline
Rules & State change & Time progress & Activities execution\\
\hline\hline
Inaction rules & $-$ & $-$ & $-$\\
\hline
Action rules & $\pm$ & $+$ & $+$ \\
(stochastic or waiting multiactions) & & & \\
\hline
Action rules & $\pm$ & $-$ & $+$ \\
(immediate multiactions) & & & \\
\hline
Empty move rule & $-$ & $+$ & $-$ \\
\hline
\end{tabular}
\end{center}
\end{table}

\subsection{Transition systems}

We now construct labeled probabilistic transition systems associated with dynamic expressions. The transition systems
are used to define the operational semantics of dynamic expressions.

Let $G$ be a dynamic expression and $s=[G]_\approx$. The set of {\em all multisets of activities executable in $s$} is
defined as $Exec(s)=\{\Upsilon\mid\exists H\in s\ \exists\widetilde{H}\
H\stackrel{\Upsilon}{\rightarrow}\widetilde{H}\}$. Here $H\stackrel{\Upsilon}{\rightarrow}\widetilde{H}$ is an
inference by the rules from Table \ref{actrulesdm.tab}.

It can be proved by induction on the structure of expressions that $\Upsilon\in Exec(s)\setminus\{\emptyset\}$ implies
$\exists H\in s\ \Upsilon\in Now(H)$. The reverse statement does not hold in general, as the next example shows.

\begin{example}
Let $H,H'$ be from Example \ref{cannow.exm} and $s=[H]_\approx =[H']_\approx$. We have
$Now(H)=\{\{(\{a\},\natural_1^0)\}\}$ and $Now(H')=\{\{(\{b\},\frac{1}{2})\}\}$. Since only rules {\bf Ci} and {\bf Bi}
can be applied to $H$ while no action rule can be applied to $H'$, we get $Exec(s)=\{\{(\{a\},\natural_1^0)\}\}$. Then,
for $H'\in s$ and $\Upsilon =\{(\{b\},\frac{1}{2})\}\in Now(H')$, we obtain $\Upsilon\not\in Exec(s)$.
\label{nowexec.exm}
\end{example}

The state $s$ is {\em s-tangible (stochastically tangible)} if $Exec(s)\subseteq\nat_{fin}^{\cal SL}$. For an s-tangible
state $s$ we always have $\emptyset\in Exec(s)$ by rule {\bf E}, hence, we may have $Exec(s)=\{\emptyset\}$. The state
$s$ is {\em w-tangible (waitingly tangible)} if $Exec(s)\subseteq\nat_{fin}^{\cal WL}\setminus\{\emptyset\}$. The state
$s$ is {\em tangible} if it is s-tangible or w-tangible, i.e. $Exec(s)\subseteq\nat_{fin}^{\cal SL}\cup\nat_{fin}^{\cal
WL}$. Again, for tangible states we may have $\emptyset\in Exec(s)$ and $Exec(s)=\{\emptyset\}$. Otherwise, the state
$s$ is {\em vanishing}, and in this case $Exec(s)\subseteq\nat_{fin}^{\cal IL}\setminus\{\emptyset\}$.

Note that if $\Upsilon\in Exec(s)$ and $\Upsilon\in\nat_{fin}^{\cal SL}\cup\nat_{fin}^{\cal IL}$ then by rules {\bf
P2s}, {\bf P2i}, {\bf Sy2s}, {\bf Sy2i} and definition of $Exec(s)\ \forall\Xi\subseteq\Upsilon ,\ \Xi\neq\emptyset$,
we have $\Xi\in Exec(s)$.

Since the inaction rules only distribute and move upper and lower bars along the syntax of dynamic expressions, all
$H\in s$ have the same underlying static expression $F$. Process expressions always have a finite length, hence, the
number of all (enumerated) activities and the number of all operations in the syntax of $F$ are finite as well. The
action rules {\bf Sy2s}, {\bf Sy2i} and {\bf Sy2w} are the only ones that generate new activities. They result from the
handshake synchronization of actions and their conjugates belonging to the multiaction parts of the first and second
constituent activity, respectively. Since we have a finite number of operators $\sy$ in $F$ and all the multiaction
parts of the activities are finite multisets, the number of the new synchronized activities is also finite. The action
rules contribute to $Exec(s)$ (in addition to the empty set, if rule {\bf E} is applicable) only the sets consisting
both of activities from $F$ and the new activities, produced by {\bf Sy2s}, {\bf Sy2i} and {\bf Sy2w}. Since we have a
finite number $n$ of all such activities, we get $|Exec(s)|\leq 2^n<\infty$. Thus, summation and multiplication by
elements from the finite set $Exec(s)$ are well-defined. Similar reasoning can be used to demonstrate that for all
dynamic expressions $H$ (not just for those from $s$), $Now(H)$ is a finite set.

\begin{definition}
The {\em derivation set} of a dynamic expression $G$, denoted by $DR(G)$, is the minimal set such that
\begin{itemize}

\item $[G]_\approx\in DR(G)$;

\item if $[H]_\approx\in DR(G)$ and $\exists\Upsilon\ H\stackrel{\Upsilon}{\rightarrow}\widetilde{H}$ then
$[\widetilde{H}]_\approx\in DR(G)$.

\end{itemize}
\end{definition}

The set of {\em all s-tangible states from $DR(G)$} is denoted by $DR_{ST}(G)$, and the set of {\em all w-tangible
states from $DR(G)$} is denoted by $DR_{WT}(G)$. The set of {\em all tangible states from $DR(G)$} is denoted by
$DR_T(G)=DR_{ST}(G)\cup DR_{WT}(G)$. The set of {\em all vanishing states from $DR(G)$} is denoted by $DR_V(G)$.
Obviously, $DR(G)=DR_T(G)\uplus DR_V(G)=DR_{ST}(G)\uplus DR_{WT}(G)\uplus DR_V(G)$, where $\uplus$ denotes disjoint
union.

Let now $G$ be a dynamic expression and $s,\tilde{s}\in DR(G)$.

Let $\Upsilon\in Exec(s)\setminus\{\emptyset\}$. The {\em probability that the multiset of stochastic multiactions
$\Upsilon$ is ready for execution in $s$} or the {\em weight of the multiset of
deterministic multiactions $\Upsilon$ which is ready for execution in $s$} is

$$PF(\Upsilon ,s)=
\left\{
\begin{array}{ll}
\prod_{(\alpha ,\rho )\in\Upsilon}\rho\cdot\prod_{\{\{(\beta ,\chi )\}\in Exec(s)\mid (\beta ,\chi
)\not\in\Upsilon\}}(1-\chi ), & s\in DR_{ST}(G);\\
\sum_{(\alpha ,\natural_l^\theta )\in\Upsilon}l, & s\in DR_{WT}(G)\cup DR_V(G).
\end{array}
\right.$$

In the case $\Upsilon =\emptyset$ and $s\in DR_{ST}(G)$ we define

$$PF(\emptyset ,s)=
\left\{
\begin{array}{ll}
\prod_{\{(\beta ,\chi )\}\in Exec(s)}(1-\chi ), & Exec(s)\neq\{\emptyset\};\\
1, & Exec(s)=\{\emptyset\}.
\end{array}
\right.$$

If $s\in DR_{ST}(G)$ and $Exec(s)\neq\{\emptyset\}$ then $PF(\Upsilon ,s)$ can be interpreted as a {\em joint}
probability of independent events (in a probability sense, i.e. the probability of intersection of these events is
equal to the product of their probabilities). Each such an event consists in the positive or the negative decision to
be executed of a particular stochastic multiaction. Every executable stochastic multiaction decides probabilistically
(using its probabilistic part) and independently (from others), if it wants to be executed in $s$. If $\Upsilon$ is a
multiset of all executable stochastic multiactions which have decided to be executed in $s$ and $\Upsilon\in Exec(s)$
then $\Upsilon$ is ready for execution in $s$. The multiplication in the definition is used because it reflects the
probability of the independent event intersection. Alternatively, when $\Upsilon\neq\emptyset ,\ PF(\Upsilon ,s)$ can
be interpreted as the probability to execute {\em exclusively} the multiset of stochastic multiactions $\Upsilon$ in
$s$, i.e. the probability of {\em intersection} of two events calculated using the conditional probability formula in
the form of ${\sf P}(X\cap Y)={\sf P}(X|Y){\sf P}(Y)$. The event $X$ consists in the execution of $\Upsilon$ in $s$.
The event $Y$ consists in the non-execution in $s$ of all the executable stochastic multiactions not belonging to
$\Upsilon$. Since the mentioned non-executions are obviously independent events, the probability of $Y$ is a product of
the probabilities of the non-executions: ${\sf P}(Y)=\prod_{\{\{(\beta ,\chi )\}\in Exec(s)\mid (\beta ,\chi
)\not\in\Upsilon\}}(1-\chi )$. The conditioning of $X$ by $Y$ makes the executions of the stochastic multiactions from
$\Upsilon$ independent, since all of them can be executed in parallel in $s$ by definition of $Exec(s)$. Hence, the
probability to execute $\Upsilon$ {\em under condition} that no executable stochastic multiactions not belonging to
$\Upsilon$ are executed in $s$ is a product of probabilities of these stochastic multiactions: ${\sf
P}(X|Y)=\prod_{(\alpha ,\rho )\in\Upsilon}\rho$. Thus, the probability that $\Upsilon$ is executed {\em and} no
executable stochastic multiactions not belonging to $\Upsilon$ are executed in $s$ is the probability of $X$
conditioned by $Y$ multiplied by the probability of $Y$: ${\sf P}(X\cap Y)={\sf P}(X|Y){\sf P}(Y)=\prod_{(\alpha ,\rho
)\in\Upsilon}\rho\cdot \prod_{\{\{(\beta ,\chi )\}\in Exec(s)\mid (\beta ,\chi )\not\in\Upsilon\}}(1-\chi )$. When
$\Upsilon =\emptyset ,\ PF(\Upsilon ,s)$ can be interpreted as the probability not to execute in $s$ any executable
stochastic multiactions, thus, $PF(\emptyset ,s)=\prod_{\{(\beta ,\chi )\}\in Exec(s)}(1-\chi )$. When only the empty
multiset of activities can be executed in $s$, i.e. $Exec(s)=\{\emptyset\}$, we take $PF(\emptyset ,s)=1$, since we
stay in $s$ in this case. Note that for $s\in DR_T(G)$ we have $PF(\emptyset ,s)\in (0;1]$, hence, we can stay in $s$
at the next time moment with a certain positive probability.

If $s\in DR_{WT}(G)\cup DR_V(G)$ then $PF(\Upsilon ,s)$ could be interpreted as the {\em overall (cumulative)} weight
of the
deterministic multiactions from $\Upsilon$, i.e. the sum of all their weights. The summation here is used since the
weights can be seen as the rewards which are collected \cite{Ros96}. This means that concurrent execution of the
deterministic multiactions has more importance than that of every single one. The weights of
deterministic multiactions can also be interpreted as bonus rewards of transitions \cite{BBr01}. The rewards are summed
when
deterministic multiactions are executed in parallel, because all of them participated in the execution. In particular,
since execution of immediate multiactions takes no time, we prefer to collect in a step (parallel execution of
activities) as many parallel immediate multiactions as possible to get more progress in behaviour. This aspect will be
used later, while while evaluating performance on the basis of the EDTMCs of expressions. Concerning waiting
multiactions, only the maximal multisets of them executable from a state occur in the next moment. Therefore, the steps
of waiting multiactions produce maximal overall weights, which are used to calculate probabilities of alternative
maximal steps rather than the cumulative bonuses. Note that this reasoning is the same as that used to define the
weight of synchronized immediate (waiting, respectively) multiactions in the rules {\bf Sy2i} and~{\bf Sy2w}.

Note that the definition of $PF(\Upsilon ,s)$ (as well as the definitions of other probability functions which we shall
present) is based on the enumeration of activities which is considered implicit.

Let $\Upsilon\in Exec(s)$. Besides $\Upsilon$, some other multisets of activities may be ready for execution in $s$,
hence, a kind of conditioning or normalization is needed to calculate the execution probability. The {\em probability
to execute the multiset of activities $\Upsilon$ in $s$} is

$$PT(\Upsilon ,s)=\frac{PF(\Upsilon ,s)}{\sum_{\Xi\in Exec(s)}PF(\Xi ,s)}.$$

If $s\in DR_{ST}(G)$ then $PT(\Upsilon ,s)$ can be interpreted as the {\em conditional} probability to execute
$\Upsilon$ in $s$ calculated using the conditional probability formula in the form of ${\sf P}(Z|W)=\frac{{\sf P}(Z\cap
W)}{{\sf P}(W)}$. The event $Z$ consists in the exclusive execution of $\Upsilon$ in $s$, hence, ${\sf
P}(Z)=PF(\Upsilon ,s)$. The event $W$ consists in the exclusive execution of any set (including the empty one) $\Xi\in
Exec(s)$ in $s$. Thus, $W=\cup_j Z_j$, where $\forall j,\ Z_j$ are mutually exclusive events (in a probability sense,
i.e. intersection of these events is the empty event) and $\exists i,\ Z=Z_i$. We have ${\sf P}(W)=\sum_j{\sf
P}(Z_j)=\sum_{\Xi\in Exec(s)}PF(\Xi ,s)$, because summation reflects the probability of the mutually exclusive event
union. Since $Z\cap W=Z_i\cap (\cup_j Z_j)=Z_i=Z$, we have ${\sf P}(Z|W)=\frac{{\sf P}(Z)}{{\sf
P}(W)}=\frac{PF(\Upsilon ,s)}{\sum_{\Xi\in Exec(s)}PF(\Xi ,s)}$. $PF(\Upsilon ,s)$ can also be seen as the {\em
potential} probability to execute $\Upsilon$ in $s$, since we have $PF(\Upsilon ,s)=PT(\Upsilon ,s)$ only when {\em
all} sets (including the empty one) consisting of the executable stochastic multiactions can be executed in $s$. In
this case, all the mentioned stochastic multiactions can be executed in parallel in $s$ and we have $\sum_{\Xi\in
Exec(s)}PF(\Xi ,s)=1$, since this sum collects the products of {\em all} combinations of the probability parts of the
stochastic multiactions and the negations of these parts. But in general, for example, for two stochastic multiactions
$(\alpha ,\rho )$ and $(\beta ,\chi )$ executable in $s$, it may happen that they cannot be executed in $s$ together,
in parallel, i.e. $\emptyset ,\{(\alpha ,\rho )\},\{(\beta ,\chi )\}\in Exec(s)$, but $\{(\alpha ,\rho ),(\beta ,\chi
)\}\not\in Exec(s)$. Note that for $s\in DR_T(G)$ we have $PT(\emptyset ,s)\in (0;1]$, hence, there is a non-zero
probability to stay in the state $s$ at the next time moment, and the residence time in $s$ is at least $1$ discrete
time unit.

If $s\in DR_{WT}(G)\cup DR_V(G)$ then $PT(\Upsilon ,s)$ can be interpreted as the weight of the set of
deterministic multiactions $\Upsilon$ which is ready for execution in $s$ {\em normalized} by the weights of
{\em all} the sets executable in $s$. This approach is analogous to that used in the EMPA definition of the
probabilities of immediate actions executable from the same process state \cite{BGo98} (inspired by way in which the
probabilities of conflicting immediate transitions in GSPNs are calculated \cite{Bal07}). The only difference is that
we have a step semantics and, for every set of
deterministic multiactions executed in parallel, we should use its cumulative weight. To get the analogy with EMPA
possessing interleaving semantics, we should interpret the weights of immediate actions of EMPA as the cumulative
weights of the sets of
deterministic multiactions of dtsdPBC.

The advantage of our two-stage approach to definition of the probability to execute a set of activities is that the
resulting probability formula $PT(\Upsilon ,s)$ is valid both for (sets of) stochastic and deterministic multiactions.
It allows one to unify the notation used later while constructing the operational semantics and analyzing performance.

Note that the sum of outgoing probabilities for the expressions belonging to the derivations of $G$ is equal to $1$.
More formally, $\forall s\in DR(G)\ \sum_{\Upsilon\in Exec(s)}PT(\Upsilon ,s)=1$. This, obviously, follows from the
definition of $PT(\Upsilon ,s)$, and guarantees that it defines a probability distribution.

The {\em probability to move from $s$ to $\tilde{s}$ by executing any multiset of activities} is

$$PM(s,\tilde{s})=\sum_{\{\Upsilon\mid\exists H\in s\ \exists\widetilde{H}\in\tilde{s}\
H\stackrel{\Upsilon}{\rightarrow}\widetilde{H}\}}PT(\Upsilon ,s).$$

The summation in the definition above reflects the probability of the mutually exclusive event union, since
$\sum_{\{\Upsilon\mid\exists H\in s,\ \exists\widetilde{H}\in\tilde{s},\
H\stackrel{\Upsilon}{\rightarrow}\widetilde{H}\}}PT(\Upsilon ,s)=\frac{1}{\sum_{\Xi\in Exec(s)}PF(\Xi ,s)}\cdot
\sum_{\{\Upsilon\mid\exists H\in s,\ \exists\widetilde{H}\in\tilde{s},\
H\stackrel{\Upsilon}{\rightarrow}\widetilde{H}\}}PF(\Upsilon ,s)$, where for each $\Upsilon ,\ PF(\Upsilon ,s)$ is the
probability of the exclusive execution of $\Upsilon$ in $s$. Note that $\forall s\in DR(G)\\
\sum_{\{\tilde{s}\mid\exists H\in s\ \exists\widetilde{H}\in\tilde{s}\ \exists\Upsilon\
H\stackrel{\Upsilon}{\rightarrow}\widetilde{H}\}}PM(s,\tilde{s})=\sum_{\{\tilde{s}\mid\exists H\in s\
\exists\widetilde{H}\in\tilde{s}\ \exists\Upsilon\ H\stackrel{\Upsilon}{\rightarrow}\widetilde{H}\}}
\sum_{\{\Upsilon\mid\exists H\in s\ \exists\widetilde{H}\in\tilde{s}\
H\stackrel{\Upsilon}{\rightarrow}\widetilde{H}\}}PT(\Upsilon ,s)=\\
\sum_{\Upsilon\in Exec(s)}PT(\Upsilon ,s)=1$.

\begin{example}
Let $E=(\{a\},\rho )\cho (\{a\},\chi )$, where $\rho ,\chi\in (0;1)$. $DR(\overline{E})$ consists of the equivalence
classes $s_1=[\overline{E}]_\approx$ and $s_2=[\underline{E}]_\approx$. We have $DR_T(\overline{E})=\{s_1,s_2\}$. The
execution probabilities are calculated as follows. Since $Exec(s_1)=\{\emptyset ,\{(\{a\},\rho )\},\{(\{a\},\chi
)\}\}$, we get $PF(\{(\{a\},\rho )\},s_1)=\rho (1-\chi ),\ PF(\{(\{a\},\chi )\},s_1)=\chi (1-\rho )$ and $PF(\emptyset
,s_1)=(1-\rho )(1-\chi )$. Then $\sum_{\Xi\in Exec(s_1)}PF(\Xi ,s_1)=\rho (1-\chi )+\chi (1-\rho )+(1-\rho )(1-\chi
)=1-\rho\chi$. Thus, $PT(\{(\{a\},\rho )\},s_1)=\frac{\rho (1-\chi )}{1-\rho\chi},\ PT(\{(\{a\},\chi
)\},s_1)=\frac{\chi (1-\rho )}{1-\rho\chi}$ and $PT(\emptyset ,s_1)=PM(s_1,s_1)=\frac{(1-\rho )(1-\chi )}{1-\rho\chi}$.
Further, $Exec(s_2)=\{\emptyset\}$, hence, $\sum_{\Xi\in Exec(s_2)}PF(\Xi ,s_2)=PF(\emptyset ,s_2)=1$ and $PT(\emptyset
,s_2)=PM(s_2,s_2)=\frac{1}{1}=1$. Finally, $PM(s_1,s_2)=PT(\{(\{a\},\rho )\},s_1)+PT(\{(\{a\},\chi )\},s_1)=\frac{\rho
(1-\chi )}{1-\rho\chi}+\frac{\chi (1-\rho )}{1-\rho\chi}=\frac{\rho +\chi -2\rho\chi}{1-\rho\chi}$.

Let $E'=(\{a\},\natural_l^0)\cho (\{a\},\natural_m^0)$, where $l,m\in\real_{>0}$. $DR(\overline{E'})$ consists of the
equivalence classes $s_1'=[\overline{E'}]_\approx$ and $s_2'=[\underline{E'}]_\approx$. We have
$DR_T(\overline{E'})=\{s_2'\}$ and $DR_V(\overline{E'})=\{s_1'\}$. The execution probabilities are calculated as
follows. Since $Exec(s_1')=\{\{(\{a\},\natural_l^0)\}, \{(\{a\},\natural_m^0)\}\}$, we get
$PF(\{(\{a\},\natural_l^0)\},s_1')=l$ and $PF(\{(\{a\},\natural_m^0)\},s_1')=m$. Then $\sum_{\Xi\in Exec(s_1')}PF(\Xi
,s_1')=l+m$. Thus, $PT(\{(\{a\},\natural_l^0)\},s_1')=\frac{l}{l+m}$ and $PT(\{(\{a\},\natural_m^0)\},s_1')=
\frac{m}{l+m}$. Further, $Exec(s_2')=\{\emptyset\}$, hence, $\sum_{\Xi\in Exec(s_2')}PF(\Xi ,s_2')=PF(\emptyset
,s_2')=1$ and $PT(\emptyset ,s_2')=PM(s_2',s_2')=\frac{1}{1}=1$. Finally, $PM(s_1',s_2')=
PT(\{(\{a\},\natural_l^0)\},s_1')+PT(\{(\{a\},\natural_m^0)\},s_1')=\frac{l}{l+m}+\frac{m}{l+m}=1$.
\label{trprob.exm}
\end{example}

\begin{definition}
Let $G$ be a dynamic expression. The {\em (labeled probabilistic) transition system} of $G$ is a quadruple
$TS(G)=(S_G,L_G,{\cal T}_G,s_G)$, where
\begin{itemize}

\item the set of {\em states} is $S_G=DR(G)$;

\item the set of {\em labels} is $L_G=\nat_{fin}^{\cal SDL}\times (0;1]$;

\item the set of {\em transitions} is ${\cal T}_G=\{(s,(\Upsilon ,PT(\Upsilon ,s)),\tilde{s})\mid
s,\tilde{s}\in DR(G),\ \exists H\in s\ \exists\widetilde{H}\in\tilde{s}\
H\stackrel{\Upsilon}{\rightarrow}\widetilde{H}\}$;

\item the {\em initial state} is $s_G=[G]_\approx$.

\end{itemize}
\end{definition}

%
\begin{example}
Let $E$ be from Example \ref{overlines.exm}. The next inferences by rule {\bf E} are possible from the elements
of~$[\overline{E}]_\approx$:

$$\begin{array}{l}
\overline{(\{a\},\natural_1^3)\cho (\{b\},\frac{1}{3})}\approx\overline{(\{a\},\natural_1^3)^3}\cho (\{b\},\frac{1}{3})
\stackrel{\emptyset}{\rightarrow}\overline{(\{a\},\natural_1^3)^2}\cho (\{b\},\frac{1}{3}),\\[1mm]

\overline{(\{a\},\natural_1^3)\cho (\{b\},\frac{1}{3})}\approx (\{a\},\natural_1^3)^3\cho\overline{(\{b\},\frac{1}{3})}
\stackrel{\emptyset}{\rightarrow}(\{a\},\natural_1^3)^2\cho\overline{(\{b\},\frac{1}{3})}.
\end{array}$$

The first and second inferences suggest the empty move transition
$[\overline{E}]_\approx\stackrel{\emptyset}{\rightarrow}[\overline{(\{a\},\natural_1^3)^2\cho
(\{b\},\frac{1}{3})}]_\approx\neq [\overline{E}]_\approx$. The intuition is that the timer of the enabled waiting
multiaction $(\{a\},\natural_1^3)$ is decremented by one time unit in the both cases, whenever it is overlined or not.
Later we shall see that in the both cases, the respective waiting transition of the LDTSDPN corresponding to
$\overline{E}$ will be enabled at a ``common'' marking (that also enables a stochastic transition, matched up to
$(\{b\},\frac{1}{3})$), so its timer should be decreased by one with a time tick while staying at {\em the same
marking}, and such a time move will lead to a {\em different state} of the LDTSDPN.

\label{determ.exm}
\end{example}

The definition of $TS(G)$ is correct, i.e. for every state, the sum of the probabilities of all the transitions
starting from it is $1$. This is guaranteed by the note after the definition of $PT(\Upsilon ,s)$. Thus, we have
defined a {\em generative} model of probabilistic processes, according to the classification from \cite{GSS95}. The
reason is that the sum of the probabilities of the transitions with all possible labels should be equal to $1$, not
only of those with the same labels (up to enumeration of activities they include) as in the {\em reactive} models, and
we do not have a nested probabilistic choice as in the {\em stratified} models.

The transition system $TS(G)$ associated with a dynamic expression $G$ describes all the steps (parallel executions)
that occur at discrete time moments with some (one-step) probability and consist of multisets of activities. Every step
consisting of stochastic (waiting, respectively) multiactions or the empty step (i.e. that consisting of the empty
multiset of activities) occurs instantly after one discrete time unit delay. Each step consisting of immediate
multiactions occurs instantly without any delay. The step can change the current state to a different one. The states
are the structural equivalence classes of dynamic expressions obtained by application of action rules starting from the
expressions belonging to $[G]_\approx$. A transition $(s,(\Upsilon ,{\cal P}),\tilde{s})\in{\cal T}_G$ will be written
as $s\stackrel{\Upsilon}{\rightarrow}_{\cal P}\tilde{s}$. It is interpreted as follows: the probability to change the
state $s$ to $\tilde{s}$ as a result of executing $\Upsilon$ is ${\cal P}$.

Note that from every s-tangible state the empty multiset of activities can always be executed by rule {\bf E}. Hence,
for s-tangible states, $\Upsilon$ may be the empty multiset, and its execution only decrements by one the timer values
(if any) of the current state (i.e. the equivalence class).
Then we may have a transition $s\stackrel{\emptyset}{\rightarrow}_{\cal P}\circlearrowleft\!s$
from an s-tangible state
$s$ to the tangible (i.e. s-tangible or w-tangible) state
$\circlearrowleft\!s=\bigcup\{[\circlearrowleft\!H]_\approx\mid H\in s\cap SatOpRegDynExpr\}$. Thus,
$\circlearrowleft\!s$ is the union of the
structural equivalence classes of all saturated operative dynamic expressions from
$s$, whose timer values have been decremented by one, prior to combining them into the equivalence classes. This
corresponds to applying the empty move rule to all saturated operative dynamic expressions from
$s$, followed by unifying the structural equivalence classes of all the resulting expressions.
We have to keep track of such executions, called the {\em empty moves}, because they affect the timers and have
non-zero probabilities. The latter follows from the definition of $PF(\emptyset ,s)$ and the fact that the
probabilities of stochastic multiactions cannot be equal to $1$ as they belong to the interval $(0;1)$.
When it holds
$\forall H\in s\cap SatOpRegDynExpr\ \circlearrowleft\!H=H$, we obtain
$\circlearrowleft\!s=s$ by definition of $\circlearrowleft\!s$. Then the empty move from $s$ is in the form of
$s\stackrel{\emptyset}{\rightarrow}_{\cal P}s$, called the {\em empty loop}.
For w-tangible and vanishing states $\Upsilon$ cannot be the empty multiset, since we must execute some immediate
(waiting, respectively) multiactions from them at the current (next, respectively) time moment.

The step probabilities belong to the interval $(0;1]$, being $1$ in the case when we cannot leave an s-tangible state
$s$ and the only transition leaving it is the empty move one $s\stackrel{\emptyset}{\rightarrow}_1\circlearrowleft\!s$,
or if there is just a single transition from a w-tangible or a vanishing state to any other one.

We write $s\stackrel{\Upsilon}{\rightarrow}\tilde{s}$ if $\exists{\cal P}\ s\stackrel{\Upsilon}{\rightarrow}_{\cal
P}\tilde{s}$ and $s\rightarrow\tilde{s}$ if $\exists\Upsilon\ s\stackrel{\Upsilon}{\rightarrow}\tilde{s}$.

The first equivalence we are going to introduce is isomorphism which is a coincidence of systems up to renaming of
their components or states.

\begin{definition}
Let $G,G'$ be dynamic expressions and $TS(G)\!=\!(S_G,L_G,{\cal T}_G,s_G),TS(G')\!=\!(S_{G'},L_{G'},{\cal
T}_{G'},s_{G'})$ be their transition systems. A mapping $\beta :S_G\rightarrow S_{G'}$ is an {\em isomorphism} between
$TS(G)$ and $TS(G')$, denoted by $\beta :TS(G)\simeq TS(G')$, if
\begin{enumerate}

\item $\beta$ is a bijection such that $\beta (s_G)=s_{G'}$;

\item $\forall s,\tilde{s}\in S_G\ \forall\Upsilon\ s\stackrel{\Upsilon}{\rightarrow}_{\cal P}\tilde{s}\
\Leftrightarrow\ \beta (s)\stackrel{\Upsilon}{\rightarrow}_{\cal P}\beta (\tilde{s})$.

\end{enumerate}
Two transition systems $TS(G)$ and $TS(G')$ are {\em isomorphic}, denoted by $TS(G)\simeq TS(G')$, if $\exists\beta
:TS(G)\simeq TS(G')$.
\end{definition}

Transition systems of static expressions can be defined as well. For $E\in RegStatExpr$, let $TS(E)=TS(\overline{E})$.

\begin{definition}
Two dynamic expressions $G$ and $G'$ are {\em equivalent with respect to transition systems}, denoted by $G=_{ts}G'$,
if $TS(G)\simeq TS(G')$.
\end{definition}

Let us now present a series of examples that demonstrate how to construct the transition systems of the dynamic
expressions that include various compositions of stochastic, waiting and immediate multiactions.

\begin{figure}
\begin{center}
\includegraphics{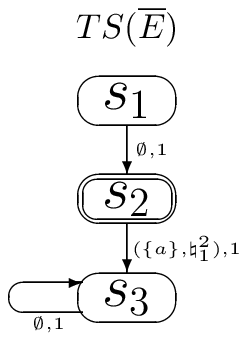}
\end{center}
\caption{The transition system of $\overline{E}$ for $E=(\{a\},\natural_1^2)\cho (\{b\},\natural_2^3)$}
\label{tschowm.fig}
\end{figure}

\begin{example}
This example demonstrates a choice
between two waiting multiactions with different delays. It shows that the waiting multiaction $(\{a\},\natural_1^2)$
with a less delay $2$ is always executed first, hence, the choice is resolved in favour of it in any case and an
absorbing state is then reached, so that the waiting multiaction $(\{b\},\natural_2^3)$ with a greater delay $3$ is
never executed.

Let $E=(\{a\},\natural_1^2)\cho (\{b\},\natural_2^3)$. $DR(\overline{E})$ consists of the equivalence classes

$$\begin{array}{ll}
s_1=[\overline{(\{a\},\natural_1^2)^2}\cho (\{b\},\natural_2^3)^3]_\approx\!=
\![(\{a\},\natural_1^2)^2\cho\overline{(\{b\},\natural_2^3)^3}]_\approx , &
s_2=[\overline{(\{a\},\natural_1^2)^1}\cho (\{b\},\natural_2^3)^2]_\approx\!=
\![(\{a\},\natural_1^2)^1\cho\overline{(\{b\},\natural_2^3)^2}]_\approx ,\\[1mm]
s_3=[\underline{(\{a\},\natural_1^2)\cho (\{b\},\natural_2^3)}]_\approx . &
\end{array}$$

We have $DR_{ST}(\overline{E})=\{s_1,s_3\},\ DR_{WT}(\overline{E})=\{s_2\}$ and $DR_V(\overline{E})=\emptyset$.

In Figure \ref{tschowm.fig}, the transition system $TS(\overline{E})$ is shown. The s-tangible and w-tangible states
are depicted in ordinary and double ovals, respectively. For simplicity of the graphical representation, the singleton
multisets of activities are written without outer braces.
\label{tschowm.exm}
\end{example}

\begin{figure}
\begin{center}
\includegraphics{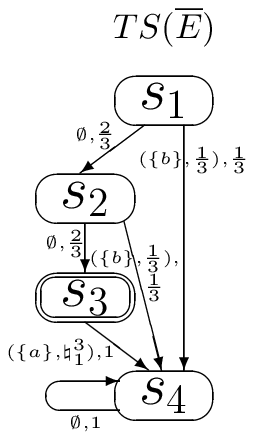}
\end{center}
\caption{The transition system of $\overline{E}$ for $E=(\{a\},\natural_1^3)\cho (\{b\},\frac{1}{3})$}
\label{tschowsm.fig}
\end{figure}

\begin{example}
This example demonstrates a choice between waiting and stochastic multiactions. It shows that the stochastic
multiaction $(\{b\},\frac{1}{3})$ can be executed until the timer value of the waiting multiaction
$(\{a\},\natural_1^3)$ beco\-mes $1$, after which only the waiting multiaction can be executed in the next moment,
leading to an absorbing~state. Thus, in our setting, a waiting multiaction that cannot be executed in the next time
moment and whose timer is still running may be interrupted (preempted) by executing a stochastic multiaction.

Let $E=(\{a\},\natural_1^3)\cho (\{b\},\frac{1}{3})$. $DR(\overline{E})$ consists of the equivalence classes

$$\begin{array}{ll}
s_1=[\overline{(\{a\},\natural_1^3)^3}\cho (\{b\},\frac{1}{3})]_\approx =
[(\{a\},\natural_1^3)^3\cho\overline{(\{b\},\frac{1}{3})}]_\approx , &
s_2=[\overline{(\{a\},\natural_1^3)^2}\cho (\{b\},\frac{1}{3})]_\approx =
[(\{a\},\natural_1^3)^2\cho\overline{(\{b\},\frac{1}{3})}]_\approx ,\\[1mm]
s_3=[\overline{(\{a\},\natural_1^3)^1}\cho (\{b\},\frac{1}{3})]_\approx =
[(\{a\},\natural_1^3)^1\cho\overline{(\{b\},\frac{1}{3})}]_\approx , &
s_4=[\underline{(\{a\},\natural_1^3)\cho (\{b\},\frac{1}{3})}]_\approx .
\end{array}$$

We have $DR_{ST}(\overline{E})=\{s_1,s_2,s_4\},\ DR_{WT}(\overline{E})=\{s_3\}$ and $DR_V(\overline{E})=\emptyset$.

In Figure \ref{tschowsm.fig}, the transition system $TS(\overline{E})$ is shown. The s-tangible and w-tangible states
are depicted in ordinary and double ovals, respectively.
\label{tschowsm.exm}
\end{example}

\begin{figure}
\begin{center}
\includegraphics{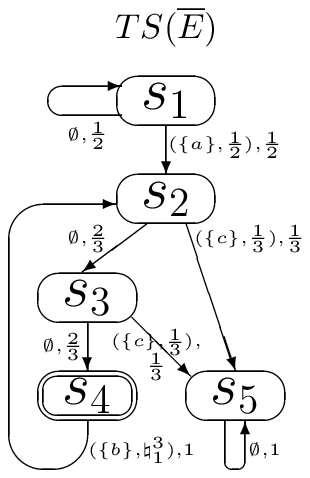}
\end{center}
\caption{The transition system of $\overline{E}$ for $E=[(\{a\},\frac{1}{2})*(\{b\},\natural_1^3)*
(\{c\},\frac{1}{3})]$}
\label{tsitswm.fig}
\end{figure}

\begin{example}
This example demonstrates an iteration loop with a waiting multiaction. The iteration initiation is modeled by a
(initiating) stochastic multiaction $(\{a\},\frac{1}{2})$. The iteration body that corresponds to the loop consists of
a (looping) waiting multiaction $(\{b\},\natural_1^3)$. The iteration termination is represented by a (terminating)
stochastic multiaction $(\{c\},\frac{1}{3})$. The terminating stochastic multiaction can be executed until the timer
value of the waiting multiaction becomes $1$, after which only the waiting multiaction can be executed in the next
moment. Thus, the iteration termination can either complete the repeated execution of the iteration body or break its
execution when the waiting multiaction timer shows some intermediate value (that is less than the initial value, being
the multiaction delay, but greater than $1$). The execution of the waiting multiaction leads to the repeated start of
the iteration body. The execution of the terminating stochastic multiaction brings to the final absorbing state of the
iteration construction.

Let $E=[(\{a\},\frac{1}{2})*(\{b\},\natural_1^3)*(\{c\},\frac{1}{3})]$. $DR(\overline{E})$ consists of the equivalence
classes

$$\begin{array}{l}
s_1=[[\overline{(\{a\},\frac{1}{2})}*(\{b\},\natural_1^3)*(\{c\},\frac{1}{3})]]_\approx ,\\[1.5mm]
s_2=[[(\{a\},\frac{1}{2})*\overline{(\{b\},\natural_1^3)^3}*(\{c\},\frac{1}{3})]]_\approx =
[[(\{a\},\frac{1}{2})*(\{b\},\natural_1^3)^3*\overline{(\{c\},\frac{1}{3})}]]_\approx ,\\[1.5mm]
s_3=[[(\{a\},\frac{1}{2})*\overline{(\{b\},\natural_1^3)^2}*(\{c\},\frac{1}{3})]]_\approx =
[[(\{a\},\frac{1}{2})*(\{b\},\natural_1^3)^2*\overline{(\{c\},\frac{1}{3})}]]_\approx ,\\[1.5mm]
s_4=[[(\{a\},\frac{1}{2})*\overline{(\{b\},\natural_1^3)^1}*(\{c\},\frac{1}{3})]]_\approx =
[[(\{a\},\frac{1}{2})*(\{b\},\natural_1^3)^1*\overline{(\{c\},\frac{1}{3})}]]_\approx ,\\[1.5mm]
s_5=[\underline{[(\{a\},\frac{1}{2})*(\{b\},\natural_1^3)*(\{c\},\frac{1}{3})]}]_\approx .
\end{array}$$

We have $DR_{ST}(\overline{E})=\{s_1,s_2,s_3,s_5\},\ DR_{WT}(\overline{E})=\{s_4\}$ and $DR_V(\overline{E})=\emptyset$.

In Figure \ref{tsitswm.fig}, the transition system $TS(\overline{E})$ is shown. The s-tangible and w-tangible states
are depicted in ordinary and double ovals, respectively.
\label{tsitswm.exm}
\end{example}

\begin{figure}
\begin{center}
\includegraphics{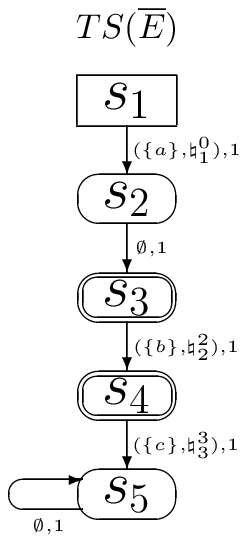}
\end{center}
\caption{The transition system of $\overline{E}$ for $E=(\{a\},\natural_1^0)\| (\{b\},\natural_2^2)\|
(\{c\},\natural_3^3)$}
\label{tspariwm.fig}
\end{figure}

\begin{example}
This example demonstrates a parallel composition of an immediate and two waiting multiactions with different delays. It
shows that the immediate multiaction $(\{a\},\natural_1^0)$ is always executed before any parallel with it waiting
multiaction. Further, from the two parallel waiting multiactions, that $(\{b\},\natural_2^2)$ with a less delay $2$
executed first in any case. Finally, the execution of the waiting multiaction $(\{c\},\natural_3^3)$ with a greater
delay $3$ leads to an absorbing state. Thus, in spite of parallelism of those three deterministic multiactions, they
are executed sequentially in fact, in the increasing order of their (different) delays. That sequence also includes the
empty set, executed after the immediate multiaction, since the waiting multiaction with a less delay will then need a
passage of one time unit (one time tick) for its timer value (RTE) become $1$ and it can be executed itself. Though the
example is not complex, it shows a transition system with all
types of states: s-tangible, w-tangible and~vanishing.

Let $E=(\{a\},\natural_1^0)\| (\{b\},\natural_2^2)\| (\{c\},\natural_3^3)$. $DR(\overline{E})$ consists of the
equivalence classes

$$\begin{array}{ll}
s_1=[\overline{(\{a\},\natural_1^0)}\|\overline{(\{b\},\natural_2^2)^2}\|
\overline{(\{c\},\natural_3^3)^3}]_\approx , &
s_2=[\underline{(\{a\},\natural_1^0)}\|\overline{(\{b\},\natural_2^2)^2}\|
\overline{(\{c\},\natural_3^3)^3}]_\approx ,\\[1.5mm]
s_3=[\underline{(\{a\},\natural_1^0)}\|\overline{(\{b\},\natural_2^2)^1}\|
\overline{(\{c\},\natural_3^3)^2}]_\approx , &
s_4=[\underline{(\{a\},\natural_1^0)}\|\underline{(\{b\},\natural_2^2)}\|
\overline{(\{c\},\natural_3^3)^1}]_\approx ,\\[1.5mm]
s_5=[\underline{(\{a\},\natural_1^0)\|(\{b\},\natural_2^2)\|(\{c\},\natural_3^3)}]_\approx . &
\end{array}$$

We have $DR_{ST}(\overline{E})=\{s_2,s_5\},\ DR_{WT}(\overline{E})=\{s_3,s_4\}$ and $DR_V(\overline{E})=\{s_1\}$.

In Figure \ref{tspariwm.fig}, the transition system $TS(\overline{E})$ is shown. The s-tangible and w-tangible states
are depicted in ordinary and double ovals, respectively, and the vanishing ones are depicted in boxes.
\label{tspariwm.exm}
\end{example}

\begin{figure}
\begin{center}
\includegraphics{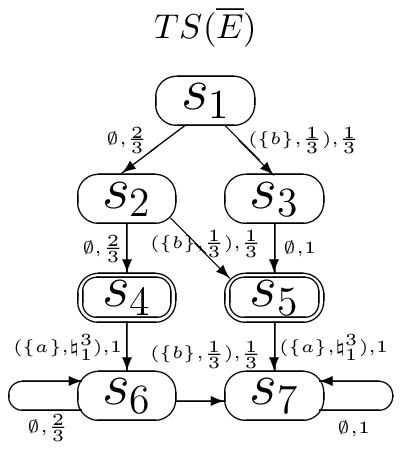}
\end{center}
\caption{The transition system of $\overline{E}$ for $E=(\{a\},\natural_1^3)\| (\{b\},\frac{1}{3})$}
\label{tsparwsm.fig}
\end{figure}

\begin{example}
This example demonstrates a parallel composition of waiting and stochastic multiactions. It shows that the stochastic
multiaction $(\{b\},\frac{1}{3})$ can be executed until the timer value of the waiting multiaction
$(\{a\},\natural_1^3)$ becomes $1$, after which only the waiting multiaction can be executed in the next moment. The
execution of the latter leads to an absorbing state either directly or indirectly, via executing a possible empty loop,
followed (via sequential composition) by the stochastic multiaction that has not been executed in the preceding states.

Let $E=(\{a\},\natural_1^3)\| (\{b\},\frac{1}{3})$. $DR(\overline{E})$ consists of the equivalence classes

$$\begin{array}{lll}
s_1=[\overline{(\{a\},\natural_1^3)^3}\|\overline{(\{b\},\frac{1}{3})}]_\approx , &
s_2=[\overline{(\{a\},\natural_1^3)^2}\|\overline{(\{b\},\frac{1}{3})}]_\approx , &
s_3=[\overline{(\{a\},\natural_1^3)^2}\|\underline{(\{b\},\frac{1}{3})}]_\approx ,\\[1.5mm]
s_4=[\overline{(\{a\},\natural_1^3)^1}\|\overline{(\{b\},\frac{1}{3})}]_\approx , &
s_5=[\overline{(\{a\},\natural_1^3)^1}\|\underline{(\{b\},\frac{1}{3})}]_\approx ,&
s_6=[\underline{(\{a\},\natural_1^3)}\|\overline{(\{b\},\frac{1}{3})}]_\approx ,\\[1.5mm]
s_7=[\underline{(\{a\},\natural_1^3)\|(\{b\},\frac{1}{3})}]_\approx . & &
\end{array}$$

We have $DR_{ST}(\overline{E})=\{s_1,s_2,s_3,s_6,s_7\},\ DR_{WT}(\overline{E})=\{s_4,s_5\}$ and
$DR_V(\overline{E})=\emptyset$.

In Figure \ref{tsparwsm.fig}, the transition system $TS(\overline{E})$ is shown. The s-tangible and w-tangible states
are depicted in ordinary and double ovals, respectively.
\label{tsparwsm.exm}
\end{example}

\begin{figure}
\begin{center}
\includegraphics{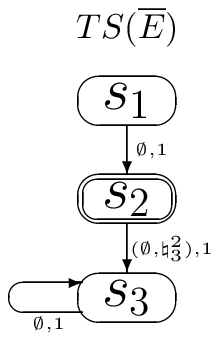}
\end{center}
\caption{The transition system of $\overline{E}$ for $E=((\{a\},\natural_1^2)\| (\{\hat{a}\},\natural_2^2))\sy a\rs a$}
\label{tsparsyrswm.fig}
\end{figure}

\begin{example}
This example demonstrates a parallel composition of two waiting multiactions $(\{a\},\natural_1^2)$ and
$(\{\hat{a}\},\natural_2^2)$, whose multiaction parts are singleton multisets with an action $a$ and its conjugate
$\hat{a}$, respectively. The resulting composition is synchronized and then restricted by that action, which (and its
conjugate) therefore ``disappears'' from the composite process behaviour. From the initial state, only the empty
multiset of activities is executed that decrements by one the values of the timers. That evolution follows by the
execution of a new waiting multiaction $(\emptyset ,\natural_3^2)$ with the empty multiaction part, resulted from
synchronization of the two waiting multiactions, which leads to an absorbing state.

Note that the timer values of the two waiting multiactions and that of the new waiting multiaction (being their
synchronous product) coincide until all of them remain
enabled with the time progress. Thus, it is very useful that the expression syntax preserves such two
enabled synchronized waiting multiactions, removed by restriction from the behaviour, since their timer values suggest
that of their synchronous product, which is not explicit in the syntax. Thus, the timer values of those two ``virtual''
enabled waiting multiactions cannot just be marked as undefined in the syntax, provided that one keeps track of the
timer value of their synchronous product being only implicit in the syntax.

If both synchronized waiting multiactions lose their
enabledness with the time progress then their synchronous product also loses its
enabledness and all of them obviously loose their timer value annotations. It may happen that one of the synchronized
waiting multiactions loses its
enabledness (for example, when a conflicting waiting multiaction is executed) while the other one keeps its
enabledness. Then their synchronous product also loses its
enabledness, together with its timer value annotation. In such a case, the timer value of the
enabled synchronized waiting multiaction does not suggest anymore that of the synchronous product. That ``saved'' timer
value merely decrements with every time tick unless it becomes equal to $1$, after which either the
enabled synchronized waiting multiaction is executed or it cannot be executed by some reason (for example, when
affected by restriction) and then the timer value $1$ remains unchanged with the time progress. To verify this, recall
the empty move rule {\bf E} from Table \ref{actrulesdm.tab} and the definition of $\circlearrowleft\!G$ with
$\max\{1,\delta -1\}=\max\{1,0\}=1$ when $\delta =1$.

Let $E=((\{a\},\natural_1^2)\| (\{\hat{a}\},\natural_2^2))\sy a\rs a$. $DR(\overline{E})$ consists of the equivalence
classes

$$\begin{array}{ll}
s_1=[(\overline{(\{a\},\natural_1^2)^2}\|\overline{(\{\hat{a}\},\natural_2^2)^2})\sy a\rs a]_\approx , &
s_2=[(\overline{(\{a\},\natural_1^2)^1}\|\overline{(\{\hat{a}\},\natural_2^2)^1})\sy a\rs a]_\approx ,\\[1mm]
s_3=[\underline{((\{a\},\natural_1^2)\| (\{\hat{a}\},\natural_2^2))\sy a\rs a}]_\approx . &
\end{array}$$

We have $DR_{ST}(\overline{E})=\{s_1,s_3\},\ DR_{WT}(\overline{E})=\{s_2\}$ and $DR_V(\overline{E})=\emptyset$.

In Figure \ref{tsparsyrswm.fig}, the transition system $TS(\overline{E})$ is shown. The s-tangible and w-tangible
states are depicted in ordinary and double ovals, respectively.
\label{tsparsyrswm.exm}
\end{example}

\begin{figure}
\begin{center}
\includegraphics{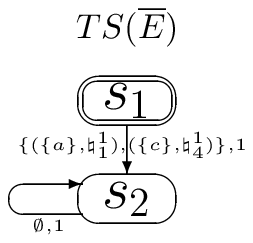}
\end{center}
\caption{The transition system of $\overline{E}$ for $E=(((\{a\},\natural_1^1);(\{b,\hat{x}\},\natural_2^0))\|
((\{x\},\natural_3^0)\cho (\{c\},\natural_4^1)))\sy x\rs x$}
\label{tsparsyrsiwm.fig}
\end{figure}

\begin{example}
This example demonstrates a parallel composition of two subprocesses, synchronized and then restricted by an auxiliary
action that (and its conjugate) hereupon ``disappears'' from the composite process behaviour. The first subprocess is a
sequential composition of the waiting $(\{a\},\natural_1^1)$ and immediate $(\{b,\hat{x}\},\natural_2^0)$ multiactions.
The second subprocess is a choice between the immediate $(\{x\},\natural_3^0)$ and waiting $(\{c\},\natural_4^1)$
multiactions. The immediate multiactions in the first and second subprocesses are synchronized via an auxiliary action
$x$ that (and its conjugate) is then removed from the behaviour by the restriction operation. Since those immediate
multiactions are within coverage of restriction by the auxiliary action, they cannot be executed. The new immediate
multiaction $(\{b\},\natural_5^0)$, resulted from that synchronization can only be executed if the waiting multiaction
(preceding it via sequential composition) in the first subprocess has occurred and the waiting multiaction (conflicting
with it via the choice composition) in the second subprocess has not occurred. Since only maximal multisets of parallel
waiting multiactions may be executed, the waiting multiactions in both the subprocesses must occur, thus preventing
execution of the new immediate multiaction, generated by synchronization.

Let $E=(((\{a\},\natural_1^1);(\{b,\hat{x}\},\natural_2^0))\| ((\{x\},\natural_3^0)\cho
(\{c\},\natural_4^1)))\sy x\rs x$. $DR(\overline{E})$ consists of the equivalence classes

$$\begin{array}{l}
s_1=[((\overline{(\{a\},\natural_1^1)^1};(\{b,\hat{x}\},\natural_2^0))\|
((\{x\},\natural_3^0)\cho\overline{(\{c\},\natural_4^1)^1}))\sy x\rs x]_\approx =\\[1mm]
\hspace{8mm}[((\overline{(\{a\},\natural_1^1)^1};(\{b,\hat{x}\},\natural_2^0))\|
(\overline{(\{x\},\natural_3^0)}\cho (\{c\},\natural_4^1)^1))\sy x\rs x]_\approx ,\\[1mm]
s_2=[(((\{a\},\natural_1^1);\overline{(\{b,\hat{x}\},\natural_2^0)})\|
\underline{((\{x\},\natural_3^0)\cho (\{c\},\natural_4^1))})\sy x\rs x]_\approx .
\end{array}$$

We have $DR_{ST}(\overline{E})=\{s_2\},\ DR_{WT}(\overline{E})=\{s_1\}$ and $DR_V(\overline{E})=\emptyset$.

In Figure \ref{tsparsyrsiwm.fig}, the transition system $TS(\overline{E})$ is shown. The s-tangible and w-tangible states
are depicted in ordinary and double ovals, respectively.
\label{tsparsyrsiwm.exm}
\end{example}

\begin{figure}
\begin{center}
\includegraphics{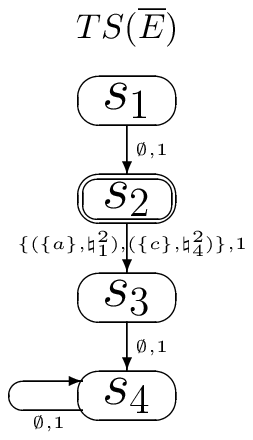}
\end{center}
\caption{The transition system of $\overline{E}$ for $E=((((\{a\},\natural_1^2);(\{b,\hat{x}\},\natural_2^2))\|
((\{x\},\natural_3^2)\cho (\{c\},\natural_4^2)))\sy x\rs x$}
\label{tsparsyrswwm.fig}
\end{figure}

\begin{example}
This example is a modification of the previous Example \ref{tsparsyrsiwm.exm} by replacing all the immediate
multiactions with the waiting ones and by setting to $2$ the delays of all the waiting multiactions from the syntax.
Thus, we examine a compound process, constructed with parallelism, synchronization and restriction operations from the
following two subprocesses. The first subprocess is a sequential composition of two waiting multiactions
$(\{a\},\natural_1^2)$ and $(\{b,\hat{x}\},\natural_2^2)$. The second subprocess is a choice between other two waiting
multiactions $(\{x\},\natural_3^2)$ and $(\{c\},\natural_4^2)$. The second waiting multiaction in the first subprocess
and the first waiting multiaction in the second subprocess are synchronized via an auxiliary action $x$ that (and its
conjugate) is then removed from the behaviour by the restriction operation. The new waiting multiaction
$(\{b\},\natural_5^2)$, resulted from that synchronization has the same delay $2$ as the two synchronized waiting
multiactions. It can only be executed if the first waiting multiaction (preceding it via sequential composition) in the
first subprocess has occurred and the second waiting multiaction (conflicting with it via the choice composition) in
the second subprocess has not occurred. Since only maximal multisets of parallel waiting multiactions may be executed,
the mentioned (``first in first'' and ``second in second'') waiting multiactions in both the subprocesses must occur,
thus preventing execution of the new waiting multiaction, generated by synchronization.

Note that the overlined second waiting multiaction in the first subprocess is within coverage of restriction by the
auxiliary action. Consider the state, reached from the initial state by execution of the empty multiset of activities,
followed by the parallel execution of the mentioned (`first in first'' and ``second in second'') waiting multiactions.
After the empty multiset execution from the considered state, the associated timer value of that overlined waiting
multiaction is decremented to $1$. Then an absorbing state is reached, from which only the empty loop is possible,
which leaves that timer value $1$ unchanged though. To verify this, recall the empty move rule {\bf E} from Table
\ref{actrulesdm.tab} and the definition of $\circlearrowleft\!G$ with $\max\{1,\delta -1\}=\max\{1,0\}=1$ when $\delta
=1$.

Let $E=(((\{a\},\natural_1^2);(\{b,\hat{x}\},\natural_2^2))\| ((\{x\},\natural_3^2)\cho
(\{c\},\natural_4^2)))\sy x\rs x$. $DR(\overline{E})$ consists of the equivalence classes

$$\begin{array}{l}
s_1=[((\overline{(\{a\},\natural_1^2)^2};(\{b,\hat{x}\},\natural_2^2))\|
((\{x\},\natural_3^2)^2\cho\overline{(\{c\},\natural_4^2)^2}))\sy x\rs x]_\approx =\\[1mm]
\hspace{8mm}[((\overline{(\{a\},\natural_1^2)^2};(\{b,\hat{x}\},\natural_2^2))\|
(\overline{(\{x\},\natural_3^2)^2}\cho (\{c\},\natural_4^2)^2))\sy x\rs x]_\approx ,\\[1mm]
s_2=[((\overline{(\{a\},\natural_1^2)^1};(\{b,\hat{x}\},\natural_2^2))\|
((\{x\},\natural_3^2)^1\cho\overline{(\{c\},\natural_4^2)^1}))\sy x\rs x]_\approx =\\[1mm]
\hspace{8mm}[((\overline{(\{a\},\natural_1^2)^1};(\{b,\hat{x}\},\natural_2^2))\|
(\overline{(\{x\},\natural_3^2)^1}\cho (\{c\},\natural_4^2)^1))\sy x\rs x]_\approx ,\\[1mm]
s_3=[(((\{a\},\natural_1^2);\overline{(\{b,\hat{x}\},\natural_2^2)^2})\|
\underline{((\{x\},\natural_3^2)\cho (\{c\},\natural_4^2))})\sy x\rs x]_\approx ,\\[1mm]
s_4=[(((\{a\},\natural_1^2);\overline{(\{b,\hat{x}\},\natural_2^2)^1})\|
\underline{((\{x\},\natural_3^2)\cho (\{c\},\natural_4^2))})\sy x\rs x]_\approx .
\end{array}$$

We have $DR_{ST}(\overline{E})=\{s_1,s_3,s_4\},\ DR_{WT}(\overline{E})=\{s_2\}$ and $DR_V(\overline{E})=\emptyset$.

In Figure \ref{tsparsyrswwm.fig}, the transition system $TS(\overline{E})$ is shown. The s-tangible and w-tangible states
are depicted in ordinary and double ovals, respectively.
\label{tsparsyrswwm.exm}
\end{example}

\begin{figure}
\begin{center}
\includegraphics{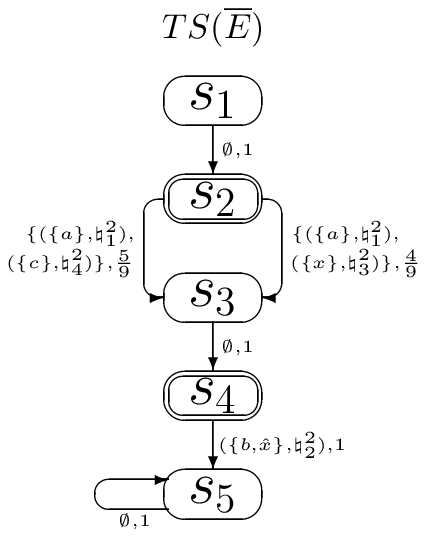}
\end{center}
\caption{The transition system of $\overline{E}$ for $E=((((\{a\},\natural_1^2);(\{b,\hat{x}\},\natural_2^2))\|
((\{x\},\natural_3^2)\cho (\{c\},\natural_4^2)))\sy x$}
\label{tsparsywwm.fig}
\end{figure}

\begin{example}
This example is a modification of the previous Example \ref{tsparsyrswwm.exm} by removing restriction from the syntax.
Thus, we examine a compound process, constructed with parallelism and synchronization operations from the two
subprocesses being a sequential composition of two waiting multiactions $(\{a\},\natural_1^2)$ and
$(\{b,\hat{x}\},\natural_2^2)$ and a choice between other two waiting multiactions $(\{x\},\natural_3^2)$ and
$(\{c\},\natural_4^2)$, respectively. All the four waiting multiactions have the same delay $2$. The second waiting
multiaction in the first subprocess and the first waiting multiaction in the second subprocess are synchronized via an
auxiliary action $x$. The new waiting multiaction $(\{b\},\natural_5^2)$, resulted from that synchronization has the
same delay $2$ as the two synchronized waiting multiactions. It can only be executed if the first waiting multiaction
(preceding it via sequential composition) in the first subprocess has occurred and the second waiting multiaction
(conflicting with it via the choice composition) in the second subprocess has not occurred. Since only maximal
multisets of parallel waiting multiactions may be executed, the mentioned (``first in first'' and ``second in second'')
waiting multiactions in the subprocesses must occur, thus preventing execution of the new waiting multiaction,
generated by synchronization. The alternative maximal multiset of parallel waiting multiactions that may be executed
from the same state consists of the ``first in first'' and ``first in second'' waiting multiactions in the
subprocesses, but the `first in second'' waiting multiaction is the second of the two synchronized waiting
multiactions, whose occurrence also prevents execution of their synchronization product.

Let $E=(((\{a\},\natural_1^2);(\{b,\hat{x}\},\natural_2^2))\| ((\{x\},\natural_3^2)\cho
(\{c\},\natural_4^2)))\sy x$. $DR(\overline{E})$ consists of the equivalence classes

$$\begin{array}{l}
s_1=[((\overline{(\{a\},\natural_1^2)^2};(\{b,\hat{x}\},\natural_2^2))\|
(\overline{(\{x\},\natural_3^2)^2}\cho (\{c\},\natural_4^2)^2))\sy x]_\approx =\\[1mm]
\hspace{8mm}[((\overline{(\{a\},\natural_1^2)^2};(\{b,\hat{x}\},\natural_2^2))\|
((\{x\},\natural_3^2)^2\cho\overline{(\{c\},\natural_4^2)^2}))\sy x]_\approx ,\\[1mm]
s_2=[((\overline{(\{a\},\natural_1^2)^1};(\{b,\hat{x}\},\natural_2^2))\|
(\overline{(\{x\},\natural_3^2)^1}\cho (\{c\},\natural_4^2)^1))\sy x]_\approx =\\[1mm]
\hspace{8mm}[((\overline{(\{a\},\natural_1^2)^1};(\{b,\hat{x}\},\natural_2^2))\|
((\{x\},\natural_3^2)^1\cho\overline{(\{c\},\natural_4^2)^1}))\sy x]_\approx ,\\[1mm]
s_3=[(((\{a\},\natural_1^2);\overline{(\{b,\hat{x}\},\natural_2^2)^2})\|
\underline{((\{x\},\natural_3^2)\cho (\{c\},\natural_4^2))})\sy x]_\approx ,\\[1mm]
s_4=[(((\{a\},\natural_1^2);\overline{(\{b,\hat{x}\},\natural_2^2)^1})\|
\underline{((\{x\},\natural_3^2)\cho (\{c\},\natural_4^2))})\sy x]_\approx ,\\[1mm]
s_5=[\underline{(((\{a\},\natural_1^2);(\{b,\hat{x}\},\natural_2^2))\|
((\{x\},\natural_3^2)\cho (\{c\},\natural_4^2)))\sy x}]_\approx .
\end{array}$$

We have $DR_{ST}(\overline{E})=\{s_1,s_3,s_5\},\ DR_{WT}(\overline{E})=\{s_2,s_4\}$ and $DR_V(\overline{E})=\emptyset$.

In Figure \ref{tsparsywwm.fig}, the transition system $TS(\overline{E})$ is shown. The s-tangible and w-tangible states
are depicted in ordinary and double ovals, respectively.
\label{tsparsywwm.exm}
\end{example}

\begin{figure}
\begin{center}
\includegraphics{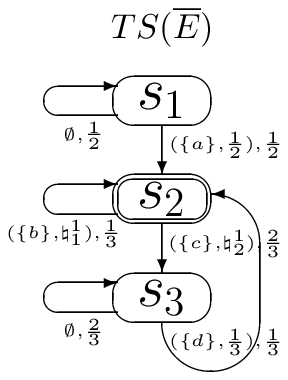}
\end{center}
\caption{The transition system of $\overline{E}$ for $E=[(\{a\},\frac{1}{2})*((\{b\},\natural_1^1)\cho
((\{c\},\natural_2^1);(\{d\},\frac{1}{3})))*{\sf Stop}]$}
\label{tsitchoswm.fig}
\end{figure}

\begin{example}
Consider the expression ${\sf Stop}=(\{g\},\frac{1}{2})\rs g$ specifying the special process that is only able to
perform empty loops with probability $1$ and never terminates. We could actually use any arbitrary action from ${\cal
A}$ and any probability belonging to the interval $(0;1)$ in the definition of ${\sf Stop}$. Note that ${\sf Stop}$ is
analogous to the one used in the examples within sPBC. The latter is a continuous time stochastic analogue of the {\sf
stop} process proposed in \cite{BDK01}. {\sf Stop} is a discrete time stochastic analogue of the {\sf stop}.

This example demonstrates an infinite iteration loop. The loop is preceded with the iteration initiation, modeled by a
(first) stochastic multiaction $(\{a\},\frac{1}{2})$. The iteration body that corresponds to the loop consists of the
choice between two conflicting waiting multiactions $(\{b\},\natural_1^1)$ and $(\{c\},\natural_2^1)$ with the same
delay
$1$, the second of them followed (via sequential composition) by a (second) stochastic multiaction
$(\{d\},\frac{1}{3})$. Hence, the iteration loop actually consists of the two alternative subloops, such that the first
one is a self-loop (one-state loop from a state to itself) with the first waiting multiaction, and the second one is a
two-state loop with an intermediate state, reached after the second waiting multiaction has been executed, and from
which the second stochastic multiaction is then started. Thus, the iteration generates the self-loop with probability
less than one (since the two-state loop from the same state has a non-zero probability) from the states in which only
waiting multiactions are executed. The iteration termination ${\sf Stop}$ demonstrates an empty behaviour, assuring
that the iteration does not reach its final state after any number of repeated executions of its body.

Let $E=[(\{a\},\frac{1}{2})*((\{b\},\natural_1^1)\cho ((\{c\},\natural_2^1);(\{d\},\frac{1}{3})))*{\sf Stop}]$.
$DR(\overline{E})$ consists of the equivalence classes

$$\begin{array}{l}
s_1=[[\overline{(\{a\},\frac{1}{2})}*((\{b\},\natural_1^1)\cho ((\{c\},\natural_2^1);(\{d\},\frac{1}{3})))*
{\sf Stop}]]_\approx ,\\[1mm]
s_2=[[(\{a\},\frac{1}{2})*(\overline{(\{b\},\natural_1^1)^1}\cho ((\{c\},\natural_2^1)^1;(\{d\},\frac{1}{3})))*
{\sf Stop}]]_\approx =\\[1mm]
\hspace{8mm}[[(\{a\},\frac{1}{2})*((\{b\},\natural_1^1)^1\cho (\overline{(\{c\},\natural_2^1)^1};(\{d\},\frac{1}{3})))*
{\sf Stop}]]_\approx ,\\[1mm]
s_3=[[(\{a\},\frac{1}{2})*((\{b\},\natural_1^1)\cho ((\{c\},\natural_2^1);\overline{(\{d\},\frac{1}{3})}))*
{\sf Stop}]]_\approx .
\end{array}$$

We have $DR_{ST}(\overline{E})=\{s_1,s_3\},\ DR_{WT}(\overline{E})=\{s_2\}$ and $DR_V(\overline{E})=\emptyset$.

In Figure \ref{tsitchoswm.fig}, the transition system $TS(\overline{E})$ is presented. The s-tangible states are
depicted in ovals and the vanishing ones are depicted in boxes.
\label{tsitchoswm.exm}
\end{example}

\begin{figure}
\begin{center}
\includegraphics{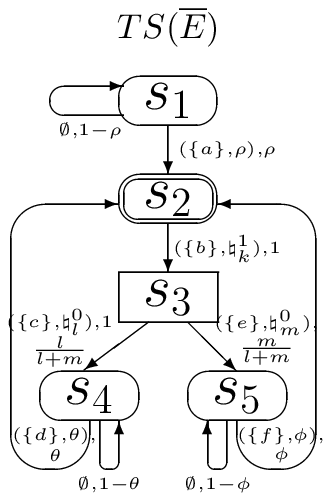}
\end{center}
\caption{The transition system of $\overline{E}$ for $E=[(\{a\},\rho )*((\{b\},\natural_k^1);(((\{c\},\natural_l^0);
(\{d\},\theta ))\cho ((\{e\},\natural_m^0);(\{f\},\phi ))))*{\sf Stop}]$}
\label{tsitchoswim.fig}
\end{figure}

\begin{example}
This example demonstrates an infinite iteration loop. The loop is preceded with the iteration initiation, modeled by a
stochastic multiaction $(\{a\},\rho )$. The iteration body that corresponds to the loop consists of a waiting
multiaction $(\{b\},\natural_k^1)$, followed (via sequential composition) by the probabilistic choice, modeled via two
conflicting immediate multiactions $(\{c\},\natural_l^0)$ and $(\{e\},\natural_m^0)$,
followed
by different stochastic multiactions $(\{d\},\theta )$ and $(\{f\},\phi )$. The iteration termination ${\sf Stop}$
demonstrates an empty behaviour, assuring that the iteration does not reach its final state after any number of
repeated executions of its body.

Let $E=[(\{a\},\rho )*((\{b\},\natural_k^1);(((\{c\},\natural_l^0);(\{d\},\theta ))\cho
((\{e\},\natural_m^0);(\{f\},\phi ))))*{\sf Stop}]$, where $\rho ,\theta ,\phi\in (0;1)$ and $k,l,m\in\real_{>0}$.
$DR(\overline{E})$ consists of the equivalence classes

$$\begin{array}{l}
s_1=[[\overline{(\{a\},\rho )}*((\{b\},\natural_k^1);(((\{c\},\natural_l^0);(\{d\},\theta ))\cho
((\{e\},\natural_m^0);(\{f\},\phi ))))*{\sf Stop}]]_\approx ,\\[1mm]
s_2=[[(\{a\},\rho )*(\overline{(\{b\},\natural_k^1)^1};(((\{c\},\natural_l^0);(\{d\},\theta ))\cho
((\{e\},\natural_m^0);(\{f\},\phi ))))*{\sf Stop}]]_\approx ,\\[1mm]
s_3=[[(\{a\},\rho )*((\{b\},\natural_k^1);((\overline{(\{c\},\natural_l^0)};(\{d\},\theta ))\cho
((\{e\},\natural_m^0);(\{f\},\phi ))))*{\sf Stop}]]_\approx =\\[1mm]
\hspace{8mm}[[(\{a\},\rho )*((\{b\},\natural_k^1);(((\{c\},\natural_l^0);(\{d\},\theta ))\cho
(\overline{(\{e\},\natural_m^0)};(\{f\},\phi ))))*{\sf Stop}]]_\approx ,\\[1mm]
s_4=[[(\{a\},\rho )*((\{b\},\natural_k^1);(((\{c\},\natural_l^0);\overline{(\{d\},\theta ))}\cho
((\{e\},\natural_m^0);(\{f\},\phi ))))*{\sf Stop}]]_\approx ,\\[1mm]
s_5=[[(\{a\},\rho )*((\{b\},\natural_k^1);(((\{c\},\natural_l^0);(\{d\},\theta ))\cho
((\{e\},\natural_m^0);\overline{(\{f\},\phi ))}))*{\sf Stop}]]_\approx .
\end{array}$$

We have $DR_{ST}(\overline{E})=\{s_1,s_4,s_5\},\ DR_{WT}(\overline{E})=\{s_2\}$ and $DR_V(\overline{E})=\{s_3\}$.

In Figure \ref{tsitchoswim.fig}, the transition system $TS(\overline{E})$ is presented. The s-tangible and w-tangible
states are depicted in ordinary and double ovals, respectively, and the vanishing ones are depicted in boxes.
\label{tsitchoswim.exm}
\end{example}

\section{Denotational semantics}
\label{denosem.sec}

In this section, we construct the denotational semantics in terms of a subclass of labeled discrete time stochastic and
deterministic PNs (LDTSDPNs), called discrete time stochastic and immediate Petri boxes (dtsd-boxes).

\subsection{Labeled DTSDPNs}

Let us introduce a class of labeled discrete time stochastic and deterministic PNs (LDTSDPNs), which are essentially a
subclass of DTSPNs \cite{Mol81,Mol85} (since we do not allow the stochastic transition probabilities to be equal to
$1$) extended with transition labeling and deterministic transitions. LDTSDPNs resemble in part discrete time
deterministic and stochastic PNs (DTDSPNs) \cite{Zij95,Zij97,ZFGH00,ZFH01}, as well as discrete deterministic and
stochastic PNs (DDSPNs) \cite{ZCH97}. DTDSPNs and DDSPNs are the extensions of DTSPNs with deterministic transitions
(having fixed delay that can be zero), inhibitor arcs, priorities and guards. In addition, while stochastic transitions
of DTDSPNs, like those of DTSPNs, have geometrically distributed delays, stochastic transitions of DDSPNs have discrete
time phase-type distributed delays. At the same time, LDTSDPNs are not subsumed by DTDSPNs or DDSPNs, by the following
reasons. First, in DTDSPNs from \cite{Zij95,Zij97}, both stochastic and deterministic (including immediate) transitions
have probabilities and weights associated, but in LDTSDPNs only stochastic transitions have probabilities and only
immediate ones have weights, hence, the state change probabilities of the underlying Markov chains for those PN classes
are calculated in two different ways. Second, LDTSDPNs have a step semantics while DTDSPNs from \cite{ZFGH00,ZFH01} and
DDSPNs have interleaving one, since in in the first PN class simultaneous transition firings are possible while in the
second and third PN classes only firings of single transitions are allowed. LDTSDPNs are somewhat similar to labeled
weighted DTSPNs (LWDTSPNs) from \cite{BT01}, but in LWDTSPNs there are no deterministic transitions, all (stochastic)
transitions have weights, the transition probabilities may be equal to $1$ and only maximal fireable subsets of the
enabled transitions are fired.

Stochastic preemptive time PNs (spTPNs) \cite{BSV05} is a discrete time model with a maximal step semantics,
where both time ticks and instantaneous parallel firings of maximal transition sets are possible, but the transition
steps in LDTSDPNs are not obliged to be maximal (excepting the steps of waiting transitions). The transition delays in
spTPNs are governed by static general discrete distributions, associated with the transitions, while the transitions of
LDTSDPNs are only associated with probabilities, used later to calculate the step probabilities after one unit (from
tangible markings) or zero (from vanishing markings) delay. Further, LDTSDPNs have just geometrically distributed or
deterministic zero delays at the markings. Moreover, the discrete time tick and concurrent transition firing are
treated in spTPNs as different events while firing every (possibly empty) set of stochastic or waiting transitions in
LDTSDPNs requires one unit time delay. spTPNs are essentially a modification and extension of unlabeled LWDTSPNs with
additional facilities, such as inhibitor arcs, priorities, resources, preemptions, schedulers etc. However, the price
of such an expressiveness of spTPNs is that the model is rather intricate and difficult to analyze.

Note also that guards in DTDSPNs and DDSPNs, inhibitor arcs and priorities in DTDSPNs, DDSPNs and spTPNs, as well as
the maximal step semantics of LWDTSPNs and spTPNs make all these models Turing powerful, resulting in undecidability of
many important behavioural properties.

First, we present a formal definition (construction, syntax) of LDTSDPNs. The set of {\em all row vectors of
$n\in\nat_{\geq 1}$ elements from a set $X$} is defined as $X^n=\{(x_1,\ldots ,x_n)\mid x_i\in X\ (1\leq i\leq n)\}$.

\begin{definition}
A {\em labeled discrete time stochastic and deterministic PN (LDTSDPN)} is a tuple\\
$N=(P_N,T_N,W_N,D_N,\Omega_N,{\cal L}_N,Q_N)$, where
\begin{itemize}

\item $P_N$ and $T_N=Ts_N\uplus Td_N$ are finite sets of {\em places} and {\em stochastic and deterministic
transitions}, respectively, such that $P_N\cup T_N\neq\emptyset$ and $P_N\cap T_N=\emptyset$;

\item $W_N:(P_N\times T_N)\cup (T_N\times P_N)\rightarrow\nat$ is a function providing the {\em weights of arcs}
between places and transitions;

\item $D_N:Td_N\rightarrow\nat$ is the {\em transition delay} function imposing delays to deterministic transitions;

An {\em immediate transition} is a deterministic transition with the delay $0$ while a {\em waiting transition} is
that with a positive delay. Then $Td_N=Ti_N\uplus Tw_N$ consists of the sets of {\em immediate and waiting
transitions}.

\item $\Omega_N$ is the {\em transition probability and weight} function such that
\begin{itemize}

\item $\Omega_N|_{Ts_N}:Ts_N\rightarrow (0;1)$ (it associates stochastic transitions with probabilities);

\item $\Omega_N|_{Td_N}:Td_N\rightarrow\real_{>0}$ (it associates deterministic transitions with weights);

\end{itemize}

\item ${\cal L}_N:T_N\rightarrow{\cal L}$ is the {\em transition labeling} function assigning multiactions to
transitions;

\item $Q_N=(M_N,V_N)$ is the {\em initial state}, where $M_N\in\nat_{fin}^{P_N}$ is the {\em initial marking}
(distribution of tokens in the places) and
$V_N:Tw_N\rightarrow\nat_{\geq 1}\cup\{*\}$ is the {\em initial timer valuation function} of the waiting
transitions (in the vector notation, $V_N\in (\nat_{\geq 1}\cup\{*\})^{|Tw_N|}$), where `$*$' denotes the undefined
value of inactive timers; we define $\forall t\in Tw_N\cap Ena(M_N)\ V_N(t)=D_N(t)$ (each enabled waiting
transition is initially valuated with its transition delay) and $\forall t\in Tw_N\setminus Ena(M_N)\ V_N(t)=*$
(each non-enabled waiting transition is initially valuated with the undefined value), where $Ena(M)$ denotes the
set of transitions enabled at the marking $M$, to be defined later.

\end{itemize}
\end{definition}

The graphical representation of LDTSDPNs is like that for standard labeled PNs, but with probabilities or delays and
weights written near the corresponding transitions. Square boxes of normal thickness depict stochastic transitions, and
those with thick borders represent deterministic transitions. In the case the probabilities or the delays and weights
are not given in the picture, they are considered to be of no importance in the corresponding examples. The weights of
arcs are depicted with them. The names of places and transitions are depicted near them when needed.

We now consider the semantics of LDTSDPNs.

Let $N$ be an LDTSDPN and $t\in T_N,\ U\in\nat_{fin}^{T_N}$. The {\em precondition} $^\bullet t$ and the {\em
postcondition} $t^\bullet$ of $t$ are the multisets of places defined as $({^\bullet}t)(p)=W_N(p,t)$ and $(t^\bullet
)(p)=W_N(t,p)$. The {\em precondition} $^\bullet U$ and the {\em postcondition} $U^\bullet$ of $U$ are the multisets of
places defined as ${^\bullet}U=\sum_{t\in U}{^\bullet}t$ and $U^\bullet =\sum_{t\in U}t^\bullet$. Note that for
$U=\emptyset$ we have ${^\bullet}\emptyset =\emptyset =\emptyset^\bullet$.

Let $N$ be an LDTSDPN and $Q=(M,V),\widetilde{Q}=(\widetilde{M},\widetilde{V})\in\nat_{fin}^{P_N}\times (\nat_{\geq
1}\cup\{*\})^{|Tw_N|}$ be its states.

Deterministic transitions have a priority over stochastic ones, and there is also difference in priorities between
immediate and waiting transitions. One can assume that all immediate transitions have (the highest) priority $2$ and
all waiting transitions have (the intermediate) priority $1$, whereas all stochastic transitions have (the lowest)
priority $0$. This means that at a marking where all kinds of transitions can occur, immediate transitions always occur
before waiting ones that, in turn, are always executed before stochastic ones.

A transition $t\in T_N$ is {\em enabled}
at a marking $M\in\nat_{fin}^{P_N}$, if ${^\bullet}t\subseteq M$.
%
%
%
%
%
In other words, a transition is enabled at a
marking if it has enough tokens in its input places (i.e. in the places from its precondition) at the
marking.
Let
$Ena(M)$ be the set of {\em all transitions enabled at
$M$}.

Firings of transitions are atomic operations, and transitions can fire in parallel by taking part in steps. We assume
that all transitions participating in a step should differ, hence, only the sets (not multisets) of transitions may
fire. Thus, we do not allow self-concurrency, i.e. firing of transitions in parallel to themselves. This restriction is
introduced to avoid some technical difficulties while calculating probabilities for multisets of transitions as we
shall see after the following formal definitions. Moreover, we do not need to consider self-concurrency, since
denotational semantics of expressions will be defined via dtsd-boxes which are safe LDTSDPNs (hence, no
self-concurrency is possible).

The following definition of fireability respects the prioritization among different types of transitions. A set of
transitions $U\subseteq Ena(M)$ is {\em fireable} in a state $Q=(M,V)$, if ${^\bullet}U\subseteq M$ (i.e. it has enough
tokens in its input places at the substituent marking $M$), and one of the following holds:
\begin{enumerate}

\item $\emptyset\neq U\subseteq Ti_N$; or

\item $\emptyset\neq U\subseteq Tw_N$ and
\begin{itemize}

\item $\forall t\in U\ V(t)=1$,

\item $Ena(M-{^\bullet}U)\cap\{u\in Tw_N\mid V(u)=1\}=\emptyset$,

\item $Ena(M)\subseteq Tw_N\cup Ts_N$; or

\end{itemize}

\item $U\subseteq Ts_N$ and
\begin{itemize}

\item $Ena(M)\subseteq Ts_N$.

\end{itemize}
\end{enumerate}
In other words, a set of transitions $U$ is fireable in a state, if has enough tokens in its input places at the
substituent marking $M$ of the state and the following holds. If $U$ consists of {\em immediate} transitions then it is
enabled, since no additional condition is needed for its fireability. If $U$ consists of {\em waiting} transitions then
the countdown timer value (called remaining time to fire or RTF) of each transition from $U$ equals one, $U$ is a
maximal (by the inclusion relation) set of the enabled at $M$ waiting transitions with the RTF equal to one and enough
tokens in its input places at $M$, and there exist no immediate transitions enabled at $M$. If $U$ is empty or it
consists of {\em stochastic} transitions then there exist no immediate or waiting transition enabled at $M$.
Note that the second condition of item $2$ of the above definition means that no waiting transition (from $Ena(M)$)
with the RTF being one can be added to $U$ so that the resulting transition set will still have enough tokens in its
input places at $M$. This condition is equivalent to the following maximality requirement (informally mentioned above):
$\forall T\subseteq Ena(M),\ (\forall u\in T\ V(u)=1)\wedge ({^\bullet}T\subseteq M)\wedge (U\subseteq T)\ \Rightarrow\
T=U$. Let $Fire(Q)$ be the set of {\em all transition sets fireable in $Q$}.

By the fireability definition, it follows that $Fire(Q)\subseteq 2^{Ti_N}\setminus\{\emptyset\}$ or $Fire(Q)\subseteq
2^{Tw_N}\setminus\{\emptyset\}$, or $Fire(Q)\subseteq 2^{Ts_N}$ (to be convinced of it, check the definition's items in
the reverse order).
The state $Q$ is {\em s-tangible (stochastically tangible)}, denoted by $stang(Q)$, if $Fire(Q)\subseteq 2^{Ts_N}$, in
particular, if $Fire(Q)=\{\emptyset\}$. The state $Q$ is {\em w-tangible (waitingly tangible)}, denoted by $wtang(Q)$,
if $Fire(Q)\subseteq 2^{Tw_N}\setminus\{\emptyset\}$. The state $Q$ is {\em tangible}, denoted by $tang(Q)$, if
$stang(Q)$ or $wtang(Q)$, i.e. $Fire(Q)\subseteq 2^{Ts_N}\cup 2^{Tw_N}$, in particular, if $Fire(Q)=\{\emptyset\}$.
Otherwise, the state $Q$ is {\em vanishing}, denoted by $vanish(Q)$, and in this case $Fire(Q)\subseteq
2^{Ti_N}\setminus\{\emptyset\}$. A transition $t\in Ena(M)$ is {\em fireable} in a state $Q$, denoted by $t\in
Fire(Q)$, if $\{t\}\in Fire(Q)$. If $stang(Q)$ then a stochastic transition $t\in Fire(Q)$ fires with probability
$\Omega_N(t)$ when no different stochastic transition in conflict with it is
fireable. By definition if fireability, if $stang(Q)$ or $vanish(Q)$ then $\forall U\in Fire(Q)\ 2^U\subseteq
Fire(Q)$.

Let $U\in Fire(Q)$ and $U\neq\emptyset$. The {\em probability that the set of stochastic transitions $U$ is ready for
firing in $Q$} or the {\em weight of the set of
deterministic transitions $U$ which is ready for firing in $Q$} is

$$PF(U,Q)=
\left\{
\begin{array}{ll}
\prod_{t\in U}\Omega_N(t)\cdot\prod_{\{u\in Fire(Q)\mid u\not\in U\}}(1-\Omega_N(u)), & stang(Q);\\
\sum_{t\in U}\Omega_N(t), & wtang(Q)\vee vanish(Q).
\end{array}
\right.$$

In the case $U=\emptyset$ and $stang(Q)$ we define

$$PF(\emptyset ,Q)=
\left\{
\begin{array}{ll}
\prod_{u\in Fire(Q)}(1-\Omega_N(u)), & Fire(Q)\neq\emptyset ;\\
1, & Fire(Q)=\emptyset .
\end{array}
\right.$$

Let $U\in Fire(Q)$ and $U\neq\emptyset$ or $U=\emptyset$ and $stang(Q)$. Besides $U$, some other sets of transitions
may be ready for firing in $Q$, hence, a kind of conditioning or normalization is needed to calculate the firing
probability. The parallel firing of the transitions from $U$ changes the state $Q=(M,V)$ to another state
$\widetilde{Q}=(\widetilde{M},\widetilde{V})$, denoted by $Q\stackrel{U}{\rightarrow}_{\cal P}\widetilde{Q}$, where
\begin{enumerate}

\item $\widetilde{M}=M-{^\bullet}U+U^\bullet$;

\item $\forall u\in Tw_N\ \widetilde{V}(u)=
\left\{
\begin{array}{ll}
*, &
u\not\in Ena(\widetilde{M});\\
V_N(u), &
u\in Ena(\widetilde{M})\setminus Ena(M-{^\bullet}U);\\
V(u), &
(u\in Ena(M-{^\bullet}U))\wedge (U\subseteq Ti_N);\\
V(u)-1, & \mbox{otherwise};
\end{array}
\right.$

\item ${\cal P}=PT(U,Q)$ is the {\em probability that the set of transitions $U$ fires in $Q$} defined as

$$PT(U,Q)=\frac{PF(U,Q)}{\sum_{V\in Fire(Q)}PF(V,Q)}.$$

\end{enumerate}

Let us explain the definition above in more detail. The first case of the item $2$ demonstrates a waiting transition
$u$ that
is not enabled at the marking $\widetilde{M}$, regardless of whether it
was enabled at the ``intermediate'' marking $M-{^\bullet}U$ (obtained by removing from $M$ the input places of all
transitions belonging to $U$, and that should be examined, especially when $N$ has structural loops), and therefore the
transition timer becomes inactive (turned off) and it is set to the undefined value $*$. The second case of the item
$2$ describes a waiting transition $u$ that
was not enabled at $M-{^\bullet}U$ and has first
been enabled at $\widetilde{M}$, hence, its timer is restored to the initial value $V_N(u)$, which is the delay of that
transition. The third case of the item $2$ explains a waiting transition $u$ that
was enabled at $M-{^\bullet}U$ and, hence, still
is enabled at $\widetilde{M}$, resulted in an firing of a set of immediate transitions $U$ instantly (in zero time), so
the transition timer does not decrement and its value stays equal to $V(u)$. The fourth case of the item $2$ corresponds
to the remaining option, i.e. a waiting transition $u$ that
was enabled at $M-{^\bullet}U$ and, hence, still
is enabled at $\widetilde{M}$, resulted in an firing of a set of stochastic (waiting) transitions $U$ at a time tick
(in one time unit), so the transition timer decrements by one and its value becomes $V(u)-1$.

We do not have to worry that for $u\in Tw_N$, such that $u\in Ena(M-{^\bullet}U)$, where $U\subseteq Ts_N\cup Tw_N$,
the value of $\widetilde{V}(u)=V(u)-1$ could become zero or negative, by the following reasons. Note that by the
definition of fireability, we have $Ena(M)\subseteq Tw_N\cup Ts_N$. If $V(u)=1$ then $u$ must fire in the next time
moment within some maximal (by the inclusion relation) set of the enabled at $M$ waiting transitions with the RTF equal
to one and enough tokens in the set's input places at $M$. Then we get $U\in Fire(Q)\subseteq
2^{Tw_N}\setminus\{\emptyset\}$, hence, $\emptyset\neq U\subseteq Tw_N$. Therefore, $\forall t\in U\ V(t)=1$ and
$Ena(M-{^\bullet}U)\cap\{w\in Tw_N\mid V(w)=1\}=\emptyset$, which contradicts to $u\in Ena(M-{^\bullet}U)\cap\{w\in
Tw_N\mid V(w)=1\}$. Thus, there exists no transition $u\in Tw_N$, such that $u\in Ena(M-{^\bullet}U)$ and $V(u)=1$. In
regard to the transitions $t\in U\subseteq Tw_N$ with $V(t)=1$, we have $\widetilde{V}(t)=*$, if $t\not\in
Ena(\widetilde{M})$, or $\widetilde{V}(t)=V_N(t)$, if $t\in Ena(\widetilde{M})\setminus Ena(M-{^\bullet}U)$.

Note that
when $U=\emptyset$ and $stang(Q)$, we get $M=\widetilde{M}$ and $\forall u\in Tw_N\ \widetilde{V}(u)=
\left\{
\begin{array}{ll}
*, &
u\not\in Ena(M);\\
V(u)-1, &
u\in Ena(M).
\end{array}
\right.$

Notice that the timers of all waiting transitions that are disabled when a marking change occurs become inactive
(turned off) and their values become undefined while the timers of all those staying enabled continue running with
their stored values. Hence, we adopt the {\em enabling memory} policy \cite{MBCDF95,AHR00,Bal01,Bal07} when the
markings are changed and the enabling of deterministic transitions is possibly modified (remember that immediate
transitions may be seen as those with the timers displaying a single value $0$, so we do not need to store their
values). Then the timer values of waiting transitions are taken as the enabling memory variables.

The advantage of our two-stage approach to definition of the probability that a set of transitions fires is that the
resulting probability formula $PT(U,Q)$ is valid both for (sets of) stochastic and deterministic transitions. It allows
one to unify the notation used later while constructing the denotational semantics and analyzing performance.

Note that for all states of an LDTSDPN $N$, the sum of outgoing probabilities is equal to $1$. More formally, $\forall
Q=(M,V)\in\nat_{fin}^{P_N}\times (\nat_{\geq 1}\cup\{*\})^{|Tw_N|}\ \sum_{\{U\subseteq Ena(M)\mid{^\bullet}U\subseteq
M\}}PT(U,Q)=1$. This obviously follows from the definition of $PT(U,Q)$ and guarantees that it defines a probability
distribution.

We write $Q\stackrel{U}{\rightarrow}\widetilde{Q}$ if $\exists{\cal P}\ Q\stackrel{U}{\rightarrow}_{\cal
P}\widetilde{Q}$ and $Q\rightarrow\widetilde{Q}$ if $\exists U\ Q\stackrel{U}{\rightarrow}\widetilde{Q}$.

The {\em probability to move from $Q$ to $\widetilde{Q}$ by firing any set of transitions} is

$$PM(Q,\widetilde{Q})=\sum_{\{U\mid Q\stackrel{U}{\rightarrow}\widetilde{Q}\}}PT(U,Q).$$

Since $PM(Q,\widetilde{Q})$ is the probability for {\em any} (including the empty one) transition set to change marking
$Q$ to $\widetilde{Q}$, we use summation in the definition. Note that $\forall Q=(M,V)\in\nat_{fin}^{P_N}\times
(\nat_{\geq 1}\cup\{*\})^{|Tw_N|}\ \sum_{\{\widetilde{Q}\mid Q\rightarrow\widetilde{Q}\}}PM(Q,\widetilde{Q})=
\sum_{\{\widetilde{Q}\mid Q\rightarrow\widetilde{Q}\}}\sum_{\{U\mid Q\stackrel{U}{\rightarrow}\widetilde{Q}\}}PT(U,Q)=
\sum_{\{U\subseteq Ena(M)\mid{^\bullet}U\subseteq M\}}PT(U,Q)=1$.

\begin{definition}
Let $N$ be an LDTSDPN. The {\em reachability set} of $N$, denoted by $RS(N)$, is the minimal set of markings such that
\begin{itemize}

\item $Q_N\in RS(N)$;

\item if $Q\in RS(N)$ and $Q\rightarrow\widetilde{Q}$ then $\widetilde{Q}\in RS(N)$.

\end{itemize}
\end{definition}

\begin{definition}
Let $N$ be an LDTSDPN. The {\em reachability graph} of $N$ is a (labeled probabilistic) transition system
$RG(N)=(S_N,L_N,{\cal T}_N,s_N)$, where
\begin{itemize}

\item the set of {\em states} is $S_N=RS(N)$;

\item the set of {\em labels} is $L_N=2^{T_N}\times (0;1]$;

\item the set of {\em transitions} is ${\cal T}_N=\{(Q,(U,{\cal P}),\widetilde{Q})\mid Q,\widetilde{Q}\in RS(N),\
Q\stackrel{U}{\rightarrow}_{\cal P}\widetilde{Q}\}$;

\item the {\em initial state} is $s_N=Q_N$.

\end{itemize}
\end{definition}

The set of {\em all s-tangible markings from $RS(N)$} is denoted by $RS_{ST}(N)$, and the set of {\em all w-tangible
markings from $RS(N)$} is denoted by $RS_{WT}(N)$. The set of {\em all tangible markings from $RS(N)$} is denoted by
$RS_T(N)=RS_{ST}(N)\cup RS_{WT}(N)$. The set of {\em all vanishing markings from $RS(N)$} is denoted by $RS_V(N)$.
Obviously, $RS(N)=RS_T(N)\uplus RS_V(N)=RS_{ST}(N)\uplus RS_{WT}(N)\uplus RS_V(N)$.

\subsection{Algebra of dtsd-boxes}

We now introduce discrete time stochastic and deterministic Petri boxes and the algebraic operations to define a net
representation of dtsdPBC expressions.

\begin{definition}
A {\em discrete time stochastic and deterministic Petri box (dtsd-box)} is a tuple\\
$N=(P_N,T_N,W_N,\Lambda_N)$, where
\begin{itemize}

\item $P_N$ and $T_N$ are finite sets of {\em places} and {\em transitions}, respectively, such that $P_N\cup
T_N\neq\emptyset$ and $P_N\cap T_N=\emptyset$;

\item $W_N:(P_N\times T_N)\cup (T_N\times P_N)\rightarrow\nat$ is a function providing the {\em weights of arcs}
between places and transitions;

\item $\Lambda_N$ is the {\em place and transition labeling} function such that
\begin{itemize}

\item $\Lambda_N|_{P_N}:P_N\rightarrow\{{\sf e},{\sf i},{\sf x}\}$ (it specifies {\em entry, internal} and {\em exit}
places, respectively);

\item $\Lambda_N|_{T_N}:T_N\rightarrow\{\varrho\mid\varrho\subseteq\nat_{fin}^{\cal SDL}\times{\cal SDL}\}$ (it associates
transitions with the {\em relabeling relations} on activities).

\end{itemize}

\end{itemize}
Moreover, $\forall t\in T_N\ {^\bullet}t\neq\emptyset\neq t^\bullet$. In addition, for the set of {\em entry} places of
$N$, defined as ${^\circ}N=\{p\in P_N\mid\Lambda_N(p)={\sf e}\}$, and for the set of {\em exit} places of $N$, defined
as $N^\circ=\{p\in P_N\mid\Lambda_N(p)={\sf x}\}$, the following conditions hold: ${^\circ}N\neq\emptyset\neq N^\circ$
and ${^\bullet}({^\circ}N)=\emptyset =(N^\circ )^\bullet$.

\end{definition}

A dtsd-box is {\em plain} if $\forall t\in T_N\ \exists (\alpha ,\kappa )\in{\cal SDL}\ \Lambda_N(t)=\varrho_{(\alpha
,\kappa )}$, where $\varrho_{(\alpha ,\kappa )}=\{(\emptyset ,(\alpha ,\kappa ))\}$ is a constant relabeling that can
be identified with the activity $(\alpha ,\kappa )$. A {\em (marked and timer-)clocked plain dtsd-box} is a pair
$(N,Q)$, where $N=(P_N,T_N,W_N,\Lambda_N)$ is a plain dtsd-box and $Q=(M,V)$ is its {\em state}. Here
$M\in\nat_{fin}^{P_N}$ is the {\em marking} of $N$ and $V:Tw_N\rightarrow\nat_{\geq 1}\cup\{*\}$ is the {\em timer
valuation function} of the waiting transitions of $N$. The set of {\em waiting transitions} of $N$ is defined as
$Tw_N=\{t\in T_N\mid\Lambda_N(t)=\varrho_{(\alpha ,\natural_l^\theta )},\ \theta\in\nat_{\geq 1},\ l\in\real_{>0}\}$. A
plain dtsd-box $N=(P_N,T_N,W_N,\Lambda_N)$ can be seen as a clocked plain dtsd-box $(N,Q^*)$ with $Q^*=(\emptyset
,V^*)$, where
for every waiting transition $t$ it holds $V^*(t)=*$, i.e. $V^*=*$.

We denote $\overline{N}=(N,Q_{\overline{N}})$, where $Q_{\overline{N}}=({^\circ}N,V_{\overline{N}})$ and
$V_{\overline{N}}:Tw_N\rightarrow\nat_{\geq 1}\cup\{*\}$ is such that $\forall t\in Tw_N\cap Ena({^\circ}N)\
V_{\overline{N}}(t)=\theta$ for $\Lambda_N(t)=\varrho_{(\alpha ,\natural_l^\theta )}$, whereas $\forall t\in
Tw_N\setminus Ena({^\circ}N)\ V_{\overline{N}}(t)=*$. We also denote $\underline{N}=(N,Q_{\underline{N}})$, where
$Q_{\underline{N}}=(N^\circ ,V^*)$.
We call ${^\circ}N$ and $N^\circ$ the {\em entry} and {\em exit markings} of $N$, respectively.

Note that a clocked plain dtsd-box $(P_N,T_N,W_N,\Lambda_N,Q_N)$ can be interpreted as the LDTSDPN\\
$(P_N,T_N,W_N,D_N,\Omega_N,{\cal L}_N,Q_N)$, where the functions $D_N,\ \Omega_N$ and ${\cal L}_N$ are defined as
follows: $\forall t\in T_N\ \Omega_N(t)=\kappa$ if $\kappa\in (0;1)$; or $D_N(t)=\theta ,\ \Omega_N(t)=l$ if $\kappa
=\natural_l^\theta ,\ \theta\in\nat ,\ l\in\real_{>0}$; and ${\cal L}_N(t)=\alpha$, where
$\Lambda_N(t)=\varrho_{(\alpha ,\kappa )}$ (we say that the transition $t$ {\em corresponds} to the activity $(\alpha
,\kappa )$ in such a case). Behaviour of the clocked dtsd-boxes follows from the firing rule of LDTSDPNs. A plain
dtsd-box $N$ is {\em $n$-bounded} ($n\in\nat$) if $\overline{N}$ is so, i.e. $\forall Q=(M,V)\in RS(\overline{N})\
\forall p\in P_N\ M(p)\leq n$, and it is {\em safe} if it is $1$-bounded. A plain dtsd-box $N$ is {\em clean} if
$\forall Q=(M,V)\in RS(\overline{N})\ {^\circ}N\subseteq M\ \Rightarrow\ M={^\circ}N$ and $N^\circ\subseteq M\
\Rightarrow\ M=N^\circ$, i.e. if there are tokens in all its entry (exit) places then no other places have tokens.

The structure of the plain dtsd-box corresponding to a static expression is constructed like in PBC \cite{BKo95,BDK01},
i.e. we use simultaneous refinement and relabeling meta-operator (net refinement) in addition to the {\em operator
dtsd-boxes} corresponding to the algebraic operations of dtsdPBC and featuring transformational transition relabelings.
Operator dtsd-boxes specify $n$-ary functions from plain dtsd-boxes to plain dtsd-boxes (we have $1\leq n\leq 3$ in
dtsdPBC). Thus, as we shall see in Theorem \ref{safeclean.the}, the resulting plain dtsd-boxes are safe and clean. In
the definition of the denotational semantics, we shall apply standard constructions used for PBC. Let $\Theta$ denote
{\em operator box} and $u$ denote {\em transition name} from the PBC setting.

The relabeling relations $\varrho\subseteq\nat_{fin}^{\cal SDL}\times{\cal SDL}$ are defined as follows:
\begin{itemize}

\item $\varrho_{id}=\{(\{(\alpha ,\kappa )\},(\alpha ,\kappa ))\mid (\alpha ,\kappa )\in{\cal SDL}\}$ is the {\em
identity relabeling} keeping the interface as it is;

\item $\varrho_{(\alpha ,\kappa )}=\{(\emptyset ,(\alpha ,\kappa ))\}$ is the {\em constant relabeling} that can be
identified with $(\alpha ,\kappa )\in{\cal SDL}$ itself;

\item $\varrho_{[f]}=\{(\{(\alpha ,\kappa )\},(f(\alpha ),\kappa ))\mid (\alpha ,\kappa )\in{\cal SDL}\}$;

\item $\varrho_{\!\!\rs a}=\{(\{(\alpha ,\kappa )\},(\alpha ,\kappa ))\mid (\alpha ,\kappa )\in{\cal SDL},\
a,\hat{a}\not\in\alpha\}$;

\item $\varrho_{\!\!\sy a}$ is the least relabeling relation containing $\varrho_{id}$ such that if $(\Upsilon
,(\alpha ,\kappa )),(\Xi ,(\beta ,\lambda ))\in\varrho_{\!\!\sy a}$ and\\
$a\in\alpha ,\ \hat{a}\in\beta$ then
\begin{itemize}

\item $(\Upsilon +\Xi ,(\alpha\oplus_a\beta ,\kappa\cdot\lambda ))\in\varrho_{\!\!\sy a}$ if $\kappa ,\lambda\in
(0;1)$;

\item $(\Upsilon +\Xi ,(\alpha\oplus_a\beta ,\natural_{l+m}^\theta ))\in\varrho_{\!\!\sy a}$ if $\kappa
=\natural_l^\theta ,\ \lambda =\natural_m^\theta ,\ \theta\in\nat ,\ l,m\in\real_{>0}$.

\end{itemize}

\end{itemize}

\begin{figure}
\begin{center}
\includegraphics{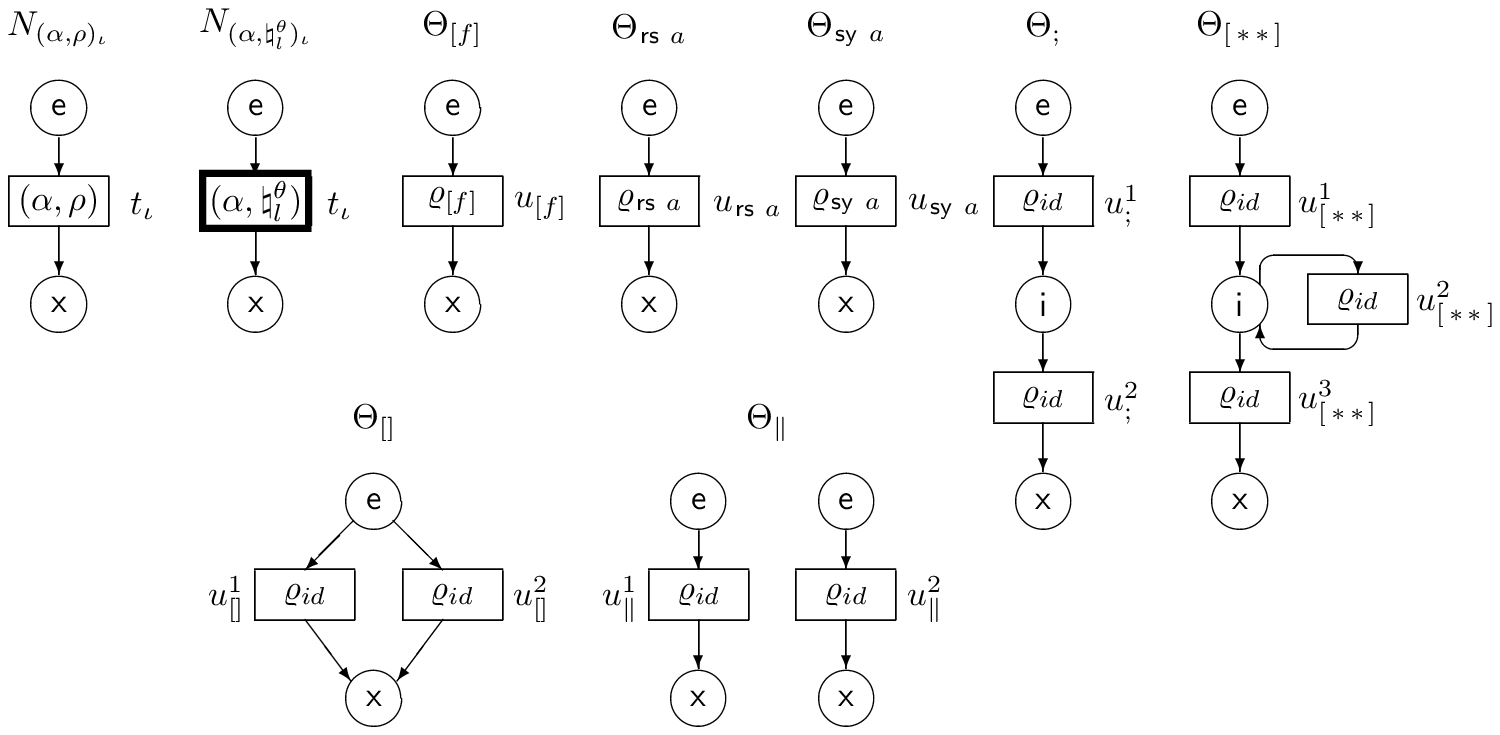}
\end{center}
\caption{The plain and operator dtsd-boxes}
\label{dtsdboxoprw.fig}
\end{figure}

The plain dtsd-boxes $N_{(\alpha ,\rho )_{\iota}},\ N_{(\alpha ,\natural_l^\theta )_{\iota}}$, where $\rho\in (0;1),\
\theta\in\nat ,\ l\in\real_{>0}$, and operator dtsd-boxes are presented in Figure \ref{dtsdboxoprw.fig}. Note that the
label {\sf i} of internal places is usually omitted.

In the case of the iteration, a decision that we must take is the selection of the operator box that we shall use for
it, since we have two proposals in plain PBC for that purpose \cite{BDK01}. One of them provides us with a safe version
with six transitions in the operator box, but there is also a simpler version, which has only three transitions. In
general, in PBC, with the latter version we may generate $2$-bounded nets, which only occurs when a parallel behavior
appears at the highest level of the body of the iteration. Nevertheless, in our case, and due to the syntactical
restriction introduced for regular terms, this particular situation cannot occur, so that the net obtained will be
always safe.

To
define a semantic function that
assigns a plain dtsd-box
to every static expression of dtsdPBC, we intro\-duce the {\em enumeration} function $Enu:T\rightarrow Num$, which
associates the numberings with transitions of a plain dtsd-box $N=(P,T,W,\Lambda )$ in accordance with those of
activities. In the case of synchronization, the function associates with the resulting new transition a concatenation
of the parenthesized numberings of the transitions it comes from.

We now define the enumeration function $Enu$ for every operator of dtsdPBC. Let $N_E=Box_{dtsd}(E)=\\
(P_E,T_E,W_E,\Lambda_E)$ be the plain dtsd-box corresponding to a static expression $E$, and
$Enu_E:T_E\rightarrow Num$ be the enumeration function for
$N_E$. We shall use the analogous notation for static expressions $F$ and $K$.
\begin{itemize}

\item $Box_{\rm dtsi}((\alpha ,\kappa )_{\iota})=N_{(\alpha ,\kappa )_{\iota}}$. Since a single transition $t_{\iota}$
corresponds to the activity $(\alpha ,\kappa )_{\iota}\in{\cal SDL}$, their numberings coincide:

$$Enu(t_{\iota})=\iota .$$

\item $Box_{dtsd}(E\circ F)=\Theta_{\circ}(Box_{dtsd}(E),Box_{dtsd}(F)),\ \circ\in\{;,\cho ,\|\}$. Since we do not
introduce new transitions, we preserve the initial numbering:

$$Enu(t)=\left\{
\begin{array}{ll}
Enu_E(t), & t\in T_E;\\
Enu_F(t), & t\in T_F.
\end{array}
\right.$$

\item $Box_{dtsd}(E[f])=\Theta_{[f]}(Box_{dtsd}(E))$. Since we only replace the labels of some multiactions by a
bijection, we preserve the initial numbering:

$$Enu(t)=Enu_E(t),\ t\in T_E.$$

\item $Box_{dtsd}(E\rs a)=\Theta_{\!\!\rs a}(Box_{dtsd}(E))$. Since we remove all transitions labeled with multiactions
containing $a$ or $\hat{a}$, this does not change the numbering of the remaining transitions:

$$Enu(t)=Enu_E(t),\ t\in T_E,\ a,\hat{a}\not\in\alpha ,\ \Lambda_E(t)=\varrho_{(\alpha ,\kappa )}.$$

\item $Box_{dtsd}(E\sy a)=\Theta_{\!\!\sy a}(Box_{dtsd}(E))$. Note that $\forall v,w\in T_E$ such that
$\Lambda_E(v)=\varrho_{(\alpha ,\kappa )},\ \Lambda_E(w)=\varrho_{(\beta ,\lambda )}$ and $a\in\alpha ,\
\hat{a}\in\beta$, the new transition $t$ resulting from synchronization of $v$ and $w$ has the label $\Lambda
(t)=\varrho_{(\alpha\oplus_a\beta ,\kappa\cdot\lambda )}$ if $t$ is a stochastic transition ($\kappa ,\lambda\in
(0;1)$); or $\Lambda (t)=\varrho_{(\alpha\oplus_a\beta ,\natural_{l+m}^\theta )}$ if $t$ is a deterministic one
($\kappa =\natural_l^\theta ,\ \lambda =\natural_m^\theta ,\ \theta\in\nat ,\ l,m\in\real_{>0}$); and the numbering
$Enu(t)=(Enu_E(v))(Enu_E(w))$.

Thus, the enumeration function is defined as

$$Enu(t)=\left\{
\begin{array}{ll}
Enu_E(t), & t\in T_E;\\
(Enu_E(v))(Enu_E(w)), & t\mbox{ results from synchronization of }v\mbox{ and }w.
\end{array}
\right.$$

According to the definition of $\varrho_{\!\!\sy a}$, the synchronization is only possible when all the transitions
in the set are stochastic (immediate or waiting, respectively). If we synchronize the same set of transitions in
different orders, we obtain several resulting transitions with the same label and probability or weight, but with
the different numberings having the same content. Then, we only consider a single transition from the resulting
ones in the plain dtsd-box to avoid introducing redundant transitions.

For example, if the transitions $t$ and $u$ are generated by synchronizing $v$ and $w$ in different orders, we have
$\Lambda (t)=\varrho_{(\alpha\oplus_a\beta ,\kappa\cdot\lambda )}=\Lambda (u)$ for stochastic transitions ($\kappa
,\lambda\in (0;1)$) or $\Lambda (t)=\varrho_{(\alpha\oplus_a\beta ,\natural_{l+m}^\theta )}=\Lambda (u)$ for
deterministic ones ($\kappa =\natural_l^\theta ,\ \lambda =\natural_m^\theta ,\ \theta\in\nat ,\ l,m\in\real_{>0}$),
but $Enu(t)=(Enu_E(v))(Enu_E(w))\neq (Enu_E(w))(Enu_E(v))=Enu(u)$, whereas $Cont(Enu(t))=Cont(Enu(v))\cup
Cont(Enu(w))=\\
Cont(Enu(u))$. Then only one transition $t$ (or $u$, symmetrically) will appear in $Box_{dtsd}(E\sy a)$.

\item $Box_{dtsd}([E*F*K])=\Theta_{[\,*\,*\,]}(Box_{dtsd}(E),Box_{dtsd}(F),Box_{dtsd}(K))$. Since we do not introduce
new transitions, we preserve the initial numbering:

$$Enu(t)=\left\{
\begin{array}{ll}
Enu_E(t), & t\in T_E;\\
Enu_F(t), & t\in T_F;\\
Enu_K(t), & t\in T_K.
\end{array}
\right.$$

\end{itemize}

We now can formally define the denotational semantics as a homomorphism.

\begin{definition}
Let $(\alpha ,\kappa )\in{\cal SDL},\ a\in Act$ and $E,F,K\in RegStatExpr$. The {\em denotational semantics} of dtsdPBC
is a mapping $Box_{dtsd}$ from $RegStatExpr$ into the domain of plain dtsd-boxes defined as follows:
\begin{enumerate}

\item $Box_{dtsd}((\alpha ,\kappa )_{\iota})=N_{(\alpha ,\kappa )_{\iota}}$;

\item $Box_{dtsd}(E\circ F)=\Theta_{\circ}(Box_{dtsd}(E),Box_{dtsd}(F)),\ \circ\in\{;,\cho ,\|\}$;

\item $Box_{dtsd}(E[f])=\Theta_{[f]}(Box_{dtsd}(E))$;

\item $Box_{dtsd}(E\circ a)=\Theta_{\circ a}(Box_{dtsd}(E)),\ \circ\in\{\!\!\rs\!\!,\!\!\sy\!\!\}$;

\item $Box_{dtsd}([E*F*K])=\Theta_{[\,*\,*\,]}(Box_{dtsd}(E),Box_{dtsd}(F),Box_{dtsd}(K))$.

\end{enumerate}
\end{definition}

The dtsd-boxes of dynamic expressions can be defined as well. For $E\in RegStatExpr$, let $Box_{dtsd}(\overline{E})=
\overline{Box_{dtsd}(E)}$ and $Box_{dtsd}(\underline{E})=\underline{Box_{dtsd}(E)}$.
Note that this definition is compositional in the sense that, for any arbitrary dynamic expression, we may decompose it
in some inner dynamic and static expressions, for which we may apply the definition, thus obtaining the corresponding
plain dtsd-boxes, which can be joined according to the term structure (by definition of $Box_{dtsd}$), the resulting
plain box being marked in the places that were marked in the argument nets. Importantly, when composing dtsd-boxes of
arbitrary dynamic expressions, we should guarantee that the operations correctly propagate the timer values from the
clocked to non-clocked operands. For that, we have to respect the time spent in the entry
markings.

Let $E,F\in RegStatExpr,\ G,H\in RegDynExpr$ and $a\in Act$.
For $N_E=Box_{dtsd}(E)=(P_E,T_E,W_E,\Lambda_E)$, the clocked plain dtsd-box of $E$ is $(N_E,(\emptyset ,V^*))$,
and analogously for $F$. Next, for
$N_G=Box_{dtsd}(\lfloor G\rfloor )=(P_G,T_G,W_G,\Lambda_G)$, the clocked plain dtsd-box of $G$ is
$Box_{dtsd}(G)=(N_G,(M_G,V_G))$ (defined by induction on the structure of $G$, as it will be descried below),
and similarly for~$H$.

Let $(P,T,W,\Lambda ,(M,V))$ be a clocked plain dtsd-box, where $T=Ts\uplus Ti\uplus Tw$ consists of stochastic,
immediate and waiting transitions.
The {\em marking age of the state $(M,V)$} is defined as $\Box (M,V)=\max\{\eta -V(u)\mid u\in Tw\cap Ena(M),\ \Lambda
(u)=\varrho_{(\beta ,\natural_m^\eta )}\}$.
The {\em minimal delay of the waiting transitions from the set $U\subseteq Tw$ that are enabled at the marking $M$}
is defined as $\theta (M,U)=\min\{\eta\mid u\in U\cap Ena(M),\ \Lambda (u)=\varrho_{(\beta ,\natural_m^\eta )}\}$. We
now
inductively define the dtsd-boxes of arbitrary dynamic expressions.
\begin{itemize}

\item $Box_{dtsd}(\overline{E})=\overline{Box_{dtsd}(E)}$ and $Box_{dtsd}(\underline{E})=\underline{Box_{dtsd}(E)}$.

\item $Box_{dtsd}(G;E)=(Box_{dtsd}(\lfloor G\rfloor ;E),(M,V))$,
where $M=M_G$, and $\forall t\in Tw_N$ with $\Lambda_N(t)=\varrho_{(\alpha ,\natural_l^\theta )}$:

$$V(t)=\left\{
\begin{array}{ll}
V_G(t), & t\in Tw_G;\\
\theta , & t\in Tw_E\cap Ena(M);\\
*, & t\in Tw_E\setminus Ena(M).
\end{array}
\right.$$
Thus, each waiting transition of $N_E$ enabled at the entry marking of it has set its timer to the transition delay
$\theta$ (the initial valuation).

\item $Box_{dtsd}(E;G)=(Box_{dtsd}(E;\lfloor G\rfloor ),(M,V))$,
where $M=M_G$, and $\forall t\in Tw_N$ with $\Lambda_N(t)=\varrho_{(\alpha ,\natural_l^\theta )}$:

$$V(t)=\left\{
\begin{array}{ll}
V_G(t), & t\in Tw_G;\\
*, & t\in Tw_E.
\end{array}
\right.$$

\item $Box_{dtsd}(G\cho E)=(Box_{dtsd}(\lfloor G\rfloor\cho E),(M,V))$,
where $M=M_G$, and $\forall t\in Tw_N$ with $\Lambda_N(t)=\varrho_{(\alpha ,\natural_l^\theta )}$:

$$V(t)=\left\{
\begin{array}{ll}
\theta -\min\{\Box (M_G,V_G),\theta (M,Tw_E)-1\}, & t\in (Tw_G\cup Tw_E)\cap Ena(M);\\
*, & t\in (Tw_G\cup Tw_E)\setminus Ena(M).
\end{array}
\right.$$
Thus, if $\zeta$ is the minimum of the time spent at the the marking $M=M_G$ of the state $(M_G,V_G)$ and the
delays of the waiting transitions from $Tw_E$, enabled at that marking, then each waiting transition, enabled at
$M$, has set its timer to $\theta -\zeta$, where $\theta$ is the delay of that transition. It is guaranteed that
the new timer value is not less than $1$, since in case when $\Box (M_G,V_G)>\theta (M,Tw_E)-1$ and $\theta =\theta
(M,Tw_E)$ we have $\theta -(\theta (M,Tw_E)-1)=1$. In that case, the subnet $N_G$ should ``wait'' for the subnet
$N_E$ by modifying appropriately (via increasing by the difference between the residence time at $M$ and minimal
delay of transitions from $Tw_E\cap Ena(M)$ minus $1$) the timer values of its waiting transitions, enabled~at~$M$.

Note that $\Box (M_G,V_G)>\theta (M,Tw_E)-1$ does not hold for any dynamic expression, obtained by applying action
rules, starting from an overlined static expression without timer value superscripts. The reason is that  all the
action rules maintain the time progress uniformity, hence, $\zeta =\Box (M_G,V_G)$ in that case. Further, the
inequality $\eta -V_G(u)<\Box (M_G,V_G)$ may only happen when the $(\beta ,\natural_m^\eta )\in{\cal WL}(G)$,
corresponding to $u\in Tw_G\cap Ena(M_G)$, is later affected by restriction, so that the timer of that waiting
multiaction stops with the value $1$ while the waiting multiaction can never be executed. Thus, if we start from an
overlined static expression without time stamps and the waiting multiaction corresponding to $t$ is not
subsequently affected by restriction then $V(t)=\theta -\Box (M_G,V_G)=V_G(t)$ for $t\in Tw_G\cap Ena(M)$ and
$V(t)=\theta -\Box (M_G,V_G)$ for $t\in Tw_E\cap Ena(M)$.

The definition of $Box_{dtsd}(E\cho G)$ is similar.

\item $Box_{dtsd}(G\| H)=(Box_{dtsd}(\lfloor G\rfloor\|\lfloor H\rfloor ),(M,V))$,
where $M=M_G\cup M_H$, and $\forall t\in Tw_N$ with\\
$\Lambda_N(t)=\varrho_{(\alpha ,\natural_l^\theta )}$:

$$V(t)=\left\{
\begin{array}{ll}
\theta -\min\{\Box (M_G,V_G),\Box (M_H,V_H)\}, & t\in (Tw_G\cup Tw_H)\cap Ena(M);\\
*, & t\in (Tw_G\cup Tw_H)\setminus Ena(M).
\end{array}
\right.$$
Thus, if $\zeta$ is the minimum of the times spent at the markings of the states $(M_G,V_G)$ and $(M_H,V_H)$ then
each waiting transi\-tion, enabled at the marking $M$, has set its timer to $\theta -\zeta$, where $\theta$ is the
delay of that transition. The idea is to ensure that the time progresses uniformly, for which the timer decrements
of all waiting transitions, enabled at $M$, should be synchronized (equalized). Hence, the subnet with the more
time spent in its local marking should ``wait'' for the other subnet by modifying appropriately (via increasing by
the difference between residence times at $M_G$ and $M_H$) the timer values of its waiting transitions, enabled at
$M$.

Note that $\Box (M_G,V_G)\neq\Box (M_H,V_H)$ cannot hold for any dynamic expression, obtained by applying action
rules, starting from an overlined static expression without timer value superscripts. The reason is that all the
action rules maintain the time progress uniformity, hence, $\zeta =\Box (M_G,V_G)=\Box (M_H,V_H)$ in that case.
Further, the inequality $\eta -V_G(u)<\Box (M_G,V_G)$ may only happen when the $(\beta ,\natural_m^\eta )\in{\cal
WL}(G)$, corresponding to $u\in Tw_G\cap Ena(M_G)$, is later affected by restriction, so that the timer of that
waiting multiaction stops with the value $1$ while the waiting multiaction can never be executed. The same holds
for $\Box (M_H,V_H)$. Thus, if we start from an overlined static expression without time stamps and the waiting
multiaction corresponding to $t$ is not subsequently affected by restriction then $V(t)=\theta -\Box
(M_G,V_G)=V_G(t)$ for $t\in Tw_G\cap Ena(M)$ and $V(t)=\theta -\Box (M_H,V_H)=V_H(t)$ for $t\in Tw_H\cap Ena(M)$.

The definition of $Box_{dtsd}(H\| G)$ is similar.

\item $Box_{dtsd}(G[f])=(Box_{dtsd}(\lfloor G\rfloor [f]),(M,V))$,
where $M=M_G$, and $\forall t\in Tw_N$ with $\Lambda_N(t)=\varrho_{(\alpha ,\natural_l^\theta )}$:

$$V(t)=\begin{array}{ll}
V_G(t), & t\in Tw_G.
\end{array}$$

\item $Box_{dtsd}(G\rs a)=(Box_{dtsd}(\lfloor G\rfloor\rs a),(M,V))$,
where $M=M_G$, and $\forall t\in Tw_N$ with $\Lambda_N(t)=\varrho_{(\alpha ,\natural_l^\theta )}$:

$$V(t)=\begin{array}{ll}
V_G(t), & t\in Tw_G,\ a,\hat{a}\not\in\alpha .
\end{array}$$

\item $Box_{dtsd}(G\sy a)=(Box_{dtsd}(\lfloor G\rfloor\sy a),(M,V))$,
where $M=M_G$, and $\forall t\in Tw_N$ with $\Lambda_N(t)=\varrho_{(\alpha ,\natural_l^\theta )}$:

$$V(t)=\left\{
\begin{array}{ll}
V_G(t), & t\in Tw_G;\\
V_G(v)=V_G(w), & t\mbox{ results from synchronization of }v,w\in Tw_G.
\end{array}
\right.$$
Thus, the timer of the synchronous product of the two waiting transitions $v$ and $w$ of $N_G$ is set to their timer
values. Those values coincide, since only waiting transitions with the same delays are synchronized and the time
progresses uniformly in every whole dynamic expression.

\item $Box_{dtsd}([G*E*F])=(Box_{dtsd}(\lfloor G\rfloor *E*F),(M,V))$,
where $M=M_G$, and $\forall t\in Tw_N$ with\\
$\Lambda_N(t)=\varrho_{(\alpha ,\natural_l^\theta )}$:

$$V(t)=\left\{
\begin{array}{ll}
V_G(t), & t\in Tw_G;\\
\theta , & t\in Tw_E\cap Ena(M);\\
*, & t\in Tw_E\setminus Ena(M).
\end{array}
\right.$$
Thus, each waiting transition of $N_E$, enabled at the entry marking of it, has set its timer to the transition delay
$\theta$ (the initial valuation).

\item $Box_{dtsd}([E*G*F])=(Box_{dtsd}(E*\lfloor G\rfloor *F),(M,V))$,
where $M=M_G$, and $\forall t\in Tw_N$ with\\
$\Lambda_N(t)=\varrho_{(\alpha ,\natural_l^\theta )}$:

$$V(t)=\left\{
\begin{array}{ll}
\theta -\min\{\Box (M_G,V_G),\theta (M,Tw_F)-1\}, & t\in (Tw_G\cup Tw_F)\cap Ena(M);\\
*, & t\in (Tw_G\cup Tw_F)\setminus Ena(M).
\end{array}
\right.$$
Thus, if $\zeta$ is the minimum of the time spent at the the marking $M=M_G$ of the state $(M_G,V_G)$ and the
delays of the waiting transitions from $Tw_F$, enabled at that marking, then each waiting transition, enabled at
$M$, has set its timer to $\theta -\zeta$, where $\theta$ is the delay of that transition. It is guaranteed that
the new timer value is not less than $1$, since in case when $\Box (M_G,V_G)>\theta (M,Tw_F)-1$ and $\theta =\theta
(M,Tw_F)$ we have $\theta -(\theta (M,Tw_F)-1)=1$. In that case, the subnet $N_G$ should ``wait'' for the subnet
$N_F$ by modifying appropriately (via increasing by the difference between the residence time at $M$ and minimal
delay of transitions from $Tw_F\cap Ena(M)$ minus $1$) the timer values of its waiting transitions, enabled~at~$M$.

The definition of $Box_{dtsd}([E*F*G])$ is similar.

\end{itemize}

\begin{theorem}
For any static expression $E,\ Box_{dtsd}(\overline{E})$ is safe and clean.
\label{safeclean.the}
\end{theorem}
{\em Proof.} The structure of the net is obtained as in PBC \cite{BKo95,BDK01}, combining both refinement and
relabeling. Consequently, the dtsd-boxes thus obtained will be safe and clean. \hfill $\eop$

Let $\simeq$ denote isomorphism between transition systems and reachability graphs that binds their initial states.
Note that the names of transitions of the dtsd-box corresponding to a static expression could be identified with the
enumerated activities of the latter.

\begin{theorem}
For any static expression $E$,

$$TS(\overline{E})\simeq RG(Box_{dtsd}(\overline{E})).$$

\label{opdensem.the}
\end{theorem}
{\em Proof.} As for the qualitative (functional) behaviour, we have the same isomorphism as in PBC \cite{BKo95,BDK01}.

The quantitative behaviour is the same by the following reasons. First, the activities of an expression have the
probability or delay and weight parts coinciding with the probabilities or delays and weights of the transitions
belonging to the corresponding dtsd-box. Second, we use analogous probability or delay and weight functions to
construct the corresponding transition systems and reachability graphs.

In general, the proof goes along the same lines as that from \cite{MVCF08}.
For instance, observe the relationships between the process expressions and their Petri net counterparts in dtsdPBC.
Let $E\in RegStatExpr$ and $\overline{N}=Box_{dtsd}(\overline{E})$.
There is the following bijection between $DR(\overline{E})$ and $RS(\overline{N})$. For a process state
$s=[G]_\approx\in DR(\overline{E})$, the corresponding net state is $Q=(M,V)\in RS(\overline{N})$ with the next
properties. First,
$M$ is the marking of the clocked dtsd-box $(N,(M,V))=Box_{dtsd}(G)$. Second, for each transition $t\in T_N$ with
$\Lambda_N(t)=\varrho_{(\alpha ,\natural_l^\theta )},\ \theta\in\nat_{\geq 1},\ l\in\real_{>0}$, we have
$V(t)=I_G((\alpha ,\natural_l^\theta ))$. Note that each expression $H\in s\cap SatOpRegDynExpr$ can be seen as a
``partial'' representation of the net marking $M$ and state $Q$, since possibly not all enabled activities are
overlined in $H$.
The transition set $U\subseteq T_N$ {\em corresponding} to a (multi)set of activities $\Upsilon\in\nat_{fin}^{\cal
SDL}$ is such that $\forall t\in U\ \Lambda_N(t)=\varrho_{(\alpha ,\kappa )}$, where $(\alpha ,\kappa )\in\Upsilon$.
Thus, here we also have a bijection. Notice that the marking $M$ is obtained from ${^\circ}N$ by firing the sequence of
transition sets corresponding to the activities (multi)sets, whose execution from $[\overline{E}]_\approx$ has led to
the state $s$.
For the corresponding process and net states $s\in DR(\overline{E})$ and $Q=(M,V)\in RS(\overline{N})$, the sets of the
enabled activities $Ena(s)$ and transitions $Ena(M)$ are constructed similarly. The sets of the executable activities
(multi)sets $Exec(s)$ and fireable transitions sets $Fire(Q)$ are analogous as well.
At last, the probability functions $PF(s,\Upsilon )$ and $PT(s,\Upsilon )$ are respectively defined in the same way as
$PF(Q,U)$ and $PT(Q,U)$, for the corresponding
(multi)set of activities $\Upsilon$ and transition set $U$. \hfill $\eop$

\begin{figure}
\begin{center}
\includegraphics{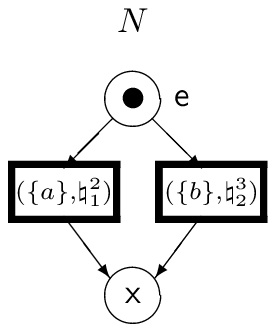}
\end{center}
\caption{The clocked dtsd-box $N=Box_{dtsd}(\overline{E})$ for $E=(\{a\},\natural_1^2)\cho (\{b\},\natural_2^3)$}
\label{boxchowm.fig}
\end{figure}

\begin{figure}
\begin{center}
\includegraphics{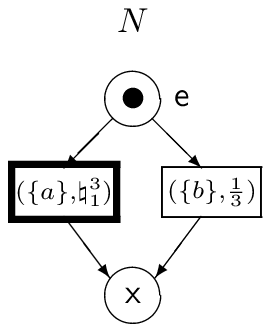}
\end{center}
\caption{The clocked dtsd-box $N=Box_{dtsd}(\overline{E})$ for $E=(\{a\},\natural_1^3)\cho (\{b\},\frac{1}{3})$}
\label{boxchowsm.fig}
\end{figure}

\begin{figure}
\begin{center}
\includegraphics{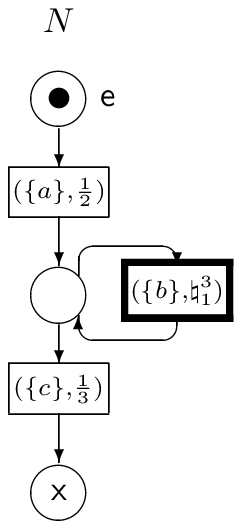}
\end{center}
\caption{The clocked dtsd-box $N=Box_{dtsd}(\overline{E})$ for $E=[(\{a\},\frac{1}{2})*(\{b\},\natural_1^3)*
(\{c\},\frac{1}{3})]$}
\label{boxitswm.fig}
\end{figure}

\begin{figure}
\begin{center}
\includegraphics{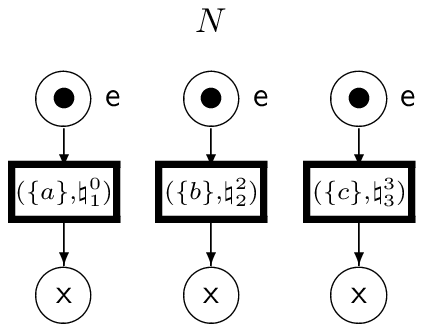}
\end{center}
\caption{The clocked dtsd-box $N=Box_{dtsd}(\overline{E})$ for $E=(\{a\},\natural_1^0)\| (\{b\},\natural_2^2)\|
(\{c\},\natural_3^3)$}
\label{boxpariwm.fig}
\end{figure}

\begin{figure}
\begin{center}
\includegraphics{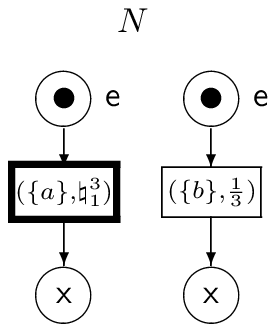}
\end{center}
\caption{The clocked dtsd-box $N=Box_{dtsd}(\overline{E})$ for $E=(\{a\},\natural_1^3)\| (\{b\},\frac{1}{3})$}
\label{boxparwsm.fig}
\end{figure}

\begin{figure}
\begin{center}
\includegraphics{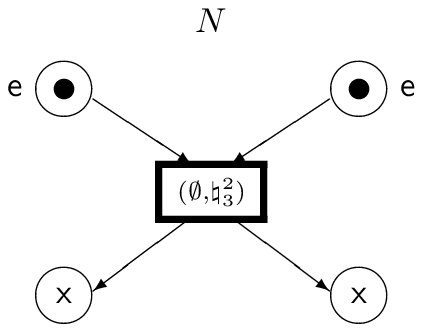}
\end{center}
\caption{The clocked dtsd-box $N=Box_{dtsd}(\overline{E})$ for $E=((\{a\},\natural_1^2)\|
(\{\hat{a}\},\natural_2^2))\sy a\rs a$}
\label{boxparsyrswm.fig}
\end{figure}

\begin{figure}
\begin{center}
\includegraphics{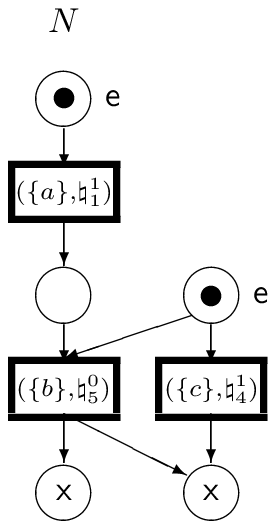}
\end{center}
\caption{The clocked dtsd-box $N=Box_{dtsd}(\overline{E})$ for $E=(((\{a\},\natural_1^1);
(\{b,\hat{x}\},\natural_2^0))\| ((\{x\},\natural_3^0)\cho (\{c\},\natural_4^1)))\sy x\rs x$}
\label{boxparsyrsiwm.fig}
\end{figure}

\begin{figure}
\begin{center}
\includegraphics{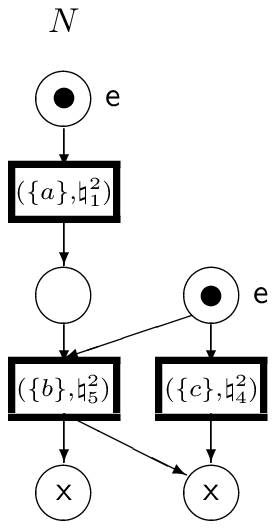}
\end{center}
\caption{The clocked dtsd-box $N=Box_{dtsd}(\overline{E})$ for $E=((((\{a\},\natural_1^2);
(\{b,\hat{x}\},\natural_2^2))\| ((\{x\},\natural_3^2)\cho (\{c\},\natural_4^2)))\sy x\rs x$}
\label{boxparsyrswwm.fig}
\end{figure}

\begin{figure}
\begin{center}
\includegraphics{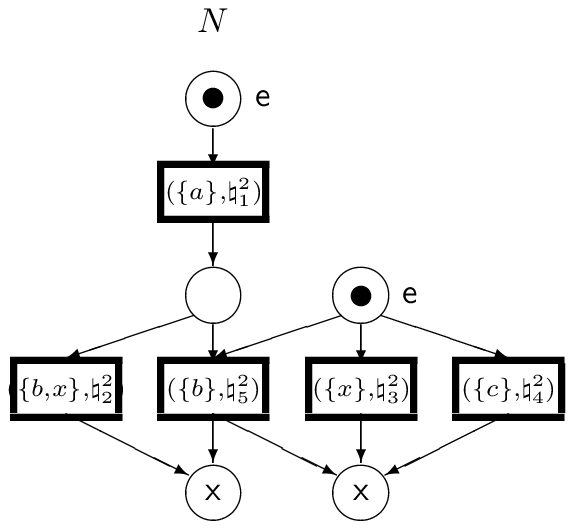}
\end{center}
\caption{The clocked dtsd-box $N=Box_{dtsd}(\overline{E})$ for $E=((((\{a\},\natural_1^2);
(\{b,\hat{x}\},\natural_2^2))\| ((\{x\},\natural_3^2)\cho (\{c\},\natural_4^2)))\sy x$}
\label{boxparsywwm.fig}
\end{figure}

\begin{figure}
\begin{center}
\includegraphics{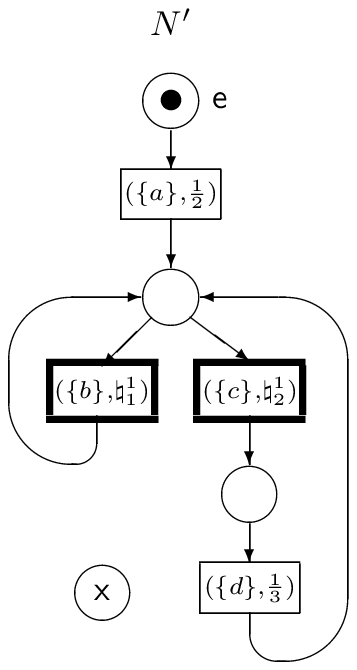}
\end{center}
\caption{The clocked dtsd-box $N=Box_{dtsd}(\overline{E})$ for $E=[(\{a\},\frac{1}{2})*((\{b\},\natural_1^1)\cho
((\{c\},\natural_2^1);(\{d\},\frac{1}{3})))*{\sf Stop}]$}
\label{boxitchoswm.fig}
\end{figure}

\begin{example}
Let $E$ be from Examples \ref{tschowm.exm}--\ref{tsitchoswm.exm}. In Figures \ref{boxchowm.fig}--\ref{boxitchoswm.fig},
the clocked dtsd-boxes $N=Box_{dtsd}(\overline{E})$ are presented. Due to the time constraints and since waiting
multiactions may be preempted by stochastic ones, some dynamic expressions can have complex transition systems and
simple clocked dtsd-boxes (Examples
\ref{tschowm.exm}--\ref{tsparwsm.exm}), or vice versa (Examples
\ref{tsparsyrswm.exm}--\ref{tsitchoswm.exm}).
\label{dtsdboxes.exm}
\end{example}

\begin{figure}
\begin{center}
\includegraphics{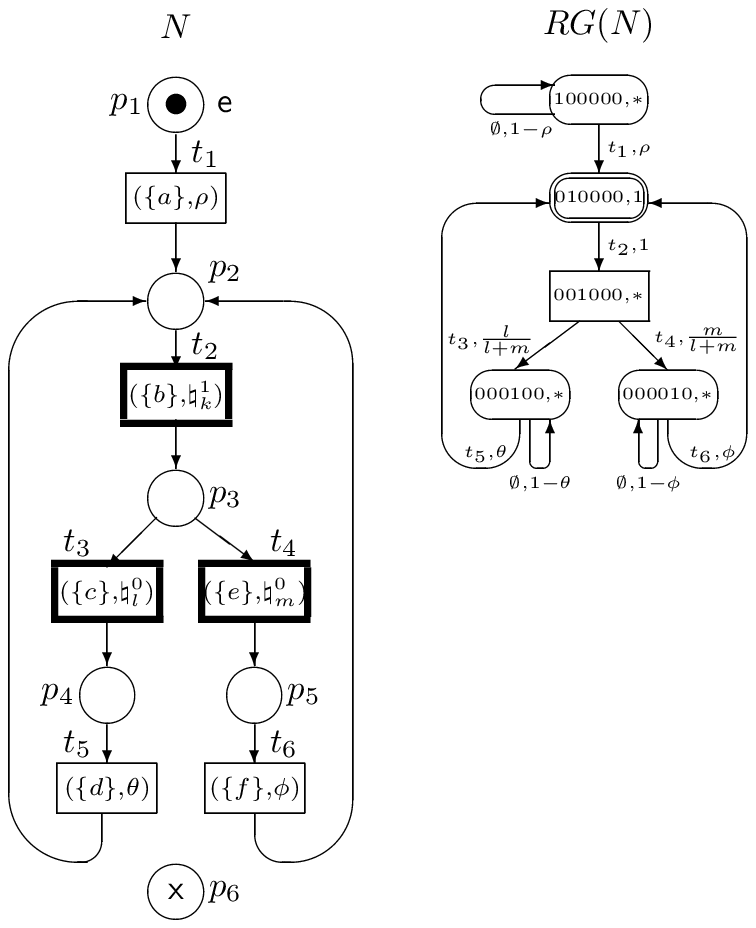}
\end{center}
\caption{The clocked dtsd-box $N=Box_{dtsd}(\overline{E})$ for $E=[(\{a\},\rho )*((\{b\},\natural_k^1);
(((\{c\},\natural_l^0);(\{d\},\theta ))\cho ((\{e\},\natural_m^0);\protect\newline
(\{f\},\phi ))))*{\sf Stop}]$ and its reachability graph}
\label{boxdrgnewrw.fig}
\end{figure}

\begin{example}
Let $E$ be from Example \ref{tsitchoswim.exm}. In Figure \ref{boxdrgnewrw.fig}, the clocked dtsd-box
$N=Box_{dtsd}(\overline{E})$ and its reach\-ability graph $RG(N)$ are presented. Since $N$ has a single waiting
transition $t_2$ that is enabled only at marking $M_2=(0,1,0,0,0,0)$ and whose delay is $1$, the timer valuation
function is defined as follows: $V(t_2)=1$ at $M_2$, and $V(t_2)=*$ at each of the four different markings. It is easy
to see that $TS(\overline{E})$ and $RG(N)$ are isomorphic.
\label{boxdrg.exm}
\end{example}

The following example demonstrates that without the syntactic restriction on regularity of expressions the
corresponding clocked dtsd-boxes may be not safe.

\begin{figure}
\begin{center}
\includegraphics{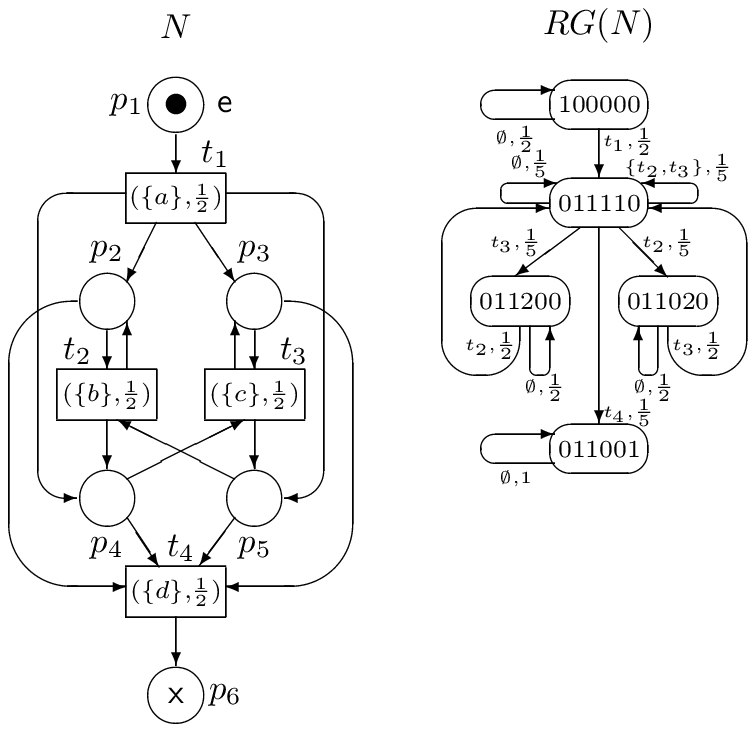}
\end{center}
\caption{The clocked dtsd-box $N=Box_{dtsd}(\overline{E})$ for $E=[((\{a\},\frac{1}{2})*((\{b\},\frac{1}{2})\|
(\{c\},\frac{1}{2}))*(\{d\},\frac{1}{2})]$ and its reachability graph}
\label{nrboxrg.fig}
\end{figure}

\begin{example}
Let $E=[((\{a\},\frac{1}{2})*((\{b\},\frac{1}{2})\| (\{c\},\frac{1}{2}))*(\{d\},\frac{1}{2})]$. In Figure
\ref{nrboxrg.fig}, the clocked dtsd-box $N=Box_{dtsd}(\overline{E})$ and its reachability graph $RG(N)$ are presented.
Since $N$ has no waiting transitions,
we may consider the substituent markings as the whole states. At the marking $(0,1,1,2,0,0)$ there are $2$ tokens in
the place $p_4$. Symmetrically, at the marking $(0,1,1,0,2,0)$ there are $2$ tokens in the place $p_5$. Thus, allowing
con\-currency in the second argument of iteration in the expression $\overline{E}$ can lead to non-safeness of the
corresponding clocked dtsd-box $N$, though, it is $2$-bounded in the worst case \cite{BDK01}. The origin of the problem
is that $N$ has as a self-loop with two subnets which can function independently. Therefore, we have decided to
consider regular expressions only, since the alternative, which is a safe version of the iteration operator with six
arguments in the corresponding dtsd-box, like that from \cite{BDK01}, is rather cumbersome and has too intricate PN
interpretation. Our motivation was to keep the algebraic and PN specifications as simple as possible.
\label{nrboxrg.exm}
\end{example}

\section{Performance evaluation}
\label{perfeval.sec}

In this section we demonstrate how Markov chains corresponding to the expressions and dtsd-boxes can be constructed and
then used for performance evaluation.

\subsection{Analysis of the underlying SMC}

For a dynamic expression $G$, a discrete random variable $\xi (s)$ is associated with every tangible state $s\in
DR_T(G)$. The variable captures the residence (sojourn) time in the state. One can interpret staying in a state at the
next discrete time moment as a failure and leaving it as a success in some trial series. It is easy to see that
$\xi (s)$ is geometrically distributed with the parameter $1-PM(s,s)$, since the probability to stay in $s$ for $k-1$
time moments and leave it at the moment $k\geq 1$, called the probability mass function (PMF) of the residence time in
$s$, is $p_{\xi (s)}(k)={\sf P}(\xi (s)=k)=PM(s,s)^{k-1}(1-PM(s,s))\ (k\in\nat_{\geq 1})$ (the residence time is $k$ in
this case).
Hence, the probability distribution function (PDF) of the residence time in $s$ is $F_{\xi (s)}(k)={\sf P}(\xi
(s)<k)=1-PM(s,s)^{k-1}\ (k\in\nat_{\geq 1})$ (the probability that the residence time in $s$ is less than $k$).
The mean value formula for the geometrical distribution allows us to calculate the average sojourn time in $s$ as
$SJ(s)=\frac{1}{1-PM(s,s)}$. The average sojourn time
in each vanishing state $s\in DR_V(G)$ is $SJ(s)=0$. Let $s\in DR(G)$.

The {\em average sojourn time in the state $s$} is

$$SJ(s)=
\left\{
\begin{array}{ll}
\frac{1}{1-PM(s,s)}, & s\in DR_T(G);\\
0, & s\in DR_V(G).
\end{array}
\right.$$

The {\em average sojourn time vector} of $G$, denoted by $SJ$, has the elements $SJ(s),\ s\in DR(G)$.

The {\em sojourn time variance in the state $s$} is

$$VAR(s)=
\left\{
\begin{array}{ll}
\frac{PM(s,s)}{(1-PM(s,s))^2}, & s\in DR_T(G);\\
0, & s\in DR_V(G).
\end{array}
\right.$$

The {\em sojourn time variance vector} of $G$, denoted by $VAR$, has the elements $VAR(s),\ s\in DR(G)$.

To evaluate performance of the system specified by a dynamic expression $G$, we should investigate the stochastic
process associated with it. The process is the underlying semi-Markov chain (SMC) \cite{Ros96,Kul10}, denoted by
$SMC(G)$, which can be analyzed by extracting from it the embedded (absorbing) discrete time Markov chain (EDTMC)
corresponding to $G$, denoted by $EDTMC(G)$. The construction of the latter is analogous to that applied in the context
of generalized stochastic PNs (GSPNs) in \cite{Mar90,Bal01,Bal07}, and also in the framework of discrete time
deterministic and stochastic PNs (DTDSPNs) in \cite{Zij95,Zij97,ZFGH00,ZFH01}, as well as within discrete deterministic
and stochastic PNs (DDSPNs) \cite{ZCH97}. $EDTMC(G)$ only describes the state changes of $SMC(G)$ while ignoring its
time characteristics. Thus, to construct the EDTMC, we should abstract from all time aspects of behaviour of the SMC,
i.e. from the sojourn time in its states. The (local) sojourn time in every state of the EDTMC is deterministic and it
is equal to one discrete time unit. It is well-known that every SMC is fully described by the EDTMC and the state
sojourn time distributions (the latter can be specified by the vector of PDFs of residence time in the states)
\cite{Hav01}.

Let $G$ be a dynamic expression and $s,\tilde{s}\in DR(G)$. The transition system $TS(G)$ can have self-loops going
from a state to itself which have a non-zero probability.
Clearly, the current state remains unchanged in this~case.

Let $s\rightarrow s$. The {\em probability to stay in $s$ due to $k\ (k\geq 1)$ self-loops} is

$$PM(s,s)^k.$$

Let $s\rightarrow\tilde{s}$ and $s\neq\tilde{s}$. The {\em probability to move from $s$ to $\tilde{s}$ by executing any
multiset of activities after possible self-loops} is

$$PM^*(s,\tilde{s})=\left\{
\begin{array}{ll}
PM(s,\tilde{s})\sum_{k=0}^{\infty}PM(s,s)^k=\frac{PM(s,\tilde{s})}{1-PM(s,s)}, & s\rightarrow s;\\
PM(s,\tilde{s}), & \mbox{otherwise};
\end{array}
\right\}=SL(s)PM(s,\tilde{s}),\mbox{ where}$$

$$SL(s)=\left\{
\begin{array}{ll}
\frac{1}{1-PM(s,s)}, & s\rightarrow s;\\
1, & \mbox{otherwise};
\end{array}
\right.$$
Here $SL(s)$ is the {\em self-loops abstraction factor in the state $s$}. The {\em self-loops abstraction vector} of
$G$, denoted by $SL$, has the elements $SL(s),\ s\in DR(G)$. The value $k=0$ in the summation above corresponds to the
case when no self-loops occur.

Let $s\in DR_T(G)$. If there exist self-loops from $s$ (i.e. if $s\rightarrow s$) then $PM(s,s)>0$ and
$SL(s)=\frac{1}{1-PM(s,s)}=SJ(s)$. Otherwise, if there exist no self-loops from $s$ then $PM(s,s)=0$ and
$SL(s)=1=\frac{1}{1-PM(s,s)}=SJ(s)$. Thus, $\forall s\in DR_T(G)\ SL(s)=SJ(s)$, hence, $\forall s\in DR_T(G)\
PM^*(s,\tilde{s})=SJ(s)PM(s,\tilde{s})$.
Note that the self-loops from tangible states are of the empty or non-empty type, the latter produced by iteration,
since empty loops are not possible from w-tangible states, but they are possible from s-tangible states, while
non-empty loops are possible from both s-tangible and w-tangible states.

Let $s\in DR_V(G)$. We have $\forall s\in DR_V(G)\ SL(s)\neq SJ(s)=0$ and $\forall s\in DR_V(G)\ PM^*(s,\tilde{s})=
SL(s)PM(s,\tilde{s})$. If there exist self-loops from $s$ then $PM^*(s,\tilde{s})=\frac{PM(s,\tilde{s})}{1-PM(s,s)}$.
Otherwise, if there exist no self-loops from $s$ then $PM^*(s,\tilde{s})=PM(s,\tilde{s})$. Note that the self-loops
from vanishing states are always of the non-empty type, produced by iteration, since empty loops are not possible from
vanishing states.

Note that after abstraction from the probabilities of transitions which do not change the states, the remaining
transition probabilities are normalized. In order to calculate transition probabilities $PT(\Upsilon ,s)$, we had to
normalize $PF(\Upsilon ,s)$. Then, to obtain transition probabilities of the state-changing steps $PM^*(s,\tilde{s})$,
we now have to normalize $PM(s,\tilde{s})$. Thus, we have a two-stage normalization as a result.

Notice that $PM^*(s,\tilde{s})$ defines a probability distribution, since $\forall s\in DR(G)$ such that $s$ is not a
terminal state, i.e. there are transitions to different states after possible self-loops from it, we have
$\sum_{\{\tilde{s}\mid s\rightarrow\tilde{s},\ s\neq\tilde{s}\}}PM^*(s,\tilde{s})=
\frac{1}{1-PM(s,s)}\sum_{\{\tilde{s}\mid s\rightarrow\tilde{s},\ s\neq\tilde{s}\}}PM(s,\tilde{s})=
\frac{1}{1-PM(s,s)}(1-PM(s,s))=1$.

We decided to consider self-loops followed only by a state-changing step just for convenience. Alternatively, we could
take a state-changing step followed by self-loops or a state-changing step preceded and followed by self-loops. In all
these three cases our sequence begins or/and ends with the loops which do not change states. At the same time, the
overall probabilities of the evolutions can differ, since self-loops have positive probabilities. To avoid
inconsistency of definitions and too complex description, we consider sequences ending with a state-changing step. It
resembles in some sense a construction of branching bisimulation \cite{Gla93} taking self-loops instead of silent
transitions.

\begin{definition}
Let $G$ be a dynamic expression. The {\em embedded (absorbing) discrete time Markov chain\\
(EDTMC)} of $G$, denoted by $EDTMC(G)$, has the state space $DR(G)$, the initial state $[G]_\approx$ and the
transitions $s\doublera_{\cal P}\tilde{s}$, if $s\rightarrow\tilde{s}$ and $s\neq\tilde{s}$, where ${\cal
P}=PM^*(s,\tilde{s})$.

The {\em underlying SMC} of $G$, denoted by $SMC(G)$, has the EDTMC $EDTMC(G)$ and the sojourn time in every $s\in
DR_T(G)$ is geometrically distributed with the parameter $1-PM(s,s)$ while the sojourn time
in every $s\in DR_V(G)$ is equal to $0$.
\end{definition}

EDTMCs and underlying SMCs of static expressions can be defined as well. For $E\in RegStatExpr$, let
$EDTMC(E)=EDTMC(\overline{E})$ and $SMC(E)=SMC(\overline{E})$.

Let $G$ be a dynamic expression. The elements ${\cal P}_{ij}^*\ (1\leq i,j\leq n=|DR(G)|)$ of the (one-step) transition
probability matrix (TPM) ${\bf P}^*$ for $EDTMC(G)$ are defined as

$${\cal P}_{ij}^*=\left\{
\begin{array}{ll}
PM^*(s_i,s_j), & s_i\rightarrow s_j,\ s_i\neq s_j;\\
0, & \mbox{otherwise}.
\end{array}
\right.$$

The transient ($k$-step, $k\in\nat$) PMF $\psi^*[k]=(\psi^*[k](s_1),\ldots ,\psi^*[k](s_n))$ for $EDTMC(G)$ is
calculated as

$$\psi^*[k]=\psi^*[0]({\bf P}^*)^k,$$
where $\psi^*[0]=(\psi^*[0](s_1),\ldots ,\psi^*[0](s_n))$ is the initial PMF defined as

$$\psi^*[0](s_i)=\left\{
\begin{array}{ll}
1, & s_i=[G]_\approx ;\\
0, & \mbox{otherwise}.
\end{array}
\right.$$

Note also that $\psi^*[k+1]=\psi^*[k]{\bf P}^*\ (k\in\nat )$.

The steady-state PMF $\psi^*=(\psi^*(s_1),\ldots ,\psi^*(s_n))$ for $EDTMC(G)$ is a solution of the equation system

$$\left\{
\begin{array}{l}
\psi^*({\bf P}^*-{\bf I})={\bf 0}\\
\psi^*{\bf 1}^T=1
\end{array}
\right.,$$
where ${\bf I}$ is the identity matrix of order $n$ and ${\bf 0}$ is a row vector of $n$ values $0,\ {\bf 1}$ is that
of $n$ values $1$.

Note that the vector $\psi^*$ exists and is unique if $EDTMC(G)$ is ergodic. Then $EDTMC(G)$ has a single steady
state, and we have $\psi^*=\lim_{k\to\infty}\psi^*[k]$.

The steady-state PMF for the underlying semi-Markov chain $SMC(G)$ is calculated via multiplication of every
$\psi^*(s_i)\ (1\leq i\leq n)$ by the average sojourn time $SJ(s_i)$ in the state $s_i$, after which we normalize the
resulting values. Remember that for each w-tangible state $s\in DR_{WT}(G)$ we have $SJ(s)=1$, and for each vanishing
state $s\in DR_V(G)$ we have $SJ(s)=0$.

Thus, the steady-state PMF $\varphi =(\varphi (s_1),\ldots ,\varphi (s_n))$ for $SMC(G)$ is

$$\varphi (s_i)=\left\{
\begin{array}{ll}
\frac{\psi^*(s_i)SJ(s_i)}{\sum_{j=1}^n\psi^*(s_j)SJ(s_j)}, & s_i\in DR_T(G);\\
0, & s_i\in DR_V(G).
\end{array}
\right.$$

Thus, to calculate $\varphi$, we apply abstraction from self-loops to get ${\bf P}^*$ and then $\psi^*$, followed by
weighting by $SJ$ and normalization. $EDTMC(G)$ has no self-loops, unlike $SMC(G)$, hence, the behaviour of $EDTMC(G)$
stabilizes quicker than that of $SMC(G)$ (if each of them has a single steady state), since ${\bf P}^*$ has only zero
elements at the main diagonal.

\begin{figure}
\begin{center}
\includegraphics{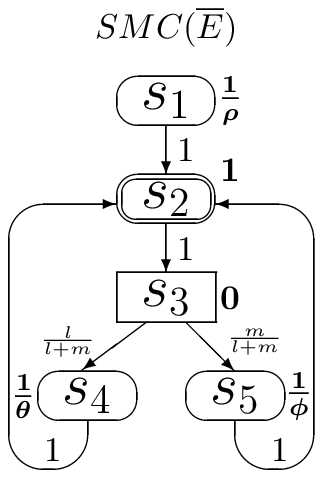}
\end{center}
\caption{The underlying SMC of $\overline{E}$ for $E=[(\{a\},\rho )*((\{b\},\natural_k^1);(((\{c\},\natural_l^0);
(\{d\},\theta ))\cho ((\{e\},\natural_m^0);(\{f\},\phi ))))*{\sf Stop}]$}
\label{exprdsmc.fig}
\end{figure}

\begin{example}
Let $E$ be from Example \ref{tsitchoswim.exm}. In Figure \ref{exprdsmc.fig}, the underlying SMC $SMC(\overline{E})$ is
presented. The average sojourn times in the states of the underlying SMC is written next to them in bold font.

The average sojourn time vector of $\overline{E}$ is

$$SJ=\left(\frac{1}{\rho},1,0,\frac{1}{\theta},\frac{1}{\phi}\right).$$

The sojourn time variance vector of $\overline{E}$ is

$$VAR=\left(\frac{1-\rho}{\rho^2},0,0,\frac{1-\theta}{\theta^2},\frac{1-\phi}{\phi^2}\right).$$

The TPM for $EDTMC(\overline{E})$ is

$${\bf P}^*=\left(\begin{array}{ccccc}
0 & 1 & 0 & 0 & 0\\
0 & 0 & 1 & 0 & 0\\
0 & 0 & 0 & \frac{l}{l+m} & \frac{m}{l+m}\\
0 & 1 & 0 & 0 & 0\\
0 & 1 & 0 & 0 & 0
\end{array}\right).$$

The steady-state PMF for $EDTMC(\overline{E})$ is

$$\psi^*=\left(0,\frac{1}{3},\frac{1}{3},\frac{l}{3(l+m)},\frac{m}{3(l+m)}\right).$$

The steady-state PMF $\psi^*$ weighted by $SJ$ is

$$\left(0,\frac{1}{3},0,\frac{l}{3\theta (l+m)},\frac{m}{3\phi (l+m)}\right).$$

It remains to normalize the steady-state weighted PMF by dividing it by the sum of its components

$$\psi^* SJ^T=\frac{\theta\phi (l+m)+\phi l+\theta m}{3\theta\phi (l+m)}.$$

Thus, the steady-state PMF for $SMC(\overline{E})$ is

$$\varphi =\frac{1}{\theta\phi (l+m)+\phi l+\theta m}(0,\theta\phi (l+m),0,\phi l,\theta m).$$

In the case $l=m$ and $\theta=\phi$ we have

$$\varphi =\frac{1}{2(1 +\theta )}(0,2\theta ,0,1,1).$$

\label{exprdsmc.exm}
\end{example}

Let $G$ be a dynamic expression and $s,\tilde{s}\in DR(G),\ S,\widetilde{S}\subseteq DR(G)$. The following standard
{\em performance indices (measures)} can be calculated based on the steady-state PMF $\varphi$ for $SMC(G)$ and the
average sojourn time vector $SJ$ of $G$ \cite{MAS85,Kat96}.
\begin{itemize}

\item The {\em average recurrence (return) time in the state $s$} (i.e. the number of discrete time units or steps
required for this) is $ReturnTime(s)=\frac{1}{\varphi (s)}$.

\item The {\em fraction of residence time in the state $s$} is $TimeFract(s)=\varphi (s)$.

\item The {\em fraction of residence time in the set of states $S$} or the {\em probability of the event determined by
a condition that is true for all states from $S$} is $TimeFract(S)=\sum_{s\in S}\varphi (s)$.

\item The {\em relative fraction of residence time in the set of states $S$ with respect to that in $\widetilde{S}$}
is\\
$RltTimeFract(S,\widetilde{S})=\frac{\sum_{s\in S}\varphi (s)}{\sum_{\tilde{s}\in\widetilde{S}}\varphi (\tilde{s})}$.

\item The {\em exit/entrance frequency (rate of leaving/entering, average number of exits/entrances per unit of time)
the state $s$} is $ExitFreq(s)=\frac{\varphi (s)}{SJ(s)}$.

\item The {\em steady-state probability to perform a step with a multiset of activities $\Xi$} is\\
$ActsProb(\Xi )=\sum_{s\in DR(G)}\varphi (s)\sum_{\{\Upsilon\mid\Xi\subseteq\Upsilon\}}PT(\Upsilon ,s)$.

\item The {\em steady-state execution frequency (throughput) of the activity $(\alpha ,\kappa )$} is\\
$ExecFreq((\alpha ,\kappa ))=\sum_{s\in DR(G)}\frac{\varphi (s)}{SJ(s)}\sum_{\{\Upsilon\mid (\alpha ,\kappa
)\in\Upsilon\}}PT(\Upsilon ,s)$.

\item The {\em probability of the event determined by a reward function $r$ on the states} is\\
$Prob(r)=\sum_{s\in DR(G)}\varphi (s)r(s)$, where $\forall s\in DR(G)\ 0\leq r(s)\leq 1$.

\end{itemize}

\begin{example}
Let us interpret $E$ be from Example \ref{tsitchoswim.exm} as a specification of the travel system. A tourist visits
regularly new cities. After seeing the sights of the current city, he goes to the next city by the nearest train or bus
available at the city station. Buses depart less frequently than trains, but the next city is quicker reached by bus
than by train. We suppose that the stay duration in every city (being a constant), the departure numbers of trains and
buses, as well as their speeds {\em do not depend} on a particular city, bus or train. The travel route has been
planned so that the distances between successive cities {\em coincide}.

The meaning of actions from the syntax of $E$ is as follows. The action $a$ corresponds to the system activation (the
travel route has been planned) that takes a time, geometrically distributed with the parameter $\rho$. The action $b$
represents the completion of looking round the current city and coming to the city station that takes a fixed time
equal to $1$ (say, one hour) for every city. The actions $c$ and $e$ correspond to the urgent getting on bus and train,
respectively, and thus model the choice between these two transport facilities. The weights of the two corresponding
immediate multiactions suggest that every $l$ departures of buses take the same time as $m$ departures of trains
($l<m$), hence, a bus departs with the probability $\frac{l}{l+m}$ while a train departs with the probability
$\frac{m}{l+m}$. The actions $c$ and $e$ correspond to the coming in a city by bus and train, respectively, that takes
a time, geometrically distributed with the parameters $\theta$ and $\phi$, respectively
($\theta >\phi$).

The meaning of states from $DR(\overline{E})$ is the following. The s-tangible state $s_1$ corresponds to staying at
home and planning the future travel. The w-tangible state $s_2$ means residence in a city for exactly one time unit
(hour). The vanishing state $s_3$ with zero residence time represents instantaneous stay at the city station,
signifying that the tourist does not wait there for departure of the transport. The s-tangible states $s_4$ and $s_5$
correspond to going by bus and train, respectively.

Using Example \ref{exprdsmc.exm}, we now calculate the performance indices, based on the steady-state PMF for
$SMC(\overline{E})\\
\varphi =\frac{1}{\theta\phi (l+m)+\phi l+\theta m}(0,\theta\phi (l+m),0,\phi l,\theta m)$ and the average sojourn time
vector of $\overline{E}\ SJ=\left(\frac{1}{\rho},1,0,\frac{1}{\theta},\frac{1}{\phi}\right)$.
\begin{itemize}

\item The average time between comings to the successive cities (mean sightseeing and travel time) is\\
$ReturnTime(s_2)=\frac{1}{\varphi (s_2)}=1+\frac{\phi l+\theta m}{\theta\phi (l+m)}$.

\item The fraction of time spent in a city (sightseeing time fraction) is $TimeFract(s_2)=\varphi (s_2)=\\
\frac{\theta\phi (l+m)}{\theta\phi (l+m)+\phi l+\theta m}$.

\item The fraction of time spent in a transport (travel time fraction) is $TimeFract(\{s_4,s_5\})=
\varphi (s_4)+\varphi (s_5)=\frac{\phi l+\theta m}{\theta\phi (l+m)+\phi l+\theta m}$.

\item The relative fraction of time spent in a city with respect to that spent in transport (sightseeing relative to
travel time fraction) is $RltTimeFract(\{s_2\},\{s_4,s_5\})=\frac{\varphi (s_2)}{\varphi (s_4)+\varphi (s_5)}=
\frac{\theta\phi (l+m)}{\phi l+\theta m}$.

\item The rate of leaving/entering a city (departure/arrival rate) is $ExitFreq(s_2)=\frac{\varphi (s_2)}{SJ(s_2)}=
\frac{\theta\phi (l+m)}{\theta\phi (l+m)+\phi l+\theta m}$.

\end{itemize}
\label{travsys.exm}
\end{example}

Let $N=(P_N,T_N,W_N,D_N,\Omega_N,{\cal L}_N,Q_N)$ be a LDTSDPN and $Q,\widetilde{Q}$ be its states. Then the average
sojourn time $SJ(Q)$, the sojourn time variance $VAR(Q)$, the probabilities $PM^*(M,\widetilde{M})$, the transition
relation $M\doublera_{\cal P}\widetilde{M}$, the EDTMC $EDTMC(N)$, the underlying SMC $SMC(N)$ and the steady-state PMF
for it are defined like the corresponding notions for dynamic expressions.

As we have mentioned earlier, every clocked plain dtsd-box could be interpreted as the LDTSDPN. Therefore, we can
evaluate performance with the LDTSDPNs corresponding to dtsd-boxes and then transfer the results to the latter.

Let $\simeq$ denote isomorphism between SMCs that binds their initial states, where two SMCs are isomorphic if their
EDTMCs are so and the sojourn times in the isomorphic states of the EDTMCs are identically distributed.

\begin{proposition}
For any static expression $E$

$$SMC(\overline{E})\simeq SMC(Box_{dtsd}(\overline{E})).$$

\label{smcs.pro}
\end{proposition}
{\em Proof.} By Theorem \ref{opdensem.the} and definitions of underlying SMCs for dynamic expressions and LDTSDPNs
taking into account the following. First, for the associated SMCs, the average sojourn time in the states is the same,
since it is defined via the analogous probability functions. Second, the transition probabilities of the associated
SMCs are the sums of those belonging to transition systems or reachability graphs.

Fore instance, observe that the probability functions $PM(s,\tilde{s})$ and $PM^*(s,\tilde{s})$ can be respectively
defined in the same way as $PM(Q,\widetilde{Q})$ and $PM^*(Q,\widetilde{Q})$, for the corresponding $s$ and $Q$, as
well as $\tilde{s}$ and $\widetilde{Q}$. \hfill $\eop$

\begin{figure}
\begin{center}
\includegraphics{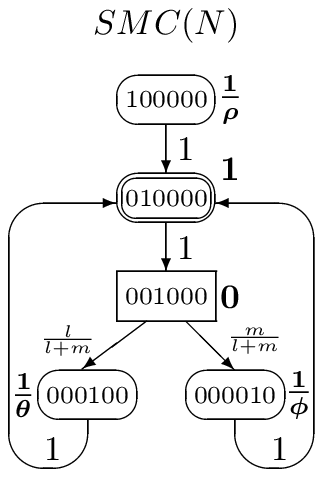}
\end{center}
\caption{The underlying SMC of $N=Box_{dtsd}(\overline{E})$ for $E=[(\{a\},\rho )*((\{b\},\natural_k^1);
(((\{c\},\natural_l^0);(\{d\},\theta ))\cho ((\{e\},\natural_m^0);\protect\newline
(\{f\},\phi ))))*{\sf Stop}]$}
\label{boxdsmc.fig}
\end{figure}

\begin{example}
Let $E$ be from Example \ref{tsitchoswim.exm}. In Figure \ref{boxdsmc.fig}, the underlying SMC $SMC(N)$ is presented. It
is easy to see that $SMC(\overline{E})$ and $SMC(N)$ are isomorphic. Thus, both the transient and steady-state PMFs for
$SMC(N)$ and $SMC(\overline{E})$ coincide.
\label{boxdsmc.exm}
\end{example}

As mentioned in \cite{Zij95,Zij97}, if is useful to consider performance measures over only the markings of DTDSPNs,
instead of their whole states, whose second components are the remaining firing time vectors. In the context of
dtsdPBC, such markings correspond to those of the dtsd-boxes of dynamic expressions, i.e. to the markings of the
respective LDTSDPNs, obtained from their states by abstracting from the second components, which are the timer
valuation functions.

Let $G$ be a dynamic expression.
The {\em underlying timer-free state} of a state $s\in DR(G)$ is defined as
$\downharpoonleft\!\!s=[\downharpoonleft\!\!H]_\approx$ for $H\in s$. Since structurally equivalent dynamic expressions
obviously remain so after removing their timer value annotations,
$\downharpoonleft\!\!s$ is unique for each $s$ and the previous definition is correct. The timer-free states (i.e. those
from $\downharpoonleft\!\!DR(G)=\{\downharpoonleft\!\!s\mid s\in DR(G)\}$) correspond to the markings of the LDTSDPN
$N=Box_{dtsd}(G)$.
Let $s\in DR(G)$ and $\bar{s}=\downharpoonleft\!\!s$. The average sojourn time vector $SJ$, sojourn time variance
vector $VAR$ and steady-state PMF for $SMC(G)$ over the timer-free states of $G$ are defined as follows:
$SJ(\bar{s})=\sum_{\{s\in DR(G)\mid\downharpoonleft\!s=\bar{s}\}}SJ(s),\ VAR(\bar{s})=\sum_{\{s\in
DR(G)\mid\downharpoonleft\!s=\bar{s}\}}VAR(s)$ and $\varphi (\bar{s})=\sum_{\{s\in DR(G)\mid
\downharpoonleft\!s=\bar{s}\}}\varphi (s)$. Then $\varphi (\bar{s})$ and $SJ(\bar{s})$ can be used to calculate the
standard {\em performance indices over the timer-free states} of $G$ (hence, over the markings of $N$), by analogy with
the standard performance indices, defined over the arbitrary states of $G$. Then also the performance measures that are
specific for LDTSDPNs can be derived, based on the numbers of tokens in the places of $N$.

\subsection{Analysis of the DTMC}

Let us consider an alternative solution method, studying the DTMCs of expressions based on the state change
probabilities $PM(s,\tilde{s})$.

\begin{definition}
Let $G$ be a dynamic expression. The {\em discrete time Markov chain (DTMC)} of $G$, denoted by $DTMC(G)$, has the
state space $DR(G)$, the initial state $[G]_\approx$ and the transitions $s\rightarrow_{\cal P}\tilde{s}$, where ${\cal
P}=PM(s,\tilde{s})$.
\end{definition}

\noindent DTMCs of static expressions can be defined as well. For $E\in RegStatExpr$, let $DTMC(E)=DTMC(\overline{E})$.

One can see that $EDTMC(G)$ is constructed from $DTMC(G)$ as follows. For each state of $DTMC(G)$, we remove a possible
self-loop associated with it and then normalize the probabilities of the remaining transitions from the state. Thus,
$EDTMC(G)$ and $DTMC(G)$ differ only by existence of self-loops and magnitudes of the probabilities of the remaining
transitions. Hence, $EDTMC(G)$ and $DTMC(G)$ have the same communication classes of states and $EDTMC(G)$ is
irreducible iff $DTMC(G)$ is so. Since both $EDTMC(G)$ and $DTMC(G)$ are finite, they are positive recurrent. Thus, in
case of irreducibility, each of them has a single stationary PMF. Note that both $EDTMC(G)$ and $DTMC(G)$ or just one
of them may be periodic, thus having a unique stationary distribution, but no steady-state (limiting) one. For example,
it may happen that $EDTMC(G)$ is periodic while $DTMC(G)$ is aperiodic due to self-loops associated with some states of
the latter. The states of $SMC(G)$ are classified using $EDTMC(G)$, hence, $SMC(G)$ is irreducible (positive recurrent,
aperiodic) iff $EDTMC(G)$ is so.

Let $G$ be a dynamic expression. The elements ${\cal P}_{ij}\ (1\leq i,j\leq n=|DR(G)|)$ of (one-step) transition
probability matrix (TPM) ${\bf P}$ for $DTMC(G)$ are defined as

$${\cal P}_{ij}=\left\{
\begin{array}{ll}
PM(s_i,s_j), & s_i\rightarrow s_j;\\
0, & \mbox{otherwise}.
\end{array}
\right.$$

The steady-state PMF $\psi$ for $DTMC(G)$ is defined like the corresponding notion $\psi^*$ for $EDTMC(G)$.

Let us determine a relationship between steady-state PMFs for $DTMC(G)$ and $EDTMC(G)$. The following theorem proposes
the equation that relates the mentioned steady-state PMFs.

First, we introduce some helpful notation. For a vector $v=(v_1,\ldots ,v_n)$, let $Diag(v)$ be a diagonal matrix of
order $n$ with the elements $Diag_{ij}(v)\ (1\leq i,j\leq n)$ defined as

$$Diag_{ij}(v)=\left\{
\begin{array}{ll}
v_i, & i=j;\\
0, & \mbox{otherwise}.
\end{array}
\right.$$

\begin{theorem}
Let $G$ be a dynamic expression and $SL$ be its self-loops abstraction vector. Then the steady-state PMFs $\psi$ for
$DTMC(G)$ and $\psi^*$ for $EDTMC(G)$ are related as follows: $\forall s\in DR(G)$

$$\psi (s)=\frac{\psi^*(s)SL(s)}{\sum_{\tilde{s}\in DR(G)}\psi^*(\tilde{s})SL(\tilde{s})}.$$

\label{pmfsdm.the}
\end{theorem}
{\em Proof.} Let $PSL$ be a vector with the elements

$$PSL(s)=\left\{
\begin{array}{ll}
PM(s,s), & s\rightarrow s;\\
0, & \mbox{otherwise}.
\end{array}\right.$$

By definition of $PM^*(s,\tilde{s})$, we have ${\bf P}^*=Diag(SL)({\bf P}-Diag(PSL))$. Further,

$$\psi^*({\bf P}^*-{\bf I})={\bf 0}\mbox{ and }\psi^*{\bf P}^*=\psi^*.$$

After replacement of ${\bf P}^*$ by $Diag(SL)({\bf P}-Diag(PSL))$ we obtain

$$\psi^*Diag(SL)({\bf P}-Diag(PSL))=\psi^*\mbox{ and }\psi^*Diag(SL){\bf P}=\psi^*(Diag(SL)Diag(PSL)+{\bf I}).$$

Note that $\forall s\in DR(G)$ we have

$$SL(s)PSL(s)+1=\left\{
\begin{array}{ll}
SL(s)PM(s,s)+1=\frac{PM(s,s)}{1-PM(s,s)}+1=\frac{1}{1-PM(s,s)}, & s\rightarrow s;\\
SL(s)\cdot 0+1=1, & \mbox{otherwise};
\end{array}\right\}=SL(s).$$

Hence, $Diag(SL)Diag(PSL)+{\bf I}=Diag(SL)$. Thus,

$$\psi^*Diag(SL){\bf P}=\psi^*Diag(SL).$$

Then, for $v=\psi^*Diag(SL)$, we have

$$v{\bf P}=v\mbox{ and }v({\bf P}-{\bf I})={\bf 0}.$$

In order to calculate $\psi$ on the basis of $v$, we must normalize it by dividing its elements by their sum, since we
should have $\psi{\bf 1}^T=1$ as a result:

$$\psi =\frac{1}{v{\bf 1}^T}v=\frac{1}{\psi^*Diag(SL){\bf 1}^T}\psi^*Diag(SL).$$

Thus, the elements of $\psi$ are calculated as follows: $\forall s\in DR(G)$

$$\psi (s)=\frac{\psi^*(s)SL(s)}{\sum_{\tilde{s}\in DR(G)}\psi^*(\tilde{s})SL(\tilde{s})}.$$

It is easy to check that $\psi$ is a solution of the equation system

$$\left\{
\begin{array}{l}
\psi({\bf P}-{\bf I})={\bf 0}\\
\psi{\bf 1}^T=1
\end{array}
\right.,$$
hence, it is indeed the steady-state PMF for $DTMC(G)$. \hfill $\eop$

The following proposition relates the steady-state PMFs for $SMC(G)$ and $DTMC(G)$.

\begin{proposition}
Let $G$ be a dynamic expression, $\varphi$ be the steady-state PMF for $SMC(G)$ and $\psi$ be the steady-state PMF for
$DTMC(G)$. Then $\forall s\in DR(G)$

$$\varphi (s)=\left\{
\begin{array}{ll}
\frac{\psi (s)}{\sum_{\tilde{s}\in DR_T(G)}\psi (\tilde{s})}, & s\in DR_T(G);\\
0, & s\in DR_V(G).
\end{array}
\right.$$

\label{pmfsmc.pro}
\end{proposition}
{\em Proof.} Let $s\in DR_T(G)$. Remember that $\forall s\in DR_T(G)\ SL(s)=SJ(s)$ and $\forall s\in DR_V(G)\ SJ(s)=0$.\\
Then, by Theorem \ref{pmfsdm.the}, we have $\frac{\psi (s)}{\sum_{\tilde{s}\in DR_T(G)}\psi
(\tilde{s})}=\frac{\frac{\psi^*(s)SL(s)}{\sum_{\tilde{s}\in DR(G)}\psi^*(\tilde{s})SL(\tilde{s})}} {\sum_{\tilde{s}\in
DR_T(G)}\left(\frac{\psi^*(\tilde{s})SL(\tilde{s})}{\sum_{\breve{s}\in
DR(G)}\psi^*(\breve{s})SL(\breve{s})}\right)}=\frac{\psi^*(s)SL(s)}{\sum_{\tilde{s}\in
DR(G)}\psi^*(\tilde{s})SL(\tilde{s})}\cdot\\
\frac{\sum_{\breve{s}\in DR(G)}\psi^*(\breve{s})SL(\breve{s})}{\sum_{\tilde{s}\in
DR_T(G)}\psi^*(\tilde{s})SL(\tilde{s})}=\frac{\psi^*(s)SL(s)}{\sum_{\tilde{s}\in
DR_T(G)}\psi^*(\tilde{s})SL(\tilde{s})}=\frac{\psi^*(s)SJ(s)}{\sum_{\tilde{s}\in
DR_T(G)}\psi^*(\tilde{s})SJ(\tilde{s})}=\frac{\psi^*(s)SJ(s)}{\sum_{\tilde{s}\in
DR(G)}\psi^*(\tilde{s})SJ(\tilde{s})}=\varphi (s)$. \hfill $\eop$

Thus, to calculate $\varphi$, one can only apply normalization to some elements of $\psi$ (corresponding to the
tangible states), instead of abstracting from self-loops to get ${\bf P}^*$ and then $\psi^*$, followed by weighting by
$SJ$ and normalization. Hence, using $DTMC(G)$ instead of $EDTMC(G)$ allows one to avoid multistage analysis, but the
payment for it is more time-consuming numerical and more complex analytical calculation of $\psi$ with respect to
$\psi^*$. The reason is that $DTMC(G)$ has self-loops, unlike $EDTMC(G)$, hence, the behaviour of $DTMC(G)$ stabilizes
slower than that of $EDTMC(G)$ (if each of them has a single steady state) and ${\bf P}$ is more dense matrix than
${\bf P}^*$, since ${\bf P}$ may additionally have non-zero elements at the main diagonal. Nevertheless, Proposition
\ref{pmfsmc.pro} is very important, since the relationship between $\varphi$ and $\psi$ it discovers will be used in
Proposition \ref{pmfsmctang.pro} to relate the steady-state PMFs for $SMC(G)$ and the reduced $DTMC(G)$.

\begin{figure}
\begin{center}
\includegraphics{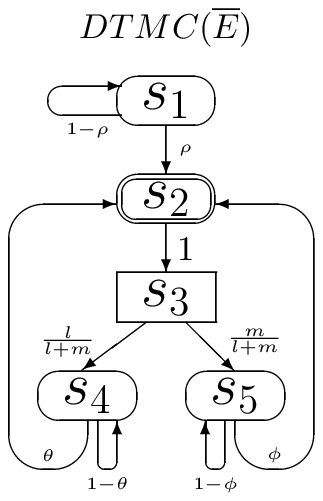}
\end{center}
\caption{The DTMC of $\overline{E}$ for $E=[(\{a\},\rho )*((\{b\},\natural_k^1);(((\{c\},\natural_l^0);
(\{d\},\theta ))\cho ((\{e\},\natural_m^0);(\{f\},\phi ))))*{\sf Stop}]$}
\label{exprddtmc.fig}
\end{figure}

\begin{example}
Let $E$ be from Example \ref{tsitchoswim.exm}. In Figure \ref{exprddtmc.fig}, the DTMC $DTMC(\overline{E})$ is
presented.

The TPM for $DTMC(\overline{E})$ is

$${\bf P}=\left(\begin{array}{ccccc}
1-\rho & \rho & 0 & 0 & 0\\
0 & 0 & 1 & 0 & 0\\
0 & 0 & 0 & \frac{l}{l+m} & \frac{m}{l+m}\\
0 & \theta & 0 & 1-\theta & 0\\
0 & \phi & 0 & 0 & 1-\phi
\end{array}\right).$$

The steady-state PMF for $DTMC(\overline{E})$ is

$$\psi =\frac{1}{2\theta\phi (l+m)+\phi l+\theta m}(0,\theta\phi (l+m),\theta\phi (l+m),\phi l,\theta m).$$

Remember that $DR_T(\overline{E})=DR_{ST}(\overline{E})\cup DR_{WT}(\overline{E})=\{s_1,s_2,s_4,s_5\}$ and
$DR_V(\overline{E})=\{s_3\}$. Hence,

$$\sum_{\tilde{s}\in DR_T(\overline{E})}\psi (\tilde{s})=\psi (s_1)+\psi (s_2)+\psi (s_4)+\psi (s_5)=
\frac{\theta\phi(l+m)+\phi l+\theta m}{2\theta\phi (l+m)+\phi l+\theta m}.$$

By Proposition \ref{pmfsmc.pro}, we have

$$\begin{array}{l}

\varphi (s_1)=0\cdot\frac{2\theta\phi (l+m)+\phi l+\theta m}{\theta\phi(l+m)+\phi l+\theta m}=0,\\[1mm]

\varphi (s_2)=\frac{\theta\phi (l+m)}{2\theta\phi (l+m)+\phi l+\theta m}\cdot
\frac{2\theta\phi (l+m)+\phi l+\theta m}{\theta\phi(l+m)+\phi l+\theta m}=
\frac{\theta\phi (l+m)}{\theta\phi(l+m)+\phi l+\theta m},\\[1mm]

\varphi (s_3)=0,\\[1mm]

\varphi (s_4)=\frac{\phi l}{2\theta\phi (l+m)+\phi l+\theta m}\cdot
\frac{2\theta\phi (l+m)+\phi l+\theta m}{\theta\phi(l+m)+\phi l+\theta m}=
\frac{\phi l}{\theta\phi(l+m)+\phi l+\theta m},\\[1mm]

\varphi (s_5)=\frac{\theta m}{2\theta\phi (l+m)+\phi l+\theta m}\cdot
\frac{2\theta\phi (l+m)+\phi l+\theta m}{\theta\phi(l+m)+\phi l+\theta m}=
\frac{\theta m}{\theta\phi(l+m)+\phi l+\theta m}.

\end{array}$$

Thus, the steady-state PMF for $SMC(\overline{E})$ is

$$\varphi =\frac{1}{\theta\phi (l+m)+\phi l+\theta m}(0,\theta\phi (l+m),0,\phi l,\theta m).$$

This coincides with the result obtained in Example \ref{exprdsmc.exm} with the use of $\psi^*$ and $SJ$.
\label{exprddtmc.exm}
\end{example}

\subsection{Analysis of the reduced DTMC}

Let us now consider the method from \cite{Chi85,CMT89,CMT91,MBCDF95,Bal01,BKr02,Bal07} that eliminates vanishing states
from the EMC (EDTMC, in our terminology) corresponding to the underlying SMC of every GSPN $N$. The TPM for the
resulting {\em reduced} EDTMC (REDTMC) has smaller size than that for the EDTMC. The method demonstrates that there
exists a transformation of the underlying SMC of $N$ into a CTMC, whose states are the tangible markings of $N$. This
CTMC, which is essentially the {\em reduced} underlying SMC (RSMC) of $N$, is constructed on the basis of the REDTMC.
The CTMC can then be directly solved to get both the transient and the steady-state PMFs over the tangible markings of
$N$. In \cite{CMT91}, the program and computational complexities of such an {\em elimination} method, based on the
REDTMC, were evaluated and compared with those of the {\em preservation} method that does not eliminate vanishing
states and based on the EDTMC. The preservation method for GSPNs corresponds in dtsdPBC to the analysis of the
underlying SMCs of expressions.

The elimination method for GSPNs can be easily transferred to dtsdPBC, hence, for every dynamic expression $G$, we can
find a DTMC (since the sojourn time in the tangible states from $DR(G)$ is discrete and geometrically distributed) with
the states from $DR_T(G)$, which can be directly solved to find the transient and the steady-state PMFs over the
tangible states. We shall demonstrate that such a {\em reduced} DTMC (RDTMC) of $G$, denoted by $RDTMC(G)$, can be
constructed from $DTMC(G)$, using the method analogous to that designed in \cite{MBCDF95,Bal01,BKr02,Bal07} in the
framework of GSPNs to transform EDTMC into REDTMC. Since the sojourn time in the vanishing states is zero, the state
changes of $RDTMC(G)$ occur in the moments of the global discrete time associated with $SMC(G)$, unlike those of
$EDTMC(G)$, which happen only when the current state changes to some {\em different} one, irrespective of the global
time. Therefore, in our case, we can skip the stages of constructing the REDTMC of $G$, denoted by $REDTMC(G)$, from
$EDTMC(G)$, and recovering RSMC of $G$, denoted by $RSMC(G)$, (which is the sought-for DTMC) from $REDTMC(G)$, since we
shall have $RSMC(G)=RDTMC(G)$.

Let $G$ be a dynamic expression and ${\bf P}$ be the TPM for $DTMC(G)$. We reorder the states from $DR(G)$ such that
the first rows and columns of ${\bf P}$ will correspond to the states from $DR_V(G)$ and the last ones will correspond
to the states from $DR_T(G)$. Let $|DR(G)|=n$ and $|DR_T(G)|=m$. The resulting matrix can be decomposed as follows:

$${\bf P}=\left(\begin{array}{cc}
{\bf C} & {\bf D}\\
{\bf E} & {\bf F}
\end{array}\right).$$

The elements of the $(n-m)\times (n-m)$ submatrix ${\bf C}$ are the probabilities to move from vanishing to vanishing
states, and those of the $(n-m)\times m$ submatrix ${\bf D}$ are the probabilities to move from vanishing to tangible
states. The elements of the $m\times (n-m)$ submatrix ${\bf E}$ are the probabilities to move from tangible to
vanishing states, and those of the $m\times m$ submatrix ${\bf F}$ are the probabilities to move from tangible to
tangible states.

The TPM ${\bf P}^{\diamond}$ for $RDTMC(G)$ is the $m\times m$ matrix, calculated as

$${\bf P}^{\diamond}={\bf F}+{\bf E}{\bf G}{\bf D},$$
where the elements of the matrix ${\bf G}$ are the probabilities to move from vanishing to vanishing states in any
number of state changes, without traversal of the tangible states.

If there are no loops among vanishing states then for any vanishing state there exists a value $l\in\nat$ such that
every sequence of state changes that starts in a vanishing state and is longer than $l$ should reach a tangible state.
Thus, $\exists l\in\nat\ \forall k>l\ {\bf C}^k={\bf 0}$ and $\sum_{k=0}^{\infty}{\bf C}^k=\sum_{k=0}^l{\bf C}^k$. If
there are loops among vanishing states then all such loops are supposed to be of ``transient'' rather than
``absorbing'' type, since the latter is treated as a specification error to be corrected, like in \cite{MBCDF95,Bal07}.
We have earlier required that $SMC(G)$ has a single closed communication (which is also ergodic) class of states.
Remember that a communication class of states is their equivalence class w.r.t. communication relation, i.e. a maximal
subset of communicating states. A communication class of states is closed if only the states belonging to it are
accessible from every its state. The ergodic class cannot consist of vanishing states only to avoid ``absorbing'' loops
among them, hence, it contains tangible states as well. Thus, any sequence of vanishing state changes that starts in
the ergodic class will reach a tangible state at some time moment. All the states that do not belong to the ergodic
class should be transient. Hence, any sequence of vanishing state changes that starts in a transient vanishing state
will some time reach either a transient tangible state or a state from the ergodic class \cite{Kul10}. In the latter
case, a tangible state will be reached as well, as argued above. Thus, every sequence of vanishing state changes in
$SMC(G)$ that starts in a vanishing state will exit the set of all vanishing states in the future. This implies that
the probabilities to move from vanishing to vanishing states in $k\in\nat$ state changes, without traversal of tangible
states, will lead to $0$ when $k$ tends to $\infty$. Then we have $\lim_{k\to\infty}{\bf C}^k=\lim_{k\to\infty}({\bf
I}-({\bf I}-{\bf C}))^k={\bf 0}$, hence, ${\bf I}-{\bf C}$ is a non-singular matrix, i.e. its determinant is not equal
to zero. Thus, the inverse matrix of ${\bf I}-{\bf C}$ exists and may be expressed by a Neumann series as
$\sum_{k=0}^{\infty}({\bf I}-({\bf I}-{\bf C}))^k=\sum_{k=0}^{\infty}{\bf C}^k=({\bf I}-{\bf C})^{-1}$. Therefore,

$${\bf G}=\sum_{k=0}^\infty{\bf C}^k=
\left\{\begin{array}{lll}
\sum_{k=0}^l{\bf C}^k, & \exists l\in\nat\ \forall k>l\ {\bf C}^k={\bf 0}, & \mbox{no loops among vanishing states};\\
({\bf I}-{\bf C})^{-1}, & \lim_{k\to\infty}{\bf C}^k={\bf 0}, & \mbox{loops among vanishing states};
\end{array}
\right.$$
where ${\bf 0}$ is the square matrix consisting only of zeros and ${\bf I}$ is the identity matrix, both of order
$n-m$.

For $1\leq i,j\leq m$ and $1\leq k,l\leq n-m$, let ${\cal F}_{ij}$ be the elements of the matrix ${\bf F},\ {\cal
E}_{ik}$ be those of ${\bf E},\ {\cal G}_{kl}$ be those of ${\bf G}$ and ${\cal D}_{lj}$ be those of ${\bf D}$. By
definition, the elements ${\cal P}_{ij}^{\diamond}$ of the matrix ${\bf P}^{\diamond}$ are calculated as

$${\cal P}_{ij}^{\diamond}={\cal F}_{ij}+\sum_{k=1}^{n-m}\sum_{l=1}^{n-m}{\cal E}_{ik}{\cal G}_{kl}{\cal D}_{lj}=
{\cal F}_{ij}+\sum_{k=1}^{n-m}{\cal E}_{ik}\sum_{l=1}^{n-m}{\cal G}_{kl}{\cal D}_{lj}=
{\cal F}_{ij}+\sum_{l=1}^{n-m}{\cal D}_{lj}\sum_{k=1}^{n-m}{\cal E}_{ik}{\cal G}_{kl},$$
i.e. ${\cal P}_{ij}^{\diamond}\ (1\leq i,j\leq m)$ is the total probability to move from the tangible state $s_i$ to
the tangible state $s_j$ in any number of steps, without traversal of tangible states, but possibly going through
vanishing states.

Let $s,\tilde{s}\in DR_T(G)$ such that $s=s_i,\ \tilde{s}=s_j$. The {\em probability to move from $s$ to $\tilde{s}$ in
any number of steps, without traversal of tangible states} is

$$PM^{\diamond}(s,\tilde{s})={\cal P}_{ij}^{\diamond}.$$

\begin{definition}
Let $G$ be a dynamic expression and $[G]_\approx\in DR_T(G)$. The {\em reduced discrete time Markov chain (RDTMC)} of
$G$, denoted by $RDTMC(G)$, has the state space $DR_T(G)$, the initial state $[G]_\approx$ and the transitions
$s\hookrightarrow_{\cal P}\tilde{s}$, where ${\cal P}=PM^{\diamond}(s,\tilde{s})$.
\end{definition}

\noindent RDTMCs of static expressions can be defined as well. For $E\in RegStatExpr$, let
$RDTMC(E)=RDTMC(\overline{E})$.

Let us now try to define $RSMC(G)$ as a ``restriction'' of $SMC(G)$ to its tangible states. Since the sojourn time in
the tangible states of $SMC(G)$ is discrete and geometrically distributed, we can see that $RSMC(G)$ is a DTMC with the
state space $DR_T(G)$, the initial state $[G]_\approx$ and the transitions whose probabilities collect all those in
$SMC(G)$ to move from the tangible to the tangible states, directly or indirectly, namely, by going through its
vanishing states only. Thus, $RSMC(G)$ has the transitions $s\hookrightarrow_{\cal P}\tilde{s}$, where ${\cal
P}=PM^{\diamond}(s,\tilde{s})$, hence, we get $RSMC(G)=RDTMC(G)$.

One can see that $RDTMC(G)$ is constructed from $DTMC(G)$ as follows. All vanishing states and all tran\-sitions to,
from and between them are removed. All transitions between tangible states are preserved. The pro\-babilities of
transitions between tangible states may become greater and new transitions between tangible states may be added, both
iff there exist moves between these tangible states in any number of steps, going through vanishing states only. Thus,
for each sequence of transitions between two tangible states in $DTMC(G)$ there exists a (possibly shorter, since the
eventual passed through vanishing states are removed) sequence between the same states in $RDTMC(G)$ and vice versa. If
$DTMC(G)$ is irreducible then all its states (including tan\-gible ones) communicate, hence, all states of $RDTMC(G)$
communicate as well and it is irreducible. Since both $DTMC(G)$ and $RDTMC(G)$ are finite, they are positive recurrent.
Thus, in case of irreducibility of $DTMC(G)$, each of them has a single stationary PMF. Note that $DTMC(G)$ and/or
$RDTMC(G)$ may be periodic, thus having a unique stationary distribution, but no steady-state (limiting) one. For
example, it may happen that $DTMC(G)$ is aperiodic while $RDTMC(G)$ is periodic due to removing vanishing states from
the former.

Let $DR_T(G)=\{s_1,\ldots ,s_m\}$ and $[G]_\approx\in DR_T(G)$. Then the transient ($k$-step, $k\in\nat$) PMF
$\psi^{\diamond}[k]=(\psi^{\diamond}[k](s_1),\ldots ,\psi^{\diamond}[k](s_m))$ for $RDTMC(G)$ is calculated as

$$\psi^{\diamond}[k]=\psi^{\diamond}[0]({\bf P}^{\diamond})^k,$$
where $\psi^{\diamond}[0]=(\psi^{\diamond}[0](s_1),\ldots ,\psi^{\diamond}[0](s_m))$ is the initial PMF defined as

$$\psi^{\diamond}[0](s_i)=\left\{
\begin{array}{ll}
1, & s_i=[G]_\approx ;\\
0, & \mbox{otherwise}.
\end{array}
\right.$$

Note also that $\psi^{\diamond}[k+1]=\psi^{\diamond}[k]{\bf P}^{\diamond}\ (k\in\nat )$.

The steady-state PMF $\psi^{\diamond}=(\psi^{\diamond}(s_1),\ldots ,\psi^{\diamond}(s_m))$ for $RDTMC(G)$ is a solution
of the equation system

$$\left\{
\begin{array}{l}
\psi^{\diamond}({\bf P}^{\diamond}-{\bf I})={\bf 0}\\
\psi^{\diamond}{\bf 1}^T=1
\end{array}
\right.,$$ where ${\bf I}$ is the identity matrix of order $m$ and ${\bf 0}$ is a row vector of $m$ values $0,\ {\bf 1}$
is that of $m$ values $1$.

Note that the vector $\psi^{\diamond}$ exists and is unique if $RDTMC(G)$ is ergodic. Then $RDTMC(G)$ has a single
steady state, and we have $\psi^{\diamond}=\lim_{k\to\infty}\psi^{\diamond}[k]$.

The zero sojourn time in the vanishing states guarantees that the state changes of $RDTMC(G)$ occur in the moments
of the global discrete time associated with $SMC(G)$, i.e. every such state change occurs after one time unit
delay. Hence, the sojourn time in the tangible states is the same for $RDTMC(G)$ and $SMC(G)$. The state change
probabilities of $RDTMC(G)$ are those to move from tangible to tangible states in any number of steps, without
traversal of the tangible states. Therefore, $RDTMC(G)$ and $SMC(G)$ have the same transient behaviour over the
tangible states, thus, the transient analysis of $SMC(G)$ is possible to accomplish using $RDTMC(G)$.

The following proposition relates the steady-state PMFs for $SMC(G)$ and $RDTMC(G)$. It proves that the steady-state
probabilities of the tangible states coincide for them.

\begin{proposition}
Let $G$ be a dynamic expression, $\varphi$ be the steady-state PMF for $SMC(G)$ and $\psi^{\diamond}$ be the
steady-state PMF for $RDTMC(G)$. Then $\forall s\in DR(G)$

$$\varphi (s)=\left\{
\begin{array}{ll}
\psi^{\diamond}(s), & s\in DR_T(G);\\
0, & s\in DR_V(G).
\end{array}
\right.$$

\label{pmfsmctang.pro}
\end{proposition}
{\em Proof.} To make the proof more clear, we use the following unified notation. ${\bf I}$ denotes the identity
matrices of any size. ${\bf 0}$ denotes square matrices and row vectors of any size and length of values $0$. ${\bf 1}$
denotes square matrices and row vectors of any size and length of values $1$.

Let ${\bf P}$ be the reordered TPM for $DTMC(G)$ and $\psi$ be the steady-state PMF for $DTMC(G)$, i.e. $\psi$ is a
solution of the equation system

$$\left\{
\begin{array}{l}
\psi({\bf P}-{\bf I})={\bf 0}\\
\psi{\bf 1}^T=1
\end{array}
\right..$$

Let $|DR(G)|=n$ and $|DR_T(G)|=m$. The decomposed ${\bf P},\ {\bf P}-{\bf I}$ and $\psi$ are

$${\bf P}=\left(\begin{array}{cc}
{\bf C} & {\bf D}\\
{\bf E} & {\bf F}
\end{array}\right),\
{\bf P}-{\bf I}=\left(\begin{array}{cc}
{\bf C}-{\bf I} & {\bf D}\\
{\bf E} & {\bf F}-{\bf I}
\end{array}\right)\mbox{ and }
\psi =(\psi_V,\psi_T),$$
where $\psi_V=(\psi_1,\ldots ,\psi_{n-m})$ is the subvector of $\psi$ with the steady-state probabilities of vanishing
states and $\psi_T=(\psi_{n-m+1},\ldots ,\psi_n)$ is that with the steady-state probabilities of tangible states.

Then the equation system for $\psi$ is decomposed as follows:

$$\left\{
\begin{array}{l}
\psi_V({\bf C}-{\bf I})+\psi_T{\bf E}={\bf 0}\\
\psi_V{\bf D}+\psi_T({\bf F}-{\bf I})={\bf 0}\\
\psi_V{\bf 1}^T+\psi_T{\bf 1}^T=1
\end{array}
\right..$$

Further, let ${\bf P}^{\diamond}$ be the TPM for $RDTMC(G)$. Then $\psi^{\diamond}$ is a solution of the equation
system

$$\left\{
\begin{array}{l}
\psi^{\diamond}({\bf P}^{\diamond}-{\bf I})={\bf 0}\\
\psi^{\diamond}{\bf 1}^T=1
\end{array}
\right..$$

We have

$${\bf P}^{\diamond}={\bf F}+{\bf E}{\bf G}{\bf D},$$
where the matrix ${\bf G}$ can have two different forms, depending on whether the loops among vanishing states exist,
hence, we consider the two following cases.
\begin{enumerate}

\item There exist {\em no loops among vanishing states}. We have $\exists l\in\nat\ \forall k>l\ {\bf C}^k={\bf 0}$
and ${\bf G}=\sum_{k=0}^l{\bf C}^k$.

Let us right-multiply the first equation of the decomposed equation system for $\psi$ by ${\bf G}$:

$$\psi_V({\bf C}{\bf G}-{\bf G})+\psi_T{\bf E}{\bf G}={\bf 0}.$$

Taking into account that ${\bf G}=\sum_{k=0}^l{\bf C}^k$, we get

$$\psi_V\left(\sum_{k=1}^l{\bf C}^k+{\bf C}^{l+1}-{\bf C}^0-\sum_{k=1}^l{\bf C}^k\right)+\psi_T{\bf E}{\bf G}={\bf
0}.$$

Since ${\bf C}^0={\bf I}$ and ${\bf C}^{l+1}={\bf 0}$, we obtain

$$-\psi_V+\psi_T{\bf E}{\bf G}={\bf 0}\mbox{ and }\psi_V=\psi_T{\bf E}{\bf G}.$$

Let us substitute $\psi_V$ with $\psi_T{\bf E}{\bf G}$ in the second equation of the decomposed equation system for
$\psi$:

$$\psi_T{\bf E}{\bf G}{\bf D}+\psi_T({\bf F}-{\bf I})={\bf 0}\mbox{ and }\psi_T({\bf F}+{\bf E}{\bf G}{\bf D}-{\bf
I})={\bf 0}.$$

Since ${\bf F}+{\bf E}{\bf G}{\bf D}={\bf P}^{\diamond}$, we have

$$\psi_T({\bf P}^{\diamond}-{\bf I})={\bf 0}.$$

\item There exist {\em loops among vanishing states}. We have $\lim_{\to\infty}{\bf C}^k={\bf 0}$ and ${\bf
G}=({\bf I}-{\bf C})^{-1}$.

Let us right-multiply the first equation of the decomposed equation system for $\psi$ by ${\bf G}$:

$$-\psi_V({\bf I}-{\bf C}){\bf G}+\psi_T{\bf E}{\bf G}={\bf 0}.$$

Taking into account that ${\bf G}=({\bf I}-{\bf C})^{-1}$, we get

$$-\psi_V+\psi_T{\bf E}{\bf G}={\bf 0}\mbox{ and }\psi_V=\psi_T{\bf E}{\bf G}.$$

Let us substitute $\psi_V$ with $\psi_T{\bf E}{\bf G}$ in the second equation of the decomposed equation system for
$\psi$:

$$\psi_T{\bf E}{\bf G}{\bf D}+\psi_T({\bf F}-{\bf I})={\bf 0}\mbox{ and }\psi_T({\bf F}+{\bf E}{\bf G}{\bf D}-{\bf
I})={\bf 0}.$$

Since ${\bf F}+{\bf E}{\bf G}{\bf D}={\bf P}^{\diamond}$, we have

$$\psi_T({\bf P}^{\diamond}-{\bf I})={\bf 0}.$$

\end{enumerate}

The third equation $\psi_V{\bf 1}^T+\psi_T{\bf 1}^T=1$ of the decomposed equation system for $\psi$ implies that if
$\psi_V$ has nonzero elements then the sum of the elements of $\psi_T$ is less than one. We normalize $\psi_T$ by
dividing its elements by their sum:

$$v=\frac{1}{\psi_T{\bf 1}^T}\psi_T.$$

It is easy to check that $v$ is a solution of the equation system

$$\left\{
\begin{array}{l}
v({\bf P}^{\diamond}-{\bf I})={\bf 0}\\
v{\bf 1}^T=1
\end{array}
\right.,$$
hence, it is the steady-state PMF for $RDTMC(G)$ and we have

$$\psi^{\diamond}=v=\frac{1}{\psi_T{\bf 1}^T}\psi_T.$$

Note that $\forall s\in DR_T(G)\ \psi_T(s)=\psi (s)$. Then the elements of $\psi^{\diamond}$ are calculated as follows:
$\forall s\in DR_T(G)$

$$\psi^{\diamond}(s)=\frac{\psi_T(s)}{\sum_{\tilde{s}\in DR_T(G)}\psi_T(\tilde{s})}=
\frac{\psi (s)}{\sum_{\tilde{s}\in DR_T(G)}\psi (\tilde{s})}.$$

By Proposition \ref{pmfsmc.pro}, $\forall s\in DR_T(G)\ \varphi (s)=\frac{\psi (s)}{\sum_{\tilde{s}\in DR_T(G)}\psi
(\tilde{s})}$.

Therefore, $\forall s\in DR_T(G)$

$$\varphi (s)=\frac{\psi (s)}{\sum_{\tilde{s}\in DR_T(G)}\psi (\tilde{s})}=\psi^{\diamond}(s).$$

\hfill $\eop$

Thus, to calculate $\varphi$, one can just take all the elements of $\psi^{\diamond}$ as the steady-state probabilities
of the tangible states, instead of abstracting from self-loops to get ${\bf P}^*$ and then $\psi^*$, followed by
weighting by $SJ$ and normalization. Hence, using $RDTMC(G)$ instead of $EDTMC(G)$ allows one to avoid such a
multistage analysis, but constructing ${\bf P}^{\diamond}$ also requires some efforts, including calculating matrix
powers or inverse matrices. Note that $RDTMC(G)$ has self-loops, unlike $EDTMC(G)$, hence, the behaviour of $RDTMC(G)$
may stabilize slower than that of $EDTMC(G)$ (if each of them has a single steady state). On the other hand, ${\bf
P}^{\diamond}$ is smaller and denser matrix than ${\bf P}^*$, since ${\bf P}^{\diamond}$ has additional non-zero
elements not only at the main diagonal, but also many of them outside it. Therefore, in most cases, we have less
time-consuming numerical calculation of $\psi^{\diamond}$ with respect to $\psi^*$. At the same time, the complexity of
the analytical calculation of $\psi^{\diamond}$ with respect to $\psi^*$ depends on the model structure, such as the
number of vanishing states and loops among them, but usually it is lower, since the matrix size reduction plays an
important role in many cases. Hence, for the system models with many immediate activities we normally have a
significant simplification of the solution. At the abstraction level of SMCs, the elimination of vanishing states
decreases their impact to the solution complexity while allowing immediate activities to specify a comprehensible
logical structure of systems at the higher level of transition systems.

\begin{figure}
\begin{center}
\includegraphics{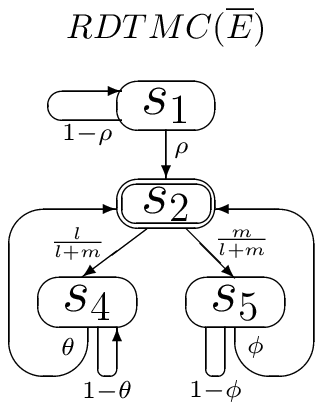}
\end{center}
\caption{The reduced DTMC of $\overline{E}$ for $E=[(\{a\},\rho )*((\{b\},\natural_k^1);(((\{c\},\natural_l^0);
(\{d\},\theta ))\cho ((\{e\},\natural_m^0);(\{f\},\phi ))))*{\sf Stop}]$}
\label{exprdrdtmc.fig}
\end{figure}

\begin{example}
Let $E$ be from Example \ref{tsitchoswim.exm}. Remember that $DR_T(\overline{E})=DR_{ST}(\overline{E})\cup
DR_{WT}(\overline{E})=\{s_1,s_2,s_4,s_5\}$ and $DR_V(\overline{E})=\{s_3\}$. We reorder the states from
$DR(\overline{E})$, by moving the vanishing states to the first positions, as follows: $s_3,s_1,s_2,s_4,s_5$.

The reordered TPM for $DTMC(\overline{E})$ is

$${\bf P}_r=\left(\begin{array}{ccccc}
0 & 0 & 0 & \frac{l}{l+m} & \frac{m}{l+m}\\
0 & 1-\rho & \rho & 0 & 0\\
1 & 0 & 0 & 0 & 0\\
0 & 0 & \theta & 1-\theta & 0\\
0 & 0 & \phi & 0 & 1-\phi
\end{array}\right).$$

The result of the decomposing ${\bf P}_r$ are the matrices

$${\bf C}=0,\
{\bf D}=\left(0,0,\frac{l}{l+m},\frac{m}{l+m}\right),\
{\bf E}=\left(\begin{array}{c}
0\\
1\\
0\\
0
\end{array}\right),\
{\bf F}=\left(\begin{array}{cccc}
1-\rho & \rho & 0 & 0\\
0 & 0 & 0 & 0\\
0 & \theta & 1-\theta & 0\\
0 & \phi & 0 & 1-\phi
\end{array}\right).$$

Since ${\bf C}^1={\bf 0}$, we have $\forall k>0\ {\bf C}^k={\bf 0}$, hence, $l=0$ and there are no loops among
vanishing states. Then

$${\bf G}=\sum_{k=0}^l{\bf C}^k={\bf C}^0={\bf I}.$$

Further, the TPM for $RDTMC(\overline{E})$ is

$${\bf P}^{\diamond}={\bf F}+{\bf E}{\bf G}{\bf D}={\bf F}+{\bf E}{\bf I}{\bf D}={\bf F}+{\bf E}{\bf D}=
\left(\begin{array}{cccc}
1-\rho & \rho & 0 & 0\\
0 & 0 & \frac{l}{l+m} & \frac{m}{l+m}\\
0 & \theta & 1-\theta & 0\\
0 & \phi & 0 & 1-\phi
\end{array}\right).$$

In Figure \ref{exprdrdtmc.fig}, the reduced DTMC $RDTMC(\overline{E})$ is presented. The steady-state PMF for
$RDTMC(\overline{E})$ is

$$\psi^{\diamond}=\frac{1}{\theta\phi (l+m)+\phi l+\theta m}(0,\theta\phi (l+m),\phi l,\theta m).$$

Note that $\psi^{\diamond}=(\psi^{\diamond}(s_1),\psi^{\diamond}(s_2),\psi^{\diamond}(s_4),\psi^{\diamond}(s_5))$. By
Proposition \ref{pmfsmctang.pro}, we have

$$\begin{array}{l}
\varphi (s_1)=0,\\[1mm]
\varphi (s_2)=\frac{\theta\phi (l+m)}{\theta\phi(l+m)+\phi l+\theta m},\\[1mm]
\varphi (s_3)=0,\\[1mm]
\varphi (s_4)=\frac{\phi l}{\theta\phi(l+m)+\phi l+\theta m},\\[1mm]
\varphi (s_5)=\frac{\theta m}{\theta\phi(l+m)+\phi l+\theta m}.
\end{array}$$

Thus, the steady-state PMF for $SMC(\overline{E})$ is

$$\varphi =\frac{1}{\theta\phi (l+m)+\phi l+\theta m}(0,\theta\phi (l+m),0,\phi l,\theta m).$$

This coincides with the result obtained in Example \ref{exprdsmc.exm} with the use of $\psi^*$ and $SJ$.
\label{exprdrdtmc.exm}
\end{example}

\begin{figure}
\begin{center}
\includegraphics{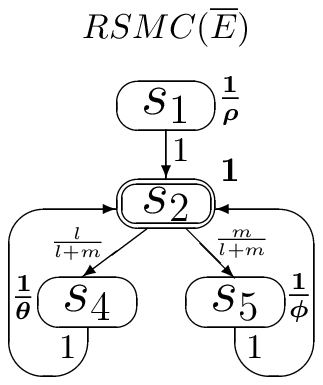}
\end{center}
\caption{The reduced SMC of $\overline{E}$ for $E=[(\{a\},\rho )*((\{b\},\natural_k^1);(((\{c\},\natural_l^0);
(\{d\},\theta ))\cho ((\{e\},\natural_m^0);(\{f\},\phi ))))*{\sf Stop}]$}
\label{exprdrsmc.fig}
\end{figure}

\begin{example}
In Figure \ref{exprdrsmc.fig}, the reduced underlying SMC $RSMC(\overline{E})$ is depicted. The average sojourn times in
the states of the reduced underlying SMC are written next to them in bold font. In spite of the equality
$RSMC(\overline{E})=RDTMC(\overline{E})$, the graphical representation of $RSMC(\overline{E})$ differs from that of
$RDTMC(\overline{E})$, since the former is based on the $REDTMC(\overline{E})$, where each state is decorated with the
{\em positive} average sojourn time of $RSMC(\overline{E})$ in it. $REDTMC(\overline{E})$ is constructed from
$EDTMC(\overline{E})$ in the similar way as $RDTMC(\overline{E})$ is obtained from $DTMC(\overline{E})$. By
construction, the residence time in each state of $RSMC(\overline{E})$ is geometrically distributed. Hence, the
associated parameter of geometrical distribution is uniquely recovered from the average sojourn time in the state.
\label{exprdrsmc.exm}
\end{example}

Note that our reduction of the underlying SMC by eliminating its vanishing states, resulting in the reduced DTMC,
resembles the reduction from \cite{MSTV09} by removing instantaneous states of stochastically discontinuous Markov
reward chains. The latter are ``limits'' of continuous time Markov chains with state rewards and fast transitions when
the rates (speeds) of these transitions tend to infinity, making them immediate. By analogy with this work, we could
consider DTMCs extended with instantaneous states instead of SMCs with geometrically distributed or zero sojourn time
in the states. However, within $dtsdPBC$, we have decided to take SMCs as the underlying stochastic process to be able
to consider not only geometrically distributed and zero residence time in the states, but arbitrary fixed discrete time
delays as well.

\section{Conclusion}
\label{conclusion.sec}

In this paper, we have proposed a discrete time stochastic extension dtsdPBC of PBC, enriched with deterministic
multiactions. The calculus has a parallel step operational semantics, based on labeled probabilistic transition systems
and a denotational semantics in terms of a subclass of LDTSDPNs. A technique of performance evaluation in the framework
of the calculus has been presented that explores the corresponding stochastic process, which is a semi-Markov chain
(SMC). It has been proven that the underlying discrete time Markov chain (DTMC) or its reduction (RDTMC) by eliminating
vanishing states may alternatively and suitably be studied for that purpose. The theory presented has been illustrated
with an extensive series of examples, among which is the travel system application example demonstrating performance
analysis within dtsdPBC.

The advantage of our framework is twofold. First, one can specify in it concurrent composition and synchronization of
(multi)actions, whereas this is not possible in classical Markov chains. Second, algebraic formulas represent processes
in a more compact way than PNs and allow one to apply syntactic transformations and comparisons. Process algebras are
compositional by definition and their operations naturally correspond to operators of programming languages. Hence, it
is much easier to construct a complex model in the algebraic setting than in PNs. The complexity of PNs generated for
practical models in the literature demonstrates that it is not straightforward to construct such PNs directly from the
system specifications. dtsdPBC is well suited for the discrete time applications, whose discrete states change with a
global time tick, such as business processes, neural and transportation networks, computer and communication systems,
timed web services \cite{VC17}, as well as for those, in which the distributed architecture or the concurrency level
should be preserved while modeling and analysis (remember that, in step semantics, we have additional transitions due
to concurrent executions). dtsdPBC is also capable to model and analyze parallel systems with fixed durations of the
typical activities (loading, processing, transfer, repair) and stochastic durations of the randomly occurring
activities (arrival, failure), including industrial, manufacturing, queueing, computing and network systems. The main
advantages of dtsdPBC are the flexible multiaction labels, deterministic multiactions, powerful operations, as well as
a step operational and a Petri net denotational semantics allowing for concurrent execution of activities
(transitions), together with an ability for analytical and parametric performance evaluation.

Future work will consist in constructing a congruence relation for dtsdPBC, i.e. the equivalence that withstands
application of all operations of the algebra. The first possible candidate is a stronger version of $\bis_{ss}$ defined
via transition systems equipped with two extra transitions {\sf skip} and {\sf redo}, like those from sPBC. Moreover,
recursion operation could be added to dtsdPBC to increase further specification power of the algebra.

\end{document}